\documentclass[a4paper,11pt]{article}
\usepackage{jheppub} 
\usepackage{amsmath,amssymb}
\usepackage{mathrsfs}
\usepackage{graphicx}
\usepackage{xcolor}
\usepackage{subfigure}
\usepackage{float}
\usepackage[utf8]{inputenc}
\usepackage{comment}

\def\be{\begin{equation}}
\def\ee{\end{equation}}
\def\ba{\begin{aligned}}
\def\ea{\end{aligned}}
\def\({\left (}
\def\){\right )}
\def\p{\partial}

\def \a {\alpha}
\def \b {\beta}
\def \g {\gamma}

\def \e {\epsilon}

\def \l {\lambda}

\def \s {\sigma}

\def \r {\rho}

\def \p {\partial}

\def \nn {\nonumber}

\def\cA{{\cal A}} \def\cB{{\cal B}} 
\def\cD{{\cal D}} \def\cE{{\cal E}} \def\cF{{\cal F}}

\def\cM{{\cal M}}  \def\cO{{\cal O}}
 \def\cQ{{\cal Q}} \def\cR{{\cal R}}
\def\cS{{\cal S}}  
 \def\cW{{\cal W}}

\numberwithin{equation}{section} 

\title{\boldmath Revisiting 3D Flat Holography: Causality Structure and Modular Flow}

\author{Yuefeng Liu}
\affiliation{Peking University,\\
Department of Physics, Peking University, No.5 Yiheyuan Rd, Beijing 100871, P.R. China}

\emailAdd{yfliu0905@pku.edu.cn}

\abstract{Flat space holography is an open and hard problem existing several different approaches, which may finally turn out to be consistent with each other, in the literature to tackle it. Focusing on how bulk emergent spacetime is encoded in quantum information of null boundaries, we choose a specific toy model called the flat$_3$/BMSFT model, which conjectures the duality between boundary BMS$_3$ invariant field theory and bulk quantum gravity in 3D asymptotic flat spacetimes (AFS), to explore. Aiming to find an entanglement wedge like quantity for single interval and a connected entanglement wedge for multi-intervals in flat$_3$/BMSFT model, we explore the bulk causality structures related to the holographic swing surface proposal through both boundary and bulk local modular flow, make a corresponding decomposition of the global Minkowski spacetime and look at the entanglement phase transition. As a byproduct, we solve the problem about the existence of partial entanglement entropy (PEE) correspondence in this model which is a bit nontrivial due to the unusual behavior of boundary modular flow in BMS$_3$ field theory. Among the literature considering quantum information aspects of flat$_3$/BMSFT model, there are several substantial, unusual but overlooked phenomena which need to be emphasized and revisited to gain more deserved attention. Thus another motivation of this paper is to find where these unusual phenomena come from, and physically show in a manifest way what they may imply. After reading we hope readers can feel sincerely what we present about the above mentioned second aim is more valuable than the mathematical results in the present paper. }

\begin{document}
\maketitle
\flushbottom

\section{Introduction}
\label{sec:intro}

The principle of holography \cite{Maldacena:1997re,Witten:1998qj,Gubser:1998bc} has been successfully used to understand theories of quantum gravity in AdS spacetime and strongly coupled field theory for more than twenty years. Starting from the proposal of Ryu-Takayanagi formula \cite{Ryu:2006bv,Hubeny:2007xt,Nishioka:2009un,Lewkowycz:2013nqa}, there has been profound progress in exploring how spacetime and gravitational dynamics can emerge from boundary quantum information theory \cite{Harlow:2018fse,Almheiri:2020cfm,Bousso:2022ntt}. There are also mixed state entanglement measures generalization of the holographic entanglement entropy, for example the correspondence of entanglement wedge cross section (EWCS) with the reflected entropy \cite{Dutta:2019gen}, entanglement negativity \cite{Kusuki:2019zsp} and balanced partial entanglement entropy (BPE) \cite{Wen:2021qgx}. At the heart of these developments is the entanglement wedge reconstruction, i.e., subregion-subregion duality \cite{Faulkner:2013ana,Jafferis:2015del,Dong:2016eik}, which states that bulk operators in the entanglement wedge $\cW_{\cE}[\cA]$ can be decoded from the operator algebra in the causal domain $D[\cA]$ of the dual CFT.
Another important approach highlighting the emergence of bulk locality and causality is the concept of modular flow \cite{Faulkner:2017vdd,Witten:2018zxz,Faulkner:2018faa}. Operator reconstruction with modular flow allow one to reach everywhere inside of entanglement wedge $\cW_{\cE}[\cA]$, which can reach far beyond the causal horizons. 
For the entanglement wedge $\cW_{\cE}[\cA]$, there are two equivalent definitions in AdS/CFT holography \cite{Headrick:2014cta}. One is the bulk domain of dependence of homology surface $\cR_{\cA}$ (defined in \eqref{adsdecop}) which interpolates between boundary interval $\cA$ and corresponding HRT surface. Another definition is the bulk region bounded by bifurcating horizons of HRT surface on one side and boundary causal domain $D[\cA]$ on the other side. Whether these properties are universal to general holographic theories beyond AdS/CFT? One motivation of the paper is exploring the story of entanglement wedge $\cW_{\cE}[\cA]$ like quantity in a toy model of 3D flat holography, i.e., the flat$_3$/BMSFT model, using mainly modular flow and other refined tools due to the complications and subtleties here. 

Although people face both practical and philosophical difficulties in formulating flat version of AdS holography, there has been interesting work on understanding holography in asymptotically flat spacetimes at the early days of AdS/CFT  \cite{Susskind:1998vk,Polchinski:1999ry,Giddings:1999jq,Arcioni:2003td,deBoer:2003vf}. In recent years, there is a delightful re-booming about this problem. One key role in this re-booming is the bottom-up approach called celestial holography \cite{Strominger:2017zoo,Pasterski:2021raf}, which proposes a correspondence between 4D gravity theories in asymptotically flat spacetimes (AFS) and 2D celestial conformal field theories (CCFT) living on the celestial sphere at null infinity due to advances in understanding the soft theorems and asymptotic structure of AFS\cite{He:2014laa,He:2015zea,Cheung:2016iub,Pasterski:2016qvg,Pasterski:2017kqt}. Bulk S-matrix elements, when written in boost eigenstate basis, can be reinterpreted as correlation functions in 2D conformal field theory. Thus very powerful CFT techniques, such as operator product expansion \cite{Fotopoulos:2019tpe,Pate:2019lpp,Banerjee:2020kaa,Costello:2022upu,Hu:2022bpa}, conformal block decomposition \cite{Atanasov:2021cje,Fan:2021isc}, crossing symmetry \cite{Mizera:2022sln}, can all be used to explore properties of celestial CFT. Also using this kind of language lead people find new $w_{1+\infty}$ symmetries \cite{Strominger:2021lvk}. However it is rather vague at this stage that how much and in which aspects the 2D CCFT would differ from the usual 2D Virasoro CFT. Moreover, the emergence of bulk flat spacetime from boundary degree of freedom of celestial CFT seems to be at least complicated. However viewing recent fascinating developments of understanding how bulk spacetime emerge from boundary and the nature of holographic map as a quantum error correction code \cite{Almheiri:2014lwa}, it is very attractive to see whether similar nature of holographic duality in AdS hold true in flat case. The most important object underlying this kind of story in AdS/CFT is the entanglement wedge $\cW_{\cE}[\cA]$ dual to specific boundary subregion, and we would like to find similar objects in flat holography. With this curiosity, we turn to another bottom-up approach called Carrollian holography, which is more similar to the usual AdS/CFT set up. Due to the matter or gravitational radiation, the gravitational charge defined at null infinity would be non-conserved \cite{Barnich:2011mi,Trautman:1958zdi,Donnay:2022wvx}. Thus we focus on 3D flat bulk with pure Einstein action, more specifically, the flat$_3$/BMSFT model. The analysis in this paper are special to 3D AFS, and we make a first but essential step in this direction. Our explorations are complementary to the main trend in literature on flat holography focusing on S-matrix elements, ward identities as well as asymptotic symmetry, and focus more on dynamical gravity aspects of holographic duality. Note another more information theoretic approach about flat holography \cite{Laddha:2020kvp} explores how quantum information is stored at null infinity by using boundary operator algebra. Also there are interesting and illuminating works trying to link Carrollian holography approach with celestial holography approach in 4D flat spacetime \cite{Bagchi:2022emh,Donnay:2022aba}.

\begin{figure}
    \centering
    \subfigure[]{
    \includegraphics[width=0.46\textwidth]{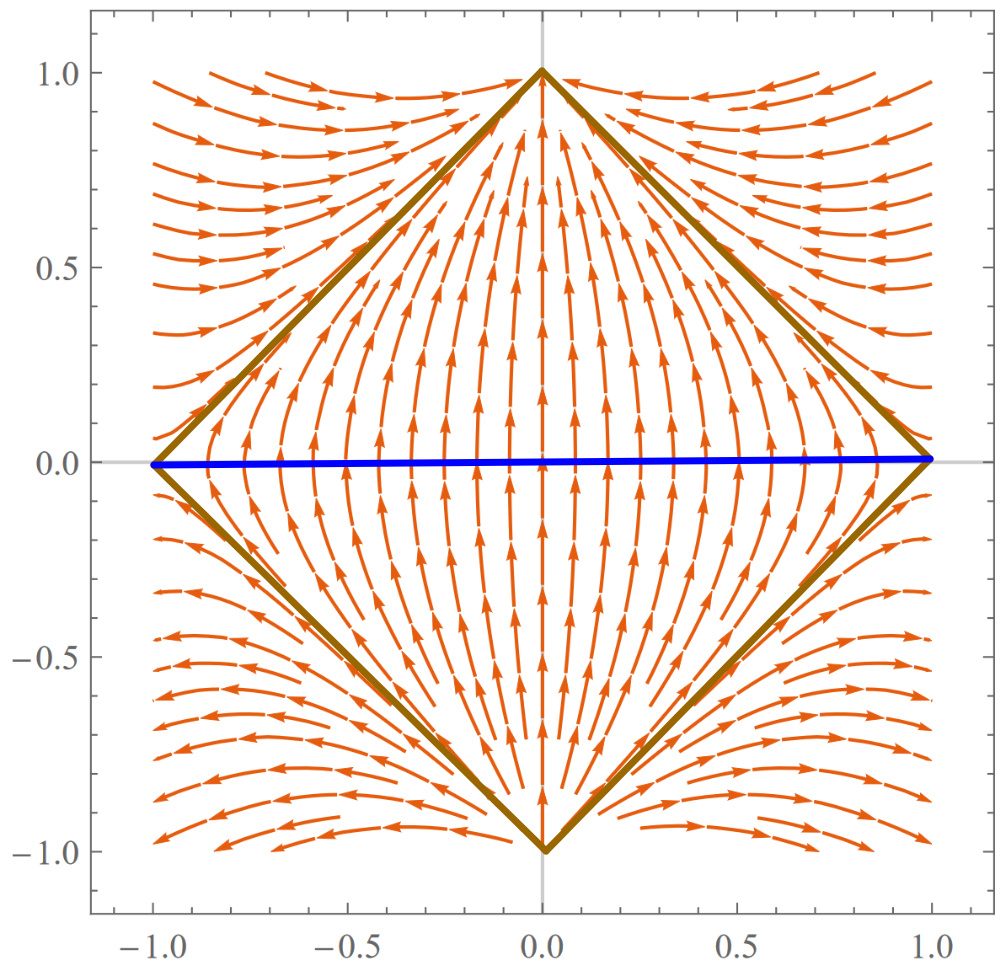} }
    \subfigure[]{
    \includegraphics[width=0.44\textwidth]{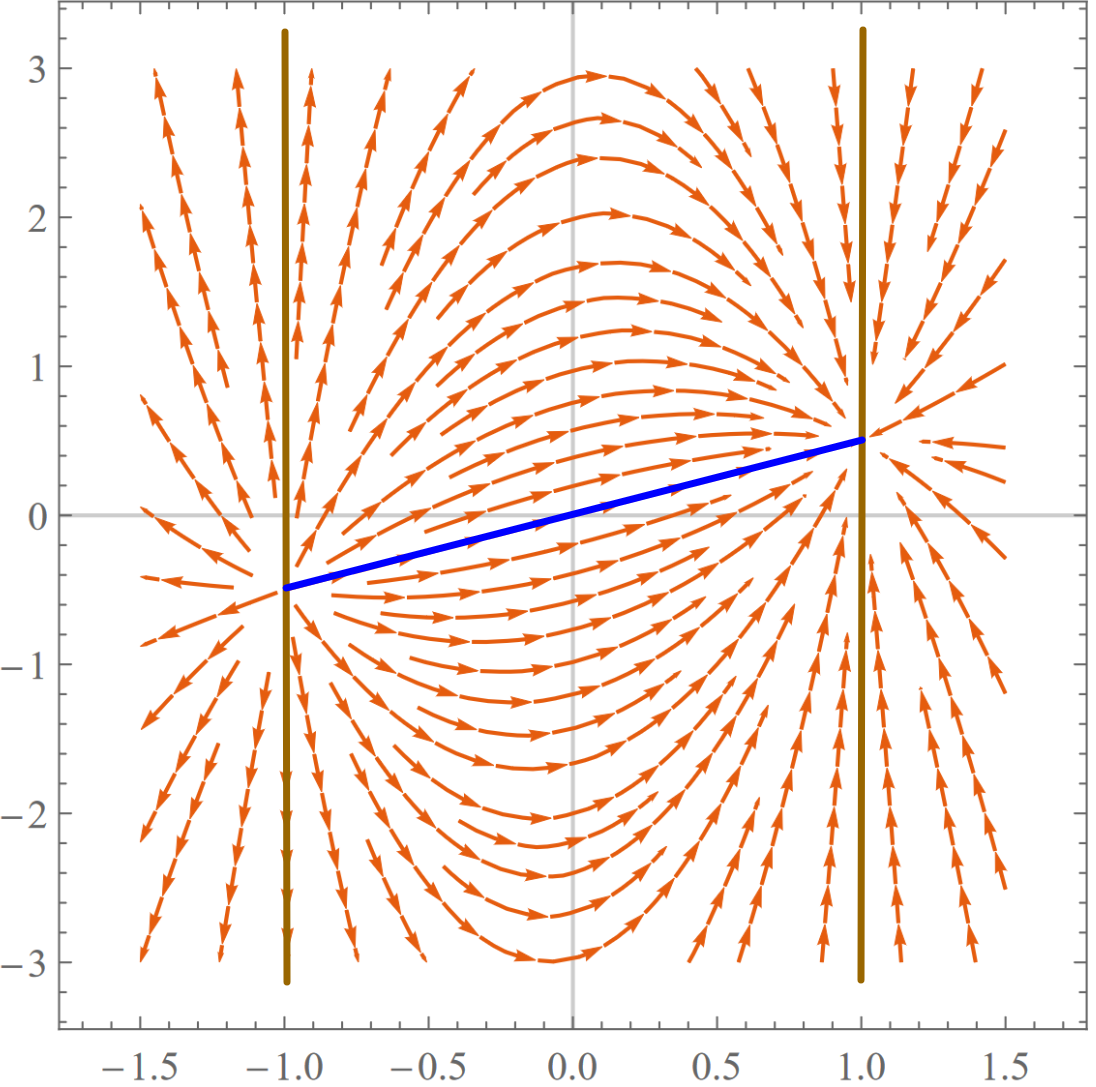} }
    \caption{The two figures show the modular flow of 2D CFT and BMS$_3$ field theory separately. Blue lines are boundary intervals $\cA$ and brown lines denote the boundary $\p D[\cA]$ of the causal domain $D[\cA]$. We can see that the direction of modular time of BMS$_3$ field theory is rather different than the ones in CFT case.}
    \label{fig:cftbmsflwo}
\end{figure}

The Carrollian holography has been proved to be successful in 3D case, and we would like to link its story to the limiting problem of AdS/CFT. Whether flat holography can be understood as a limiting case of AdS/CFT is still an open problem. Although there are plenty of research working on extracting perturbative S-matrix elements from AdS correlators \cite{Penedones:2010ue,Fitzpatrick:2011ia,deGioia:2022fcn}, they are limited to very special state in the Hilbert space of quantum gravity in AFS (if it exists!). At the level of asymptotic symmetry algebra (ASA) in the machinery of holography, \cite{Bagchi:2010zz,Bagchi:2012cy} made an interesting observation that ASA of 3D AFS, i.e., the BMS$_3$ algebra \cite{Bondi:1962px,Sachs:1962wk}, can be obtained as a ultra-relativistic limit of
2D conformal algebra. Starting from these works,
flat$_3$/BMSFT model go through several non-trivial checks, such as reproducing thermal entropy in the bulk from a Cardy-like formula at the boundary \cite{Barnich:2012xq,Bagchi:2012xr}, reproducing characters of the BMS$_3$ group from one loop partition function of 3D flat gravity \cite{Barnich:2015mui}, reproducing BMS$_3$ blocks from bulk geodesic Feynman diagrams \cite{Hijano:2018nhq} and reproducing boundary entanglement entropy from bulk swing surface \cite{Jiang:2017ecm}. Considering the holographic entanglement entropy, \cite{Apolo:2020bld,Apolo:2020qjm} updated the generalized Rindler method used in \cite{Jiang:2017ecm} for limited regions and states to more general cases using approximated modular flow method and a general swing surface proposal. Note that in flat$_3$/BMSFT model, the vacuum state in the Hilbert space of quantum gravitational theory of AFS is assumed to be unique. This may contradict with the lessons learned from celestial holography and soft theorems that the vacuum state of 4D AFS is infinitely degenerated due to supertranslation and soft gravitons \cite{Bagchi:2016bcd}. 

In purpose of exploring how boundary information are related to bulk subregion, similar to the aim of \cite{Laddha:2020kvp} but working in a more concrete model, we use various tools developed in AdS/CFT to try to find the analogue of the entanglement wedge $\cW_{\cE}[\cA]$ in flat holography. In the literature there are some checks about matching of holographic reflected entropy and balanced partial entanglement entropy (BPE) \cite{Basu:2021awn, Basak:2022cjs,Camargo:2022mme} in flat$_3$/BMSFT model, however the calculations and physical conclusions need to be reconsidered. The necessity of this revisiting originates from the following facts: 1). The flat$_3$/BMSFT model has a Lorentzian bulk spacetime and no Euclidean path integral apply here. So we should consult to a coordinate invariant codimension zero bulk region to define the entanglement wedge not a coordinate non-invariant codimension one bulk surface, which should be different from the usual AdS/CFT case; 2). Also in the literature, they only studied very limited symmetric boundary two intervals, which gave them an unrealistic illusion about their results. Although these works are interesting, actually no well defined connected entanglement wedge has ever been established in works \cite{Basu:2021awn, Basak:2022cjs,Camargo:2022mme} and the related ones. It turns out that for generic boundary non-symmetric two intervals the entanglement wedge cross section (EWCS) can totally locate outside the naive expected connected entanglement wedge, see Figure \ref{fig:ewcstwo}, although the numerical values can mysterious match with each other. Then the natural question is what's the entanglement wedge related to single boundary interval? What's the connected entanglement wedge related to multi-boundary intervals? If we can not specify the parameter range of intervals related to the connected entanglement wedge, what is the meaning of bulk EWCS we compute? Actually according to the results obtained in this paper, finding accurate answers to the above questions are a rather non-trivial task. Even the existence of normal entanglement wedge should be reasonably questioned because of the existence of negative holographic entanglement entropy noticed already in \cite{Apolo:2020bld}. Actually not only can the holographic entanglement entropy be negative, all entanglement measures calculated in flat$_3$/BMSFT model including the holographic reflected entropy, holographic entanglement negativity and balanced partial entanglement entropy (BPE) can have negative values. This is in fact a general and unique property of flat$_3$/BMSFT model, which has not been given sufficient attention in the literature. Viewing from field theory, the negative value may come from non-unitary property. From bulk side, we make the key observation that the negativeness is just a reflection about the unique structure of local modular flow of BMS$_3$ field theory. More intuitively we can see from Figure \ref{fig:cftbmsflwo}, the modular evolution along local modular flow of BMS field theory are quite different from the modular evolution of CFT which is consistent with the global time defined on the whole 2D plane. Part of the results in this paper can be viewed as the bulk manifestation of this unusual boundary modular flow behavior. 

Another tool we use to find the analogue of entanglement wedge in flat holography is the PEE (partial entanglement entropy) correspondence \cite{Wen:2018whg}, which proposes to give a fine version of RT formula. We don't comment on the physical foundation of this proposal, but rather view it as a useful tool to manifest various aspects of holographic duality when we have a local modular flow. From PEE correspondence, people can derive the balanced partial entanglement entropy (BPE) and EWCS correspondence \cite{Wen:2021qgx}. However in literature \cite{Camargo:2022mme,Basu:2022nyl} people just observed the match of BPE and EWCS without giving a more basic proof about PEE correspondence in flat$_3$/BMSFT model. The reason can again be traced back to the curious behavior of modular flow of BMS$_3$ field theory, which makes the finding of corresponding bulk point from modular flow method rather unclear with less physical intuition. As a byproduct, we solve the existence of PEE correspondence in flat$_3$/BMSFT model by using the intersection of swing surfaces (first method) and rewriting the original modular flow correspondence (second method). We find exact match between these two methods and these are solid mathematical results in this paper. The way we solve the above existing problem on modular flow method in flat$_3$/BMSFT model is to explicitly manifest the degree of freedom using our rewriting, and this is also a good place to see the subtlety in flat$_3$/BMSFT model.

Although we find more structures of the correspondence between boundary and bulk modular flow, as well as make a bulk decomposition of global flat$_3$ related to single boundary interval $\cA$, we fail to specify which bulk subregion is the most natural entanglement wedge in this model. Especially in two intervals case considering the connected entanglement wedge, the confirmed results are only made on the field side. Also we can't integrate the implications of general lesson learned from negative entanglement entropy and swing surface penetrating phenomena into the exploration of entanglement wedge. We hope that bringing these fundamental issues to researchers in a clearer and more obvious way is more valuable than the mathematical results presented in this paper. This is our second motivation for writing this paper. 

The structure of the paper is organized as follows. In section \ref{sect2}, we review the flat$_3$/BMSFT model, the general prescription of swing surface proposal for single interval holographic entanglement entropy and the PEE correspondence in AdS/CFT with useful comments at the end of each subsection. In section \ref{sectthree}, we explicitly draw the Penrose diagram of the quotient manifolds, i.e., the zero mode solutions, in AFS and show a subtle issue of the order of taking infinite limit. Then we show where the bulk negative sign of holographic entanglement entropy come from using Noether charge formalism. Finally we present observations about EWCS for general boundary intervals which manifest the loopholes of arguments in the literature. In section \ref{section4}, we mathematically and pictorially analyze the behavior of bifurcating horizons related to both finite bench $\g$ and the infinite bifurcating surface $\g_{
\xi}$. Then we decompose the global flat$_3$ spacetime into four disconnected parts using both the past and future bifurcating horizons. Intersection of swing surface method and bulk boundary modular flow correspondence method for deriving the PEE correspondence in flat$_3$/BMSFT model are presented. In the last by comparing with the entanglement wedge $\cW_{\cE}[\cA]$ in AdS/CFT case, we show the subtleties of  the flat$_3$/BMSFT model. In section \ref{section5}, we analyze the entanglement phase transition of two intervals on the boundary side and entanglement wedge nesting (EWN) property in the bulk side. In section \ref{section6}, we discuss two important open questions unique to flat$_3$/BMSFT model which are observed in section \ref{sect2}. We collect several additional results in Appendices. In Appendix \ref{appendixa}, we give a complete derivation of two disjoint interval reflected entropy in BMS$_3$ field theory with explicit calculations about the three point coefficient. This is a necessary but missing part of the calculations about reflected entropy in \cite{Basak:2022cjs}, which sincerely pointed out by \cite{Basu:2022nyl}.
Appendix \ref{appendixb} repeat the analytic analysis of the Poincar\'e vacuum for the $M>0$ zero mode backgrounds including the bifurcating horizon in Penrose diagram and the entanglement phase transition.

\section{flat$_3$/BMSFT model and PEE correspondence}
\label{sect2}

In 3D asymptotically flat spacetimes (AFS) Einstein gravity admits consistent boundary conditions at future null infinity $\mathscr{I}^{+}$, where the finite dimensional Poincar\'e isometry group is enhanced to infinite dimensional asymptotic symmetry group, i.e., the BMS$_3$ group \cite{Bondi:1962px,Sachs:1962wk}. These facts lead people to conjecture that there is a toy model of flat holography, dubbed flat$_{3}$/BMSFT model, which maps between Einstein gravity in 3D AFS and BMS invariant field theories at 2D conformal boundary. Intuitively the topology of the null boundary of 3D AFS is $S^{1}\times \mathbb{R}$ with $\mathbb{R}$ the null direction. And BMS$_3$ group include super-translation which is coordinate dependent translation along the null direction and super-rotation which is the diffeomorphism of $S^{1}$. This section includes a self-contained review of 2D BMS$_3$ invariant field theory with more emphasize on entanglement entropy, the development of the general swing surface proposal and the PEE correspondence in AdS/CFT holography. At the end of each subsection, useful comments on the subtleties are presented. 

\subsection{BMS$_3$ invariant field theory}
  
BMS invariant field theory is a class of 2D ultra relativistic quantum field theories invariant under following spacetime reparametrizations  \cite{Bagchi:2010zz,Barnich:2012xq}, 
\be  \Tilde{x} =f(x), \qquad  \Tilde{y}=y f'(x)+g(x)  \label{bmst}   \ee
where $f(x)$ and $g(x)$ are arbitrary functions, and $(x,y)$ are coordinates of the plane the field theory lives.  The infinitesimal BMS transformations are generated by following Fourier modes, 
\be
    l_{n}=-x^{n+1} \partial_{x}-(n+1)y x^{n} \partial_{y} \quad  m_{n}=-x^{n+1} \partial_{y} \label{bmscoord}
\ee
Under Lie bracket they form the BMS$_3$ algebra without the centrally extension term. While the generators $L_{m}$ and $M_{n}$ implementing local coordinate transformations \eqref{bmscoord} on quantum fields form the centrally extended BMS$_3$ algebra,
\begin{align}
   & [L_{n},L_{m}] = (n-m)L_{m+n}+\frac{c_{L}}{12}n(n^2-1)  \delta_{m+n,0} \notag  \\
    & [L_{n},M_{m}] = (n-m)M_{m+n}+\frac{c_{M}}{12}n(n^2-1) \delta_{m+n,0} \notag  \\
    &[M_{n},M_{m}] =0 \label{bmsalge}
\end{align}
where $c_{L}$ and $c_{M}$ are the central charges. The Einstein-Hilbert gravity in flat holography are expected to be dual to a BMS field theory with central charges $c_{L}=0, c_{M}=\frac{3}{G}$, while the field theory with more general value of central charges could be constructed by adding a Chern-Simons term to the Einstein-Hilbert action.

The generators $L_{m}$ and $M_{n}$ are also called BMS charges on the plane, which are the Fourier modes of the conserved currents $T(x)$ and $M(x)$,
\begin{align}
    L_{n}&=\frac{1}{2\pi i} \oint \left( 
     x^{n+1} T(x)+(n+1) x^{n} y M(x)     \right)\\
    M_{n}&=\frac{1}{2\pi i} \oint   
     x^{n+1} M(x)  
\end{align} 
where $\oint$ can be seen as the contour integral of the complexified $x$ coordinates. The conserved currents $T(x)$ and $M(x)$ generating the coordinate transformations (\ref{bmst}) transform under the transformations as   
\begin{align}
   \Tilde{M}(x) &= f'^{2}M(\Tilde{x})+\frac{c_{M}}{12} \{f,x\}  \label{bmst1} \\
   \Tilde{T}(x,y)&=f'^{2} T(\Tilde{x},\Tilde{y})+ 2f'(g'+y f'') M(\Tilde{x})+\frac{c_{L}}{12} \{f,x \}+\frac{c_{M}}{12}\left( y\frac{d}{dx} \{f,x\}+f'^{2} \frac{\partial^{3}g}{\partial f^3} \right) \notag
\end{align}
where $\{,\}$ denotes the ordinary Schwarzian derivative and the last term denotes the BMS Schwarzian derivative 
\begin{align}
    & \{f,x \}=\frac{f'''}{f'}-\frac{3}{2}\left(  \frac{f''}{f'}\right)^2 \\
    & f'^{2}\frac{\partial^{3} g}{\partial f^{3}}=f'^{-1} \left( g'''-g'\frac{f'''}{f'}-3f'' \left(\frac{g'}{f'} \right)' \right)
\end{align}
The infinite dimensional BMS$_3$ algebra not only have the singlet version of the highest weight representation (HWR), but also the multiplet version of the HWR \cite{Hao:2021urq}. In the singlet version of HWR, a local primary operator $\mathcal{O}(0,0)$ at the origin is labelled by the eigenvalues of generators $L_{0}$ and $M_{0}$ which are the center of the BMS$_3$ symmetry algebra \eqref{bmsalge},
\be  [L_{0},\mathcal{O}]=\Delta \mathcal{O}, \quad [M_{0},\mathcal{O}]=\xi \mathcal{O} \label{wetchg} \ee
where $\Delta$ denotes the conformal weight and $\xi$ denotes the boost charge. The HWR respect the following conditions,
\be [L_{n},\mathcal{O}]=0, \quad [M_{n},\mathcal{O}]=0 ,\quad   n>0\ee
The singlet primary operators transform under finite transformation (\ref{bmst}) as follows,
\be \Tilde{O}(\Tilde{x},\Tilde{y})=|f'|^{-\Delta} e^{-\xi}\frac{g'+y f''}{f'} O(x,y) \label{bmst2} \ee
By requiring the vacuum to be invariant under the global symmetry of BMS$_3$ field theory, the correlation functions on the plane have the following form,
\begin{align}
    \langle \phi(x_{1},y_{1})& \phi(x_{2},y_{2}) \rangle = \delta_{\Delta_{1},\Delta_{2}} \delta_{\xi_{1},\xi_{2}} |x_{21}|^{-2\Delta_{1}} e^{-2 \xi_{1} \frac{y_{21}}{x_{21}}} \label{twopcl}  \\
    \langle \phi_{1} \phi_{2} \phi_{3} \rangle &= \frac{c_{123}}{ |x_{12}|^{\Delta_{123}} |x_{23}|^{\Delta_{231}} |x_{31}|^{\Delta_{312}} }
    e^{-\xi_{123} \frac{y_{12}}{x_{12}} -\xi_{312} \frac{y_{13}}{x_{13}} -\xi_{231} \frac{y_{23}}{x_{23}}}  \\
    \langle \phi_{1} \phi_{2} \phi_{3} \phi_{4} \rangle&=e^{\xi_{12} \left( \frac{t_{24}}{x_{24}}- \frac{t_{14}}{x_{14}} \right)+\xi_{34} \left( \frac{t_{14}}{x_{14}}- \frac{t_{13}}{x_{13}}\right) -(\xi_{1}+\xi_{2}) \frac{t_{12}}{x_{12}}-(\xi_{3}+\xi_{4}) \frac{t_{34}}{x_{34}}   }  \notag \\
     \qquad \qquad &\times \left| \frac{x_{24}}{x_{14}} \right|^{\Delta_{12}} \left| \frac{x_{14}}{x_{13}} \right|^{\Delta_{34}} \frac{\mathcal{F}(x,t)}{\left|x_{12}\right|^{\Delta_{1}+\Delta_{2}} \left|x_{34}\right|^{\Delta_{3}+\Delta_{4}} } \label{ffourpint}
\end{align}
where two point function of primary operators are properly normalized, $c_{123}$ is the coefficient of three-point function encoding dynamical information of the BMS$_3$ field theory and 
\be x_{ij}=x_{i}-x_{j}, \quad  y_{ij}=y_{i}-y_{j}, \quad  \Delta_{ijk}=\Delta_{i}+\Delta_{j}-\Delta_{k} ,\quad \xi_{ijk}=\xi_{i}+\xi_{j}-\xi_{k}.  \ee  The $x$ and $t$ appearing in function $\cF(x,t)$ are BMS invariant cross ratios
\be x=\frac{x_{12}x_{34}}{x_{13}x_{24}}, \quad \frac{t}{x}=\frac{t_{12}}{x_{12}}+\frac{t_{34}}{x_{34}}-\frac{t_{13}}{x_{13}}-\frac{t_{24}}{x_{24}} \label{bmscrt} \ee

Entanglement entropy of BMS$_{3}$ field theory was first considered in \cite{Bagchi:2014iea} using algebraic twist operator method \cite{Calabrese:2009qy}. By generalizing the Rindler method to BMS$_3$ invariant field theory, \cite{Jiang:2017ecm} not only gets the consistent entanglement entropy through an explicitly local modular flow expression, but also extends the calculation into the bulk getting the swing surface picture. We list some results related to entanglement entropy here for later convenience. BMS$_3$ field theory is not Lorentz invariant, thus a general spatial interval $ \{ (x_1,y_1)$ $(x_2,y_2)\} $ instead of an equal time interval is need to show the dependence of entanglement entropy on the choice of frame. In the plane vacuum state, the conformal weight $\Delta$ and boost charge $\xi$ in cyclic orbifold $\mathbf{Z}_{n}$ are,
\be \Delta_{n}=\frac{c_L}{24}(n-\frac{1}{n}), \quad \xi_{n}=\frac{c_M}{24}(n-\frac{1}{n}).  \ee 
Then the partition function of the replica manifold $\Sigma_n$ and the entanglement entropy of single interval $\cA$ are,
\begin{align}
    & \text{Tr} \rho_{A}^{n} = k_n \langle \sigma_{n}(x_1,y_1) \Tilde{\sigma}_{n}(x_2,y_2) \rangle^{plane}_{\text{BMS}^{\otimes n}}
    =k_n |x_{21}|^{-\frac{c_L}{12}(n-\frac{1}{n})} e^{-\frac{c_M}{12}(n-\frac{1}{n}) \frac{y_{21}}{x_{21}}} \\
    & S_{EE;vac}^{BMS} =-\lim_{n\to 1}\partial_n \text{Tr} \rho_{A}^{n}= \frac{c_L}{6} \log{\frac{\left|x_{21}\right|}{\delta_{x}}}+ \frac{c_M}{6} \left(\frac{y_{21}}{x_{21}} \right)    \label{eevac}
\end{align}
where $\delta_{x}>0$ is the $x$ direction UV regulator introduced by $k_{n}$ relating to the regularization of the divergent partition function $\text{Tr} \rho_{A}^{n}$. We can see from \eqref{eevac} that for the bulk correspondence of Einstein gravity with $c_L=0$, the entanglement entropy $S_{EE;vac}^{BMS}$ can be negative due to possible different sign of $y_{21}$ and $x_{21}$. When considering the finite temperature state on the plane, we use the following general thermal periodicity, 
\be (\phi,u) \sim (\phi+i\beta_{\phi}, u-i \beta_{u} ) \label{bmscylinder} \ee 
where $\{\phi,u\}$ denote the coordinates on the thermal cylinder. We can use the BMS conformal transformation to map from plane to cylinder \cite{Jiang:2017ecm}, 
\be  x  =e^{\frac{2 \pi \phi}{\beta_{\phi}}}, \quad y =\frac{2 \pi }{\beta_{\phi}} e^{\frac{2 \pi \phi}{\beta_{\phi}}} \left( \phi \frac{\beta_{u}}{\beta_{\phi}}+u  \right) \label{bmsfiniteT} \ee
The two point function of twist operators evaluated on this cylinder then is given by 
\be
     \langle \sigma_{n}(\phi_1,u_1) \Tilde{\sigma}_{n}(\phi_2,u_2) \rangle^{cylinder}_{\text{BMS}^{\otimes n}} =k_{n} \big( \frac{\beta_{\phi}}{\pi \delta_{\phi} } \sinh{\frac{\pi \left|\phi_{21}\right| }{\beta_{\phi}}} \big)^{-2 \Delta_{n}} e^{-2 \xi_{n} \left( \frac{ \pi(u_{21}+\frac{\beta_{u}}{\beta_{\phi}}\phi_{21})}{\beta_{\phi}} \coth{\frac{\pi \phi_{21}}{\beta_{\phi}}}-\frac{\beta_{u}}{\beta_{\phi}} \right) } \nn
\ee
Thus the entanglement entropy of single interval $\cA$ in the thermal state is,  
\begin{align}
    & S_{EE;thermal}^{BMS}
     = \frac{c_L}{6} \log{ \big( \frac{\beta_{\phi} }{\pi \delta_{\phi}} \sinh{\frac{\pi \left|\phi_{21}\right|}{\beta_{\phi}
   }} \big)}+\frac{c_M}{6} \bigg[ \frac{\pi}{\beta_{\phi}}\big( u_{21}+\frac{\beta_{u}}{\beta_{\phi}} \phi_{21} \big)\coth{\big(\frac{\pi \phi_{21}}{\beta_{\phi}} \big)} -\frac{\beta_{u}}{\beta_{\phi}}   \bigg] \label{eethermal}
\end{align}   

\textbf{Comments:} A key assumption in the above calculations is that the twist operators $\sigma_{n}$ and $\Tilde{\sigma}_{n}$ belong to the singlet version of HWR of the BMS$_3$ algebra. It was noticed and proved in \cite{Hao:2021urq} that primary fields can also be organized in a Jordan chain and form a multiplet which is a reducible but indecomposable module together with their descendants. Cyclic $\mathbf{Z}_{n}$ Orbifold theory of BMS field on replicated Carrollian geometry is a much unexplored area and could go beyond the usual expectations. For example, see \cite{Chen:2022fte} for the subtleties about the Orbifold theory of 2D WCFT living in Newton-Cartan geometry. It is possible that the twist operators in BMS orbifold theory belong to the multiplet version of HWR, thus affect the final answer of entanglement entropy \eqref{eevac} and \eqref{eethermal}.

\subsection{Swing Surface Proposal \label{subsecswing} }

Instead of directly extend the HRT formula into the flat$_3$/BMSFT model, \cite{Jiang:2017ecm} derive the swing surface configuration by the exact correspondence between  boundary local modular flow generators and bulk killing vector fields. The advantage of this method is the holographic dictionary of the entanglement entropy is automatically consistent. While the disadvantage of this method is that the local modular flow can only exist for special entangled regions and special states. Due to the above reasons, \cite{Apolo:2020bld,Apolo:2020qjm} update the above method and propose a more general prescription to get the swing surface $\g_{\cA}$ for holographic entanglement entropy by using the approximate modular flow in both the boundary and the bulk. Let us summarize the main steps of these developments in flat$_3$/BMSFT model following \cite{Apolo:2020bld,Apolo:2020qjm} closely. 

For an interval $\cA$ on the vacuum state of BMS$_3$ field theory, which is dual to a spacetime in the bulk invariant under the same set of symmetries, we can find a consistent boundary flow generator $\zeta$ and the corresponding bulk Killing field $\xi$,
\be \zeta=\sum_{i} a_{i} h_i \equiv \partial_{\tau_{B}}, \quad \xi=\sum_{i} a_{i} H_i \equiv \partial_{\tau_{b}} \ee 
where $a_{i}$ are parameters depending on the entangling region $\cA$, $\tau_{B},\tau_{b}$ are boundary and bulk Rindler time respectively satisfying periodicity conditions
\be \tau_{B,b} \sim \tau_{B,b}+2\pi i , \label{perdbb} \ee 
$h_i$ are the vacuum symmetry generators defined on the boundary, and $H_i$ are the corresponding bulk Killing vectors under the dictionary of flat$_3$/BMSFT holography satisfying $H_{i} \lvert_{\partial \mathcal{M}}=h_{i}$. The boundary modular flow generator $\zeta$ need satisfy following conditions: 1). The transformation $x\to \Tilde{x}=f(x)$ is a symmetry of the field theory where the domain of $f(x)$ is the causal domain $D[\cA]$; 2). The transformation $x\to \Tilde{x}$ is invariant under a pure imaginary (thermal) identification $ \(\Tilde{x}^{1},\Tilde{x}^{2} \) \sim \( \Tilde{x}^{1}+ i \Tilde{\beta}^{1}  ,\Tilde{x}^{2}+ i \Tilde{\beta}^{2} \)$; 3). The one parameter flow $\Tilde{x}^{i}[s]$ generated by $\zeta$ through the exponential map $e^{s\zeta}$ leave the causal domain $D[\cA]$ and its boundary $\p D[\cA] $ invariant when $s$ is real. 

The periodicity \eqref{perdbb} is considered as a thermal identification, which implies that the bulk modular flow generator $\xi$ features bifurcating Killing horizons with surface gravity $2\pi$. We denote the bifurcating surface as $\g_{\xi}$ and two Killing horizons as $N_{l,r}$, which satisfy
\begin{align}
    & \xi \lvert_{\g_{\xi}}=0 \label{bfc1} \\
 & \nabla^{\mu}\xi^{\nu} \lvert_{\g_{\xi}}=2 \pi n^{\mu \nu } \label{bfc2} \\
 & \xi^{\nu} \nabla_{\nu}\xi^{\mu} \lvert_{N_{l,r}}=\pm 2 \pi \xi^{\mu}  \label{bfc3} \\
 & \xi_{[\mu}\nabla_{\nu}\xi_{\lambda]}\lvert_{N_{l,r}}=0 \label{bfc4}
\end{align}
where $n^{\mu}=n_{1}^{\mu}n_{2}^{\nu}-n_{2}^{\mu}n_{1}^{\nu} $ is the unit vector binormal to $\g_{\xi}$. \eqref{bfc1} follows from the fact that $\g_{\xi}$ is an extremal surface; \eqref{bfc2} shows that $\xi$ is the boost generator in the local Rindler frame near $\g_{\xi}$; \eqref{bfc3} means the surface gravity is indeed a constant value $2\pi$; \eqref{bfc4} is the Frobenius’ theorem guaranteeing the vector field is hypersurface orthogonal. Finally in this special case, the ropes $\g_{(p)}$ of the swing surface $\g_{\cA}$ are null geodesics generated by bulk modular flow while the bench $\g$ of the swing surface is the set of fixed points of bulk modular flow generator $\xi$ that extremizes the distance between the ropes. 

For more general states and boundary configurations, we need to consult to the approximate modular flow $\zeta^{(p)}$, which on the 2D boundary can be obtained from the expressions for single intervals on the vacuum by sending the other endpoint to infinity. For each end point of interval, it is possible to find the null geodesic whose tangent vector is an asymptotic Killing vector reducing to $\zeta^{(p)}$ at the conformal boundary. Then the general swing surface is the minimal extremal surface bounded by these null geodesics. One major difference compared to standard HRT surface in AdS/CFT is that, in flat$_3$/BMSFT model the fixed points of the boundary modular  flow $\zeta$ are not the fixed points of the bulk modular flow $\xi$ meaning the bifurcating surface $\g_{\xi}$ is not attached to the interval $\cA$ at the boundary.\\

\textbf{comments:} \begin{itemize}
    \item In \cite{Apolo:2020bld} the authors propose that the holographic dictionary of entanglement entropy in flat$_3$/BMSFT model is the area of swing surface $\g_{\cA}$ (or bench $\g$)
   \be S_{\cA}=\frac{\text{Area}(\g_{\cA})}{4G}=\text{min} \, \underset{X_{\mathcal{A}} \sim \mathcal{A}}{\text{ext}} \frac{ \text{Area}(X_{\mathcal{A})} }{4G},  \quad X_{\mathcal{A}}= X \cup \gamma_{b \partial} \ee 
   However as also been noticed by the same paper, the problem is how can the area term have negative value \eqref{eevac}. In the next section, we would find that the holographic dictionary of $S_{\cA}$ need more elements not just the pure gravity area property. 
  \item The descriptions of bifurcating horizons \footnote{Note our notations are different from those in \cite{Apolo:2020bld}. } $N_{l,r}$ in section $(2.3)$ of \cite{Apolo:2020bld} are not precise. According to the results in section \ref{sectthree}, the bifurcating horizons $N_{l,r}$ connected to boundary interval $\cA$ are both future directed and the killing horizons emitted from the finite bench $\g$ only touch the future null infinity $\mathscr{I}^{+}$ at two single points. Note that these unusual features seem to be unique for flat$_3$/BMSFT model, which is not due to the swing surface construction, see \cite{Chen:2022fte,Wen:2018whg} for comparison.
\end{itemize}

\subsection{PEE correspondence}

\begin{figure}
    \centering
    \subfigure[]{
    \includegraphics[width=0.5\textwidth]{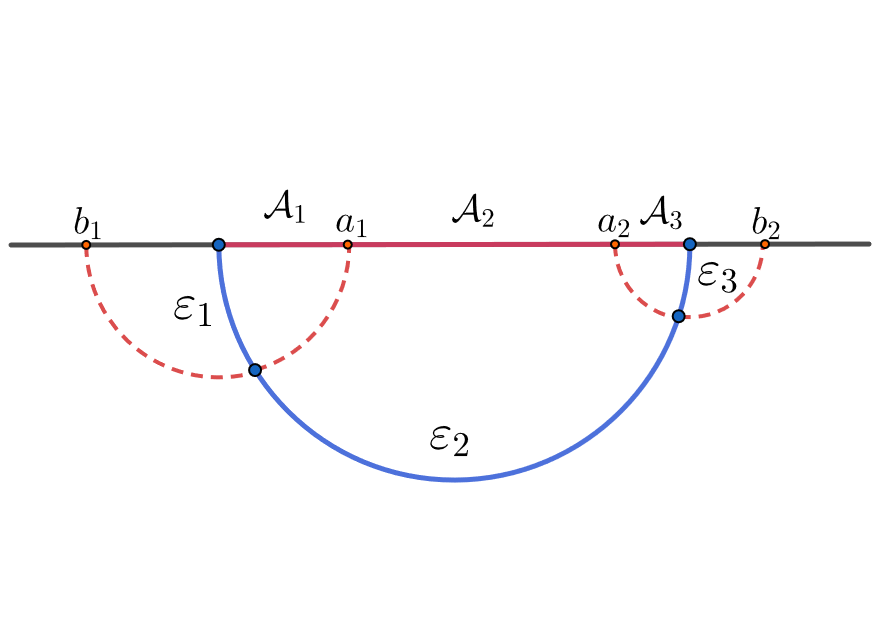} \label{fig:adspeecd} }
    \subfigure[]{
    \includegraphics[width=0.38\textwidth]{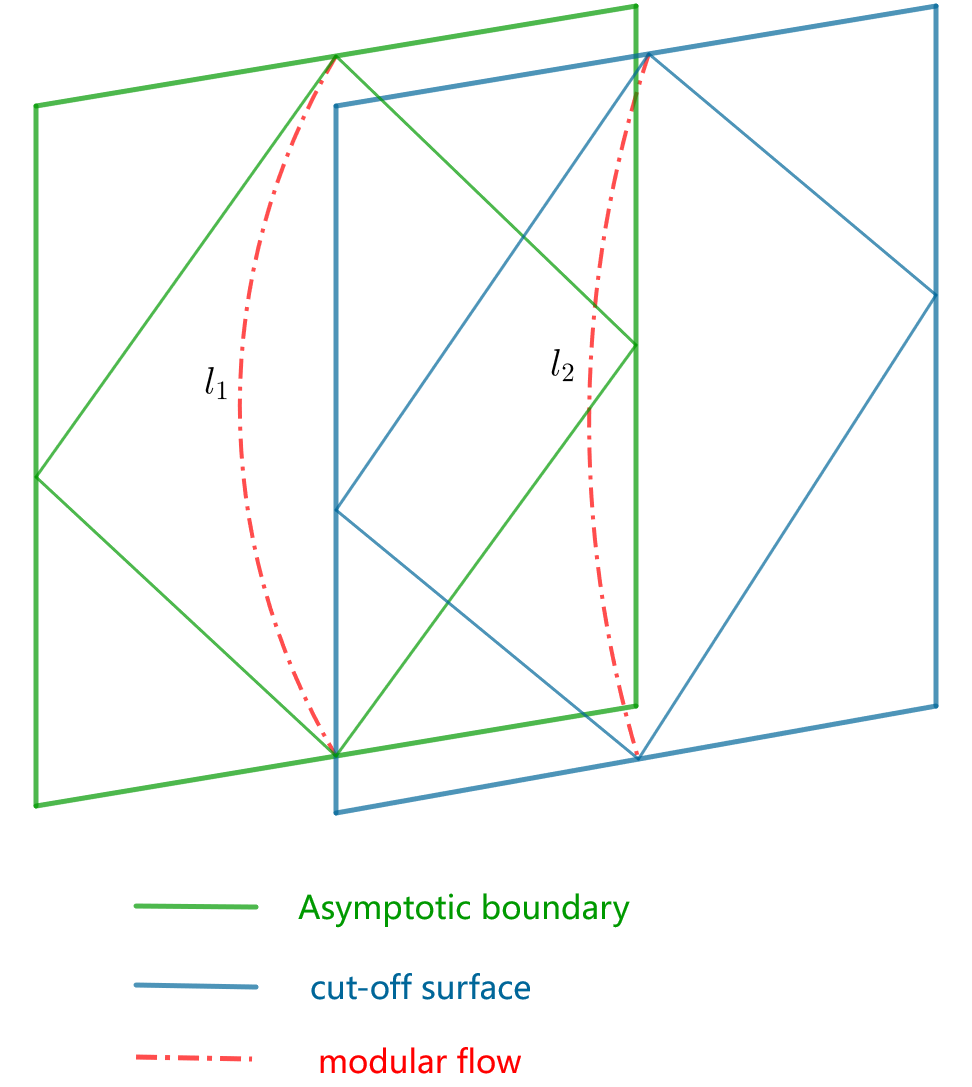} \label{fig:peefrdom}   }
    \caption{Figure \ref{fig:adspeecd} shows the PEE correspondence in AdS/CFT set up. Dashed red lines connecting $a_1,b_1$ and $a_2,b_2$ are the  corresponding RT surfaces normal to the one related to interval $\cA=\cA_1 \cup \cA_2 \cup \cA_3 $. Points in $\cA_{i}$ correspond to points in $\varepsilon_{i}$. Figure \ref{fig:peefrdom} shows one explicitly way of manifesting the degree of freedom in determining the PEE bulk corresponding point from the modular flow method. The modular flow line $l_{2}$ on the cut-off surface which corresponds to the specific modular flow line $l_1$ on asymptotic boundary has to be chosen to fix the freedom.}
\end{figure}

Since in this paper we just take the partial entanglement entropy (PEE) correspondence as a useful tool, so we only present the most basic elements of them. Please see \cite{Wen:2018whg,Wen:2021qgx} for more physical interpretations.

In \cite{Wen:2018whg}, the author made two proposals about the holographic dictionary (PEE correspondence) for the entanglement contour of a single interval in the context of AdS$_3$/CFT$_2$. The first proposal states that the partial entanglement entropy $S_{A}(A_{2})$, see Figure \ref{fig:adspeecd}, is given by a linear combination of entanglement entropies of relevant subsets inside interval $\cA$  for general 2D theories 
\be S_{A}(A_2)=\frac{1}{2} \( S_{A_1 \cup A_2}+S_{A_2 \cup A_3}-S_{A_1}-S_{A_3}  \)  \ee 
The second proposal is a fine structure analysis about the entanglement wedge through boundary and bulk modular flow, which is used in this paper as a way to explore the "entanglement wedge" of flat$_3$/BMSFT model. This bulk and boundary one-to-one correspondence can also be obtained by intersection of RT surfaces, see Figure \ref{fig:adspeecd}. Finally the holographic dictionary about PEE says that
\be  S_{A}(A_{i})=\frac{\text{Length}\( \varepsilon_{i} \)}{4G}  \ee 

\textbf{Comments:} As rigorously said by \cite{Wen:2021qgx} the bulk modular flows exactly settle at the boundary when they approach the boundary, so there are no orbits in the bulk. Thus to really find a boundary and bulk correspondence through local modular flow method, we should choose a cut-off surface, see Figure \ref{fig:peefrdom}. Then there is a degree of freedom in choosing which modular flow line in the chosen cut-off surface correspond to a specific modular flow line at the asymptotic boundary. This freedom not only can affect the bulk point of PEE correspondence, i.e., $\epsilon_{i}$ in Figure \ref{fig:peefrdom}, but also can affect the shape of the line between boundary and bulk corresponding points. \cite{Wen:2018whg} make a good proposal on how to fix this freedom in AdS$_3$/CFT$_2$, but how to fix this freedom in flat$_3$/BMSFT model is not clear. As a byproduct in this paper, we find there is a consistent way to fix the d.o.f. in flat case although the underlying physical reasons need further study. In any case, this is not the focus of this paper and the intersection of RT like surfaces way turn out to be more general and less uncertain.

\section{Quotient manifolds and observations }
\label{sectthree}

After a summary of the phase space of Einstein gravity solutions under the consistent asymptotic boundary conditions \eqref{asybc} in flat$_3$/BMSFT model, we give the exact Penrose diagrams (not cartoon pictures) of the zero mode solutions, which are quotient manifolds of global Minkowski spacetime (the global flat$_3$). To gain more intuition, a subtle issue about drawing boundary causal domain $D[\cA]$ on compact Penrose diagram of the covering global flat$_3$ is shown. Then two key observations about holographic entanglement entropy (swing surface) and holographic reflected entropy related (EWCS) are presented. One is about how to derive the "negative" sign of holographic entanglement entropy and reflected entropy in the bulk, the other one is about whether the finite bench or the infinite bifurcating surface is more fundamental, or at least more useful, in finding the "entanglement wedge" of this model. The first two subsections are preliminary to understand the explorations in this paper, the last two subsections are a revisit of the results in \cite{Apolo:2020bld,Basu:2021awn}. We try to extract some general lessons from these new observations about flat holography.  

More precisely, the above mentioned asymptotic boundary conditions near future null infinity \cite{Barnich:2010eb} in the retarded Bondi coordinates $(u,r,\phi)$ is
\be g_{rr}=0, \;\;  g_{ru}=-1+\cO \left(\frac{1}{r} \right), \;\; g_{r\phi}=0, \;\; g_{u \phi}=\cO(1), \;\;  g_{uu}=\cO(1), \;\; g_{\phi \phi}=r^2   \label{asybc}  \ee 
where $\phi \sim \phi +2 \pi$. The phase space of solutions to pure Einstein's equations in Bondi gauge is parametrized by two periodic functions $\Theta(\phi)$ and $\Xi(\phi)$ such that 
\be ds^2=\Theta(\phi) du^2-2 du dr +2 \left[ \Xi(\phi)+\frac{1}{2}u\partial_{\phi} \Theta(\phi) \right] du d\phi+r^2 d\phi^2,\label{flat3}\ee
where the null infinity is located at $r\to \infty$. The zero mode solutions with constant $\Theta(\phi)=M$ and $\Xi(\phi)=J/2$ describe some classical backgrounds of spacetime and are our main interest. With the convention $8G=1$, the parameters $M$ and $J$ correspond to the canonical energy and the angular momentum of the spacetime. In particular, the $M=-1,$ $J=0$ solution corresponds to the global Minkowski vacuum, the $-1<M<0$ solutions correspond to the conical defect geometries, and the $M=J=0$ solution, called the null orbifold, is supposed to be the analogue of zero temperature BTZ. Solutions with $M>0$ is usually referred to as flat cosmological solutions (FSC) and have Cauchy horizons. This fact can be seen clearly in the ADM form \cite{Barnich:2012aw} of the zero mode metric
\be ds^2=-\left(-M+\frac{J^2}{4r^2} \right)^2 dt^2+ \left(-M+\frac{J^2}{4r^2} \right)^{-2} dr^2+r^2 \left( d\varphi+ \frac{J}{2 r^2}dt \right)^{2} \ee 
which implies that the Cauchy horizon is located at 
\be r_{H} \equiv |r_c|=\frac{|J|}{2 \sqrt{M}} \label{cauhor} \ee 
We are also interested in the vacuum state in flat Poincar\'e coordinates with metric
\be ds^2=-2 du dr +r^2 dz^2, \quad \quad r \ge0, \; u\in (-\infty, \infty), \; z\in (-\infty, \infty) \label{poincarevacuum} \ee
which can be obtained by decompactify the angular direction $\phi$ of $M=J=0$ null orbifold solution. This is the flat limit of the Poincar\'e patch of AdS$_3$.

\subsection{Boundaries and Horizons\label{bdhorz} }

In 3D pure Einstein gravity, there is no propagating degree of freedom. The only way to construct different solutions is by taking quotient. Like the BTZ black holes are the discrete quotient manifolds of global AdS$_3$, the above mentioned zero mode backgrounds, i.e., the $M=0,J=0$ (the Poincar\'e vacuum), $M>0$ (FSC) and $M<0$ (including global Minkowski) zero mode backgrounds, are also the discrete quotient manifolds of global flat$_3$. 

\begin{figure}
  \label{bdhz}
    \centering
    \subfigure[]{
    \includegraphics[width=0.46\textwidth]{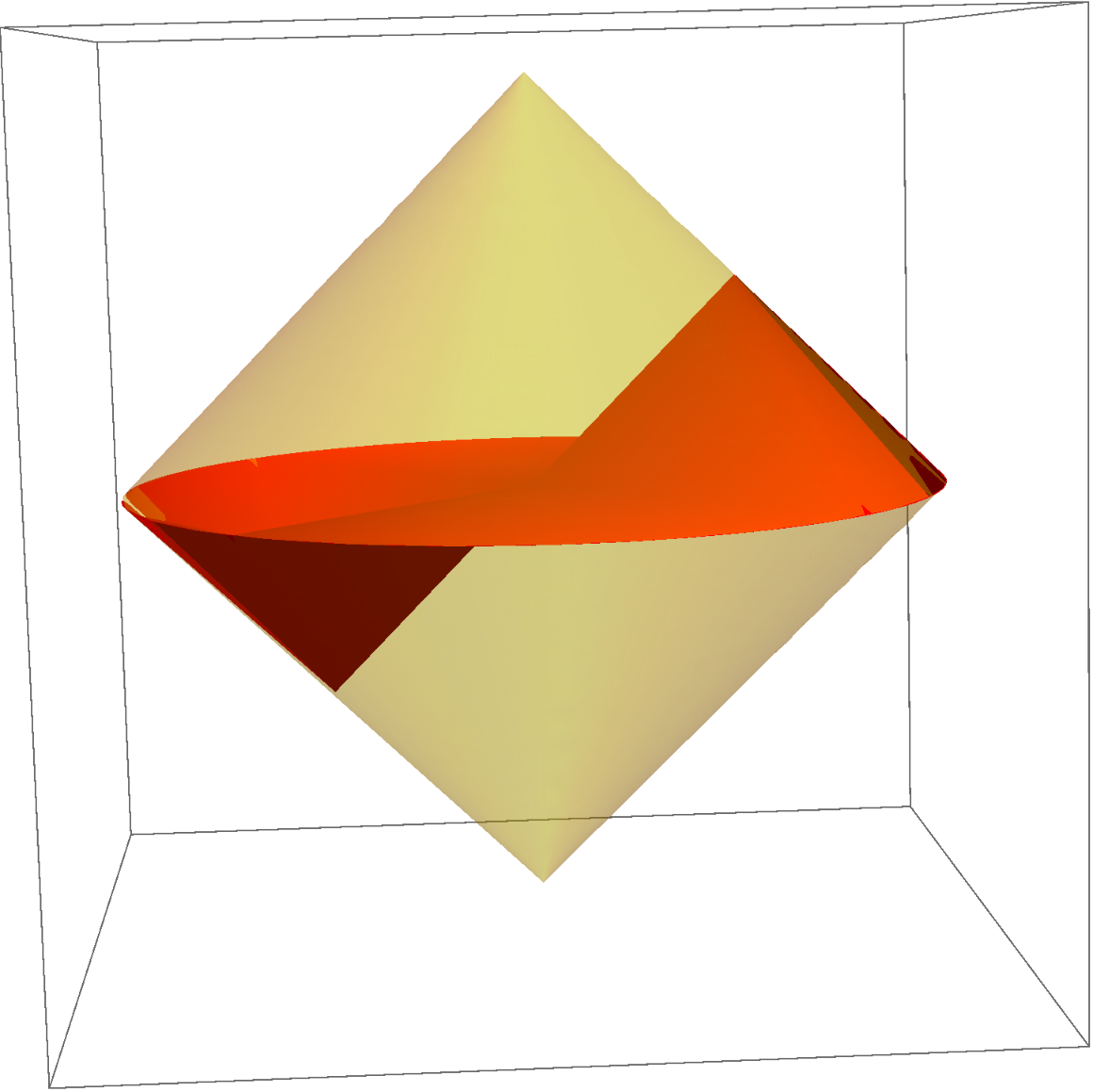} \label{fig:bdhz1} }
    \subfigure[]{
    \includegraphics[width=0.44\textwidth]{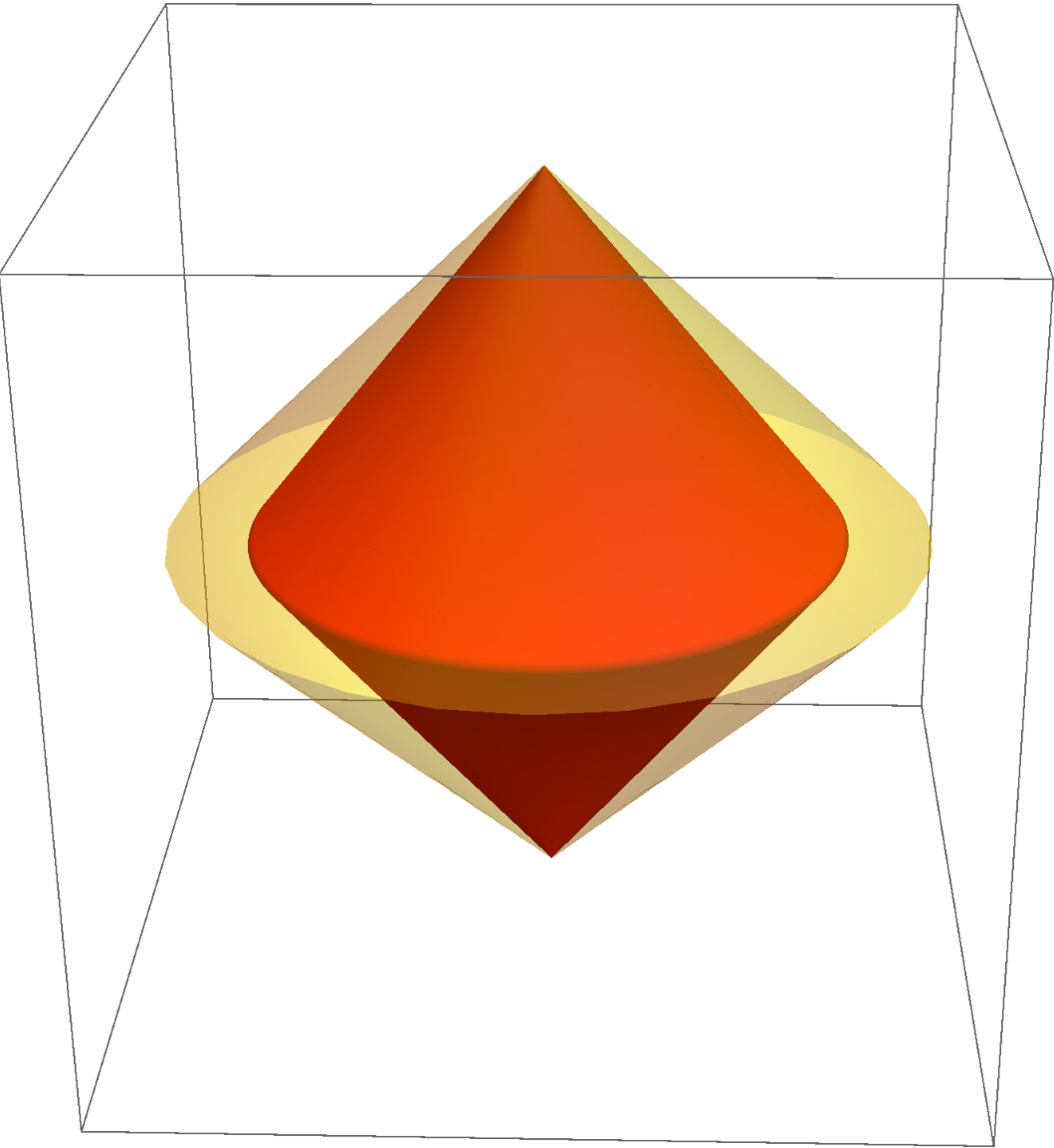} \label{fig:bdhz2} }
     \subfigure[]{
    \includegraphics[width=0.45\textwidth]{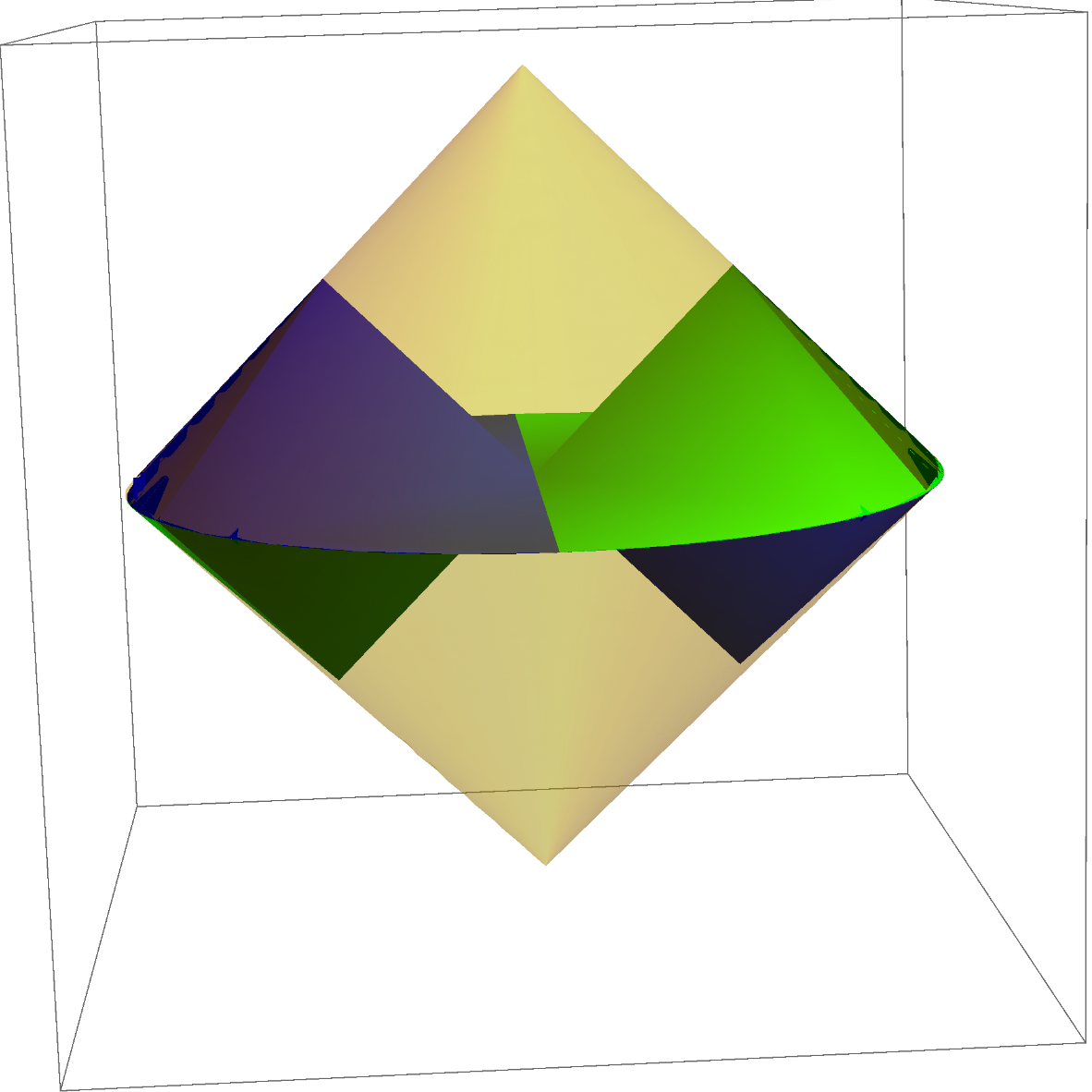} \label{fig:bdhz3} }
    \caption{The figures show the Penrose diagrams of the quotient manifolds, i.e., the Poincar\'e vacuum, $M<0$ zero mode backgrounds and $M>0$ zero mode backgrounds, in 3D global Minkowski spacetime respectively. All yellow light cones are asymptotic boundaries of the global flat$_3$; null red surface in figure \ref{fig:bdhz1} denotes the boundary $t+y \ge 0$ of the Poincar\'e vacuum; red surface (not null) in figure \ref{fig:bdhz2} denotes the boundary $x^2+y^2=2 r_{c}^2/(-M)$ of $M<0$ zero mode background; null green surface in figure \ref{fig:bdhz3} denotes Cauchy horizon $t-x>0$ and null purple surface denotes Cauchy horizon $t+x>0$ of $M>0$ zero mode background.}
\end{figure}

For each case we first give the coordinate transformations \cite{Apolo:2020bld} that map that zero mode background to the global flat3 with coordinates $(t,x,y)$, then point out the corresponding boundaries or horizons of these quotient manifolds.

\begin{itemize}
    \item \textbf{The Poincar\'e vacuum} The coordinate transformations \footnote{Note the transformations here are different with \cite{Apolo:2020bld,Jiang:2017ecm}, which depend on the boundary interval $\cA$.} are,
    \be t= \frac{(\alpha^{2} + 4 z^2)r}{4 \alpha} + \frac{2 u}{\alpha}, \quad x= z r+ \frac{\beta}{\alpha}, \quad y= \frac{(\alpha^{2}-4z^{2})r}{4 \alpha}-\frac{2 u}{\alpha} \label{pointranf} \ee 
   for any value of $\alpha$ and $\beta$. Without loss of generality we choose $\alpha=1,\beta=0$ in this paper. In order to see the boundary of spacetime clearly, we need an inverse coordinate transformations  
    \be u=\frac{t^2-x^2-y^2}{4(t+y)},\quad r=2(t+y), \quad z=\frac{x}{2(t+y)}. \ee
    So the Poincar\'e vacuum cover only the $t+y \ge 0$ part of the global Minkowski spacetime, see Figure \ref{fig:bdhz1} .   

   \item \textbf{$M<0$ zero mode backgrounds} The coordinate transformations are: 
   \be
   \ba
    & t=\frac{1}{\sqrt{-M}}\left( r-M u-\sqrt{-M} r_c \phi \right)  \\
    & x=\frac{1}{\sqrt{-M}}\left[ r \cos{ \sqrt{-M} \phi }-r_c \sin{ \sqrt{-M} \phi } \right] \label{mlessztr} \\
    & y=\frac{1}{\sqrt{-M}}\left[ r \sin{ \sqrt{-M} \phi }-r_c \cos{ \sqrt{-M} \phi } \right] 
   \ea \ee 
    So we have the relation $x^2+y^2=(r^2+r_{c}^{2})/(-M)$, which leads to the boundary location $x^2+y^2=2 r_{c}^{2}/(-M)$, See Figure \ref{fig:bdhz2}. Note that if we have $J=r_c=0$, i.e., the whole Minkowski spacetime, then the codimension one boundary in Figure \ref{fig:bdhz2} would shrink to one dimensional line with $ x=0,y=0$ excluding nothing from the global flat3 and consistent with the expectation.

    \item \textbf{$M>0$ zero mode backgrounds}  The coordinate transformations are:
    \begin{align}
        u=&\frac{1}{M} \left( r-\sqrt{M} y -\sqrt{M}r_{c} \phi \right), \quad r=\pm \sqrt{M(t^2-x^2)+r_{c}^{2} } \nn \\
        & \phi=-\frac{1}{M} \log{\left[  \frac{ \sqrt{M}(t-x) }{r+r_c}  \right] }=\frac{1}{M} \log{\left[  \frac{ \sqrt{M}(t+x) }{r-r_c}  \right] } \label{mgztranf}
    \end{align}
  The spacetime region with $r>r_c$ exterior to Cauchy horizon locating at $r=r_c$ cover the parameter range
  \be  t-x>0, \quad \quad t+x>0, \ee
 while the interior of the Cauchy horizon $0<r<r_c$ cover the parameter range
  \be  t-x>0, \quad \quad t+x<0. \ee 
 In Figure \ref{fig:bdhz3}, the exterior of Cauchy horizon is above both the green and blue surfaces, and the interior of Cauchy horizon is the right part of the region enclosed by both the green and blue surfaces.

\end{itemize}

Note that if we draw the swing surface in the above compact Penrose diagrams, the finite bench would always penetrate the boundaries or horizons of the original spacetime. This curious phenomena is discussed in the last section.

\subsection{Order of taking the Infinity Limit}

The order of taking the infinite limit is a subtle issue in mathematics. Here is a good example due to the infinite range of both $r$ and $u$ coordinates in Bondi gauge. For the above coordinate transformations, if we take the limit $r > |u|\to \infty$ of Bondi coordinate, which is equivalent to keep coordinate $u$ a finite but arbitrary value and taking $r$ to infinity, the BMS field theory would always live on the future null infinity $\mathscr{I}^+$ for all $M=0$, $M>0$ and $M<0$ cases. However, we know that the global Minkowski spacetime with $M=-1,J=0$ contain not only the future null infinity $\mathscr{I}^+$ but also the past null infinity $\mathscr{I}^{-}$, which actually comes from another limit \footnote{For simplicity in this subsection, we omit the mathematical proof of the statements, which can be obtained by following the same route as \eqref{rinfty} and \eqref{endpoint}.} $ u< -r \to  -\infty $. With the above observation, we explore the following limits 
\be 1). \;\; u<-r \to -\infty, \quad \quad 2). \;\; u> r \to \infty \label{unulimt} \ee 
in $M=0$, $M>0$ and $M<0$ solutions separately and summarize the new phenomena. 

 \begin{figure}
    \centering
    \subfigure[]{
    \includegraphics[width=0.42\textwidth]{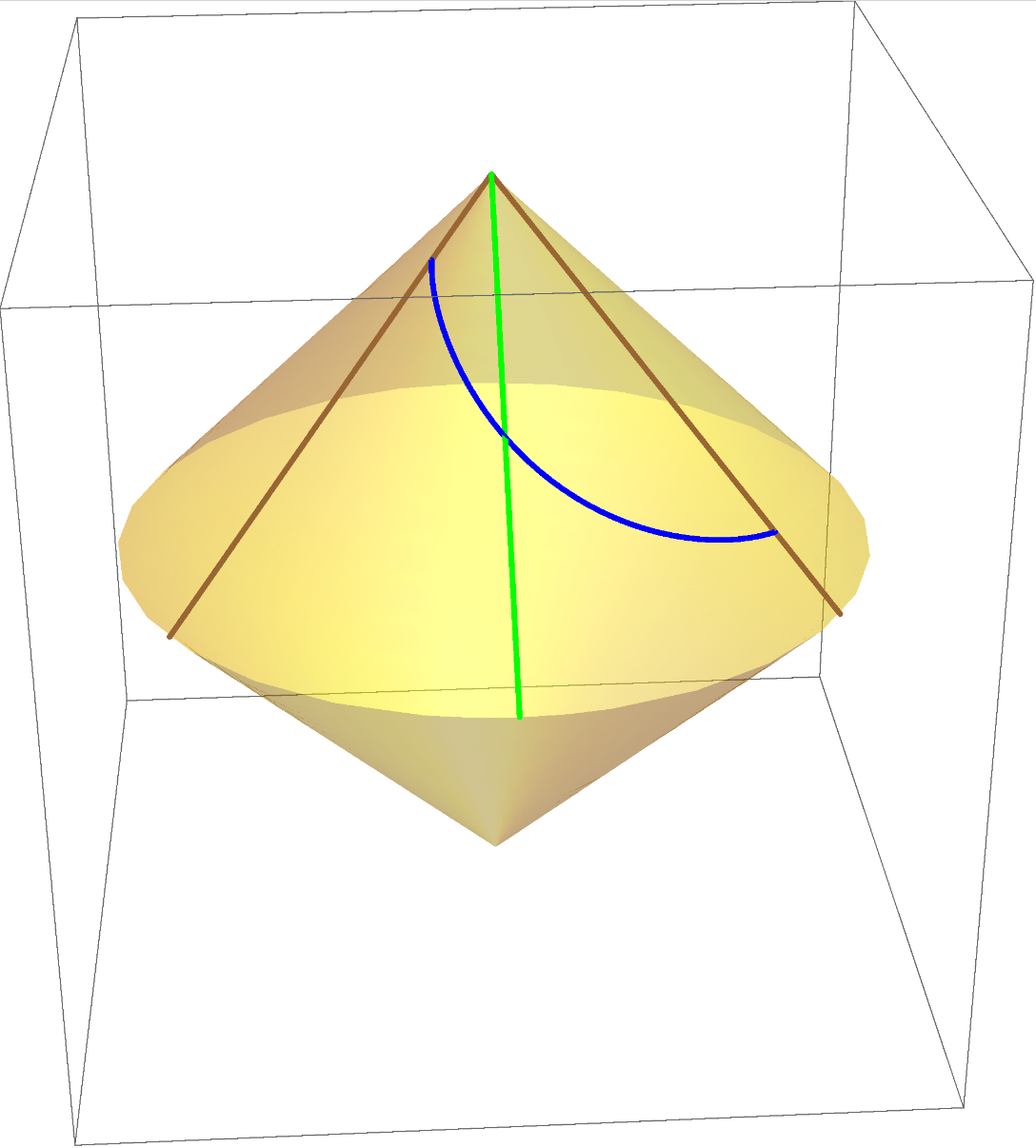} \label{fig:limit1} }
    \subfigure[]{
    \includegraphics[width=0.48\textwidth]{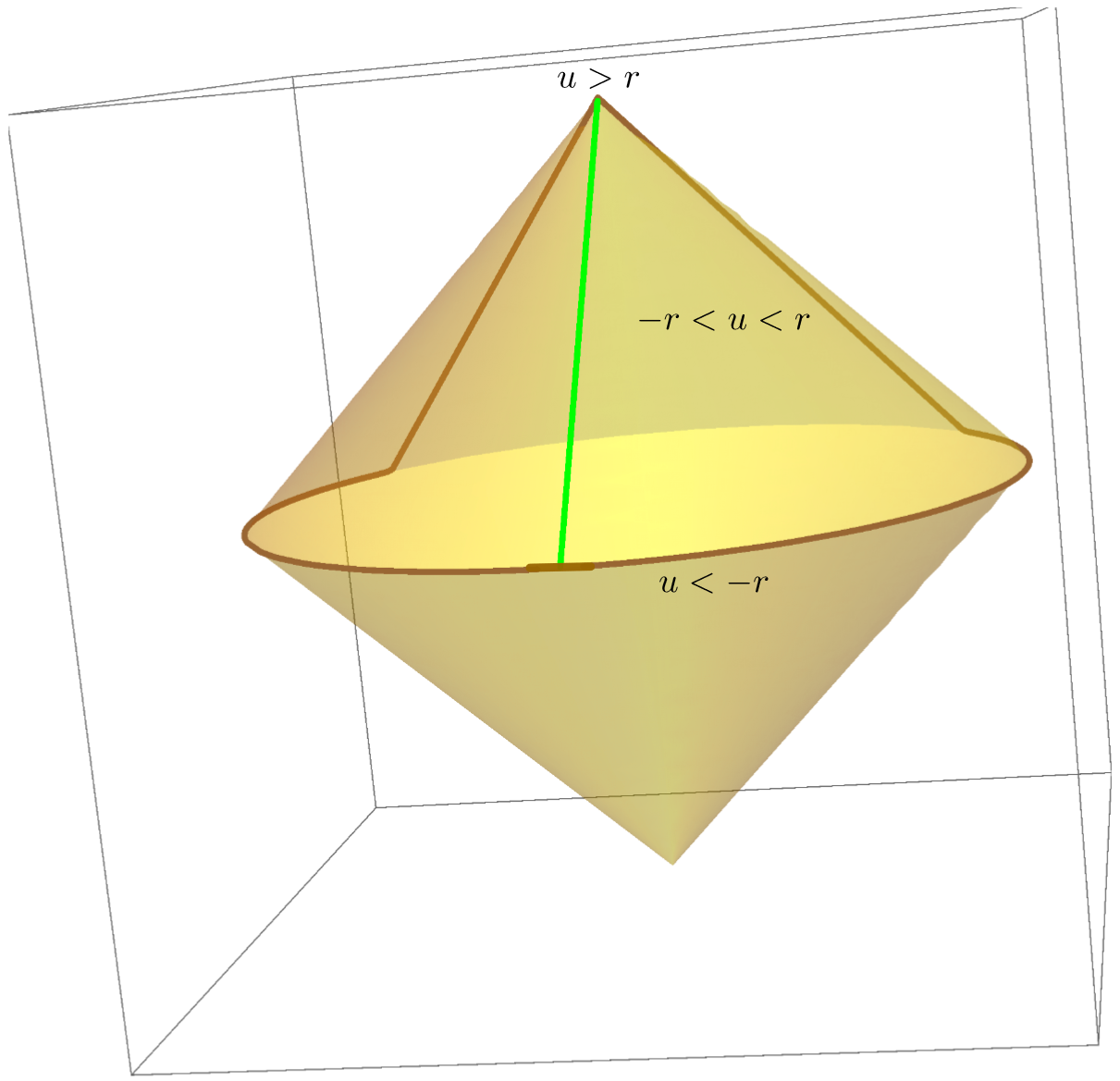} \label{fig:limit2}}
    \subfigure[]{
    \includegraphics[width=0.43\textwidth]{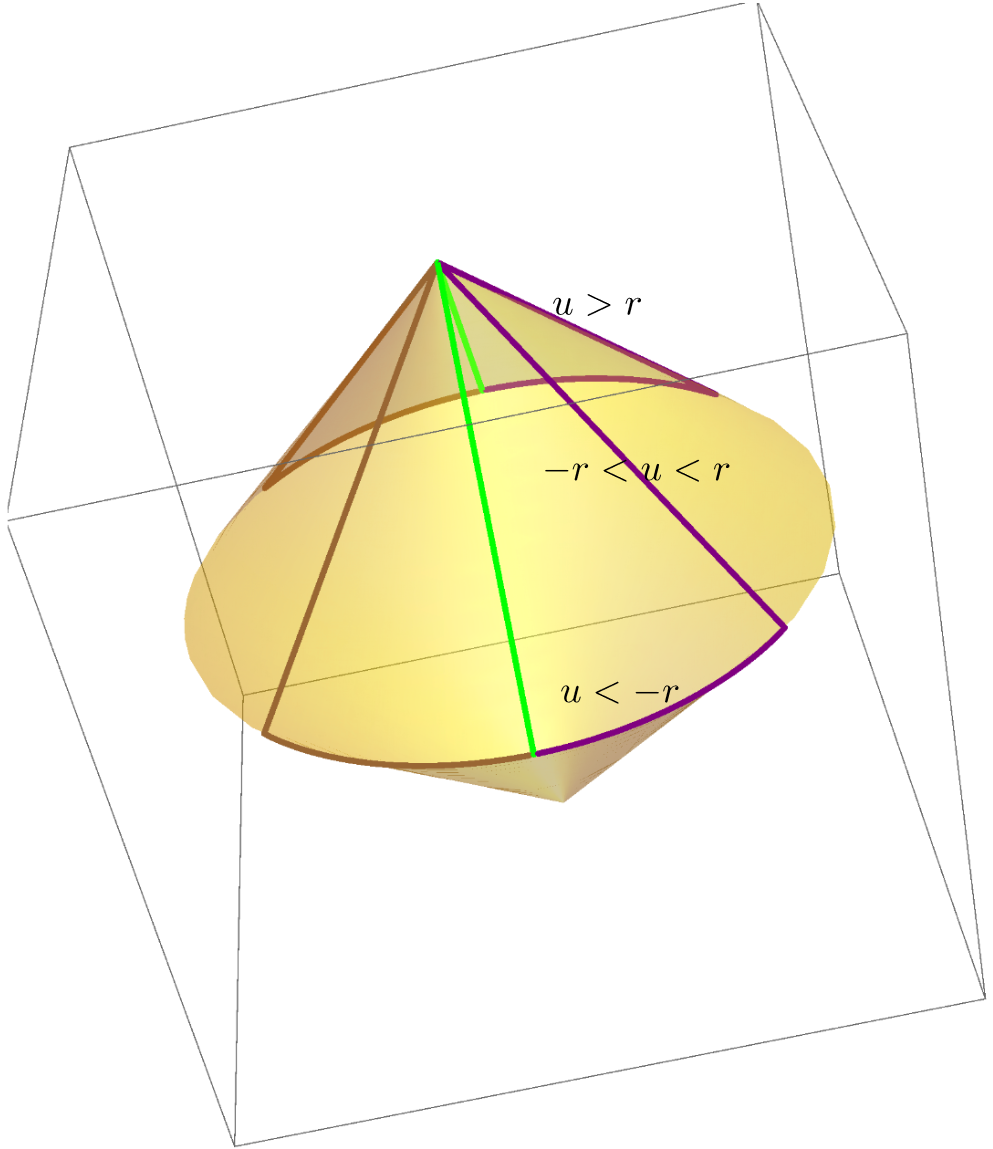}\label{fig:limit3} }
     \subfigure[]{
    \includegraphics[width=0.47\textwidth]{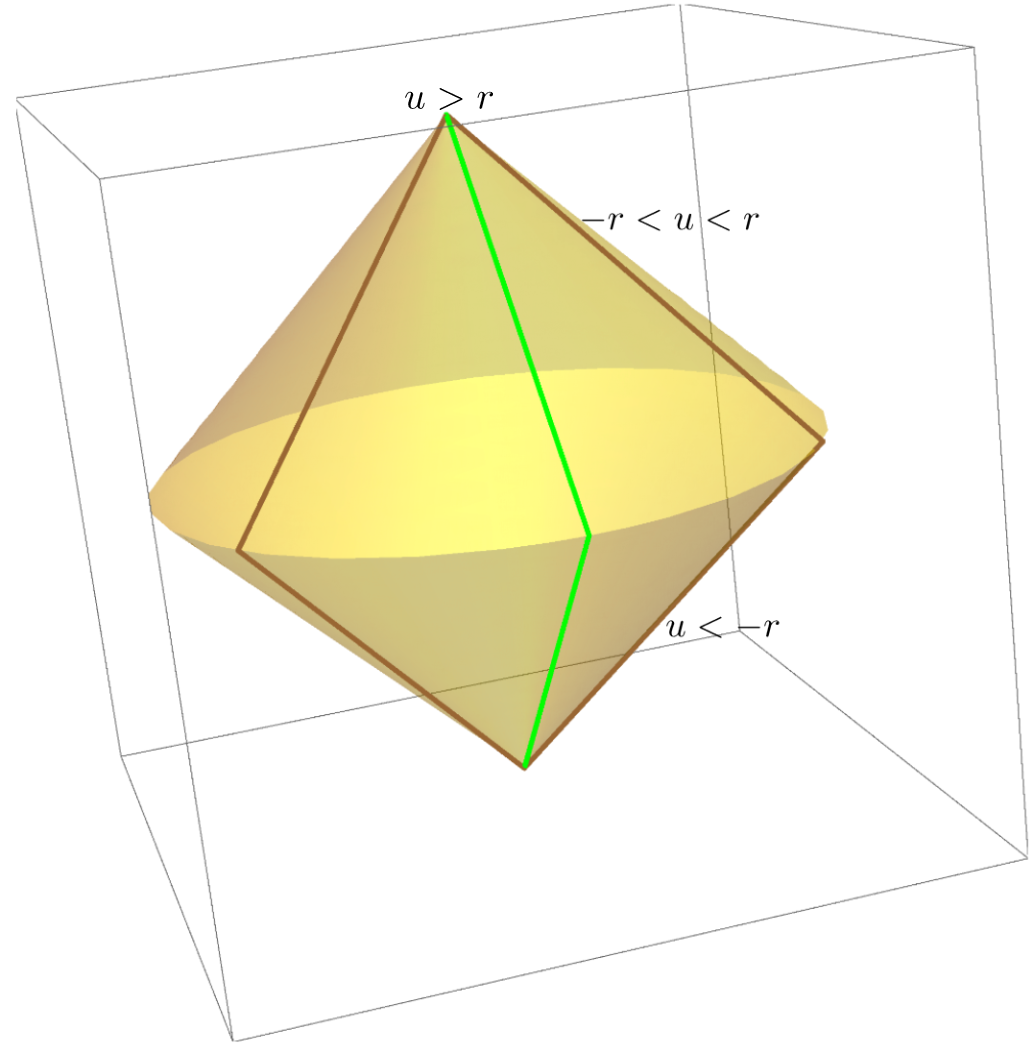}\label{fig:limit4} }
    \caption{These figures show the usual and unusual limits \eqref{unulimt} of boundary $\p D[\cA]$ of causal domain $D[\cA]$ in $M=0$, $M>0$ and $M<0$ zero mode backgrounds. Brown/Purple lines are the boundaries $\p D[\cA]$, and green lines are image of ordinate $z=0$ or $\phi=0$ of original coordinates. Figure \ref{fig:limit1} shows the expected configuration of usual limit $r > |u|\to \infty$, and figure \ref{fig:limit2} to \ref{fig:limit4} show the unusual limits of the Poincar\'e vacuum, $M>0$ zero mode backgrounds and $M<0$ zero mode backgrounds respectively with explicitly marked parameter ranges.  See the detailed descriptions in the main context.}
\end{figure}
 
\begin{itemize}

    \item For the usual $r > |u|\to \infty$ limit, we have the expected Penrose diagram as \ref{fig:limit1} for all the zero mode solutions. 
    
    \item \textbf{The Poincar\'e vacuum} New phenomena only happen in the first limit of \eqref{unulimt}. As shown in \ref{fig:limit2}, the $u<-r$ part of the boundary $\p D[\cA]$ of causal domain $D[\cA]$ go around the spacelike infinity $i^{0}$ making now the boundary $\p D[\cA]$ a closed curve.
    
    \item \textbf{$M>0$ zero mode backgrounds} New phenomena happen in both limits of \eqref{unulimt}. When $u<-r$ the $\p D[\cA]$ go around the spacelike infinity $i^{0}$ as in the case of the Poincar\'e vacuum; when $u>r$ the $\p D[\cA]$ go through the timelike infinity $i^{+}$, see \ref{fig:limit3}, to a similar configuration symmetric about $\Phi=\frac{\pi}{2}$ axis. Thus the causal domain $D[\cA]$ contains two disconnected parts, which is quite unusual. 
    
   \item \textbf{$M<0$ zero mode backgrounds} New phenomena happen only in the first limit of \eqref{unulimt}. As shown in \ref{fig:limit4}, the $u<-r$ part of the $\p D[\cA]$ would plot a similar configuration on past null infinity $\mathscr{I}^{-}$ as the one on future null infinity $\mathscr{I}^{+}$. This is the only case that the field theory can touch $\mathscr{I}^{-}$ which is consistent with boundary of zero mode backgrounds in the last subsection.
   
\end{itemize}

If we consider the configurations of boundary interval $\cA$ or the corresponding swing surface $\g_{\cA}$ in the unusual limits \eqref{unulimt}, they are the limiting ones of the usual cases and do not affect our main conclusions. 

\subsection{Negative pure and mixed state entanglement measures \label{subsecnegt}}
 
We already observed that the entanglement entropy can be negative \eqref{eevac} in flat$_3$/BMSFT model. From the BMS field theory point of view, the reason and meaning of negative entanglement entropy need further solid explorations. However from the Einstein gravity point of view, the negative holographic entanglement entropy is already annoying enough and may ruin the correspondence of swing surface proposal. In this subsection, we give the mathematical derivation of negative sign of holographic entanglement entropy by identifying the entanglement entropy as a Noether surface charge \footnote{Thank Wei Song for pointing out this viewpoint to us and Boyang Yu for early cooperation on this subsection.} and explicitly using the swing surface construction. Also we would give the physical intuition about why the situations in flat$_3$/BMSFT model are different from the ones in AdS/CFT. 

The holographic entanglement entropy can be viewed as an Noether surface charge evaluated along the HRT surface in AdS/CFT \cite{Faulkner:2013ica} ,
\be \cS_{\cA}=\cQ_{\xi}^{\g_{\cA}}= -\frac{1}{16\pi G} \int_{\text{HRT}} \nabla^{\mu}\xi^{\nu} \e_{\mu \nu \r} dx^{\r}=-\frac{1}{8}\int_{\text{HRT}} n^{\mu \nu} \e_{\mu \nu \r} dx^{\r}  \label{surcharge} \ee 
where $dx^{\r}$ denotes the unit vector along the HRT surface, and $\xi^{\nu}$ denotes the bulk modular flow vector. We used the fact \eqref{bfc2} in the third equality. The surface charge \eqref{surcharge} is actually a line integral in 3D bulk, and to do the computation we should embed it into a specific coordinate system. Take the Poincar\'e coordinate of AdS/CFT as an example. If we fix the sign of $\e_{txy}=1$ in Poincar\'e coordinates $(t,x,y)$ and integrate from the left endpoint of interval $\cA$ to the right one, we would always have the following formulas
\be n^{\mu \nu} \e_{\mu \nu \r}\lvert_{\g_{\cA}}=-2 \hat{e}_{\r}, \quad \cQ_{\xi}^{\g_{\cA}}=\frac{1}{4G} \int_{left}^{right} \hat{e}_{\r} dx^{\r}=\frac{ \text{Area}(\g_{\cA})}{4G}. \ee 

In CFT the modular flow of a general boundary interval always have positive component along the positive direction of ordinate. While in BMS$_3$ field theory when fixing the abscissa, we can change the $u$ coordinate of the boundary interval to change the relative sign of the modular flow to the positive direction of ordinate. This global degree of freedom of boundary interval in flat$_3$/BMSFT model is the key to understand the negative sign of holographic entanglement entropy. Mathematically, using \eqref{bbdu4} we can get the
parametrization equations of the bifurcating surface in Bondi coordinates of the Poincar\'e vacuum \footnote{Note there are typos in (3.38) and (3.39) of \cite{Apolo:2020bld}.}
\be  u(z)=\frac{-u_r(z-z_l)^2+ u_l(z-z_r)^2 }{ (z_l -z_r)(2 z-z_l-z_r)  },\quad r(z)=- \frac{2(u_l -u_r)}{(z_l -z_r) (2 z-z_l-z_r)}   \ee 
then the normalized directional vector $dx^{\r}$ along the bench can be obtained as
\be dx^{\r}= \text{sign}(u
_l-u_r) \left( \frac{(z-z_{l})(z-z_{r})}{z_{l}-z_{r}}, \frac{2}{z_{l}-z_{r}}, \frac{(z_{l}+z_{r}-2z)^2}{u_{l}-u_{r}} \right) \ee 
where we can see explicitly the sign of $dx^{\r}$ depend on the relative value of $u_l, u_r$ when we fix the values of $z_{l}$,$z_{r}$. The vector $\nabla^{\mu}\xi^{\nu} \e_{\mu \nu \r}$ can also be computed using \eqref{bbdu4}
\be \hat{e}_{\r}=\nabla^{\mu}\xi^{\nu} \e_{\mu \nu \r}= \left( \frac{8\pi}{z_{l}-z_{r}}, \frac{4\pi (z-z_{l})(z-z_{r})}{z_{l}-z_{r}} , -\frac{8\pi(u_{l}-u_{r})}{(z_{l}-z_{r})^2}  \right) \ee 
thus we have
\be \cQ_{\xi}^{\g_{\cA}}=\frac{1}{4G} \int_{z_l}^{z_r} \hat{e}_{\r} dx^{\r}=\text{sign}(u_l-u_r) \frac{\text{Area}(\g_{\cA})}{4G}  \ee 
which is indeed the expected form of holographic entanglement entropy in the Poincar\'e vacuum of the flat$_3$/BMSFT model \cite{Apolo:2020qjm},
\be \cS_{\cA}=\frac{u_{lr}}{2G z_{lr}}= \text{sign}(u_l-u_r) \frac{|2 l_{u}|}{4G l_{z}}= \cQ_{\xi}^{\g_{\cA}} \label{matchP} \ee 
where $u_{lr}=u_l-u_r$ and $z_{lr}=z_l-z_r$. 

We emphasize that not only the entanglement entropy \cite{Apolo:2020bld}, but also the reflected entropy \cite{Basak:2022cjs}, entanglement negativity \cite{Basu:2021awn} and PEE in flat$_3$/BMSFT model all can be negative. These key observations imply us that this is actually a general character of flat$_3$/BMSFT model. Although for the mixed state entanglement measures we do not have a local modular flow to mathematically prove the above statement. 

In flat$_3$/BMSFT model the bulk theory is pure Einstein gravity, however we need other property of swing surface to match the expectation of being holographic entanglement entropy and have to consult to the Noether charge formalism. This is unusual to what we learned in AdS/CFT. We leave the discussion in the last section.

\subsection{Finite bench or Infinity bifurcating surface?}

Due to the phenomena that swing surface always penetrate the boundary or horizon of the original spacetime, in the following we explore the causality structure in global flat$_3$ and overlook the quotient manifolds stuff momentarily. In order to specify which bulk region in global flat$_3$ have similar properties of entanglement wedge $\cW_{\cE}[\cA]$ in AdS/CFT holography, we are facing two unavoidable questions. The first one is whether this region is a closed co-dimension zero bulk region. The other one is which part of the bifurcating surface is more fundamental, or at least more useful, the finite bench $\g$ or the infinite bifurcating surface $\g_{\xi}$. 

\begin{figure}
    \centering
    \subfigure[]{
    \includegraphics[width=0.46 \textwidth]{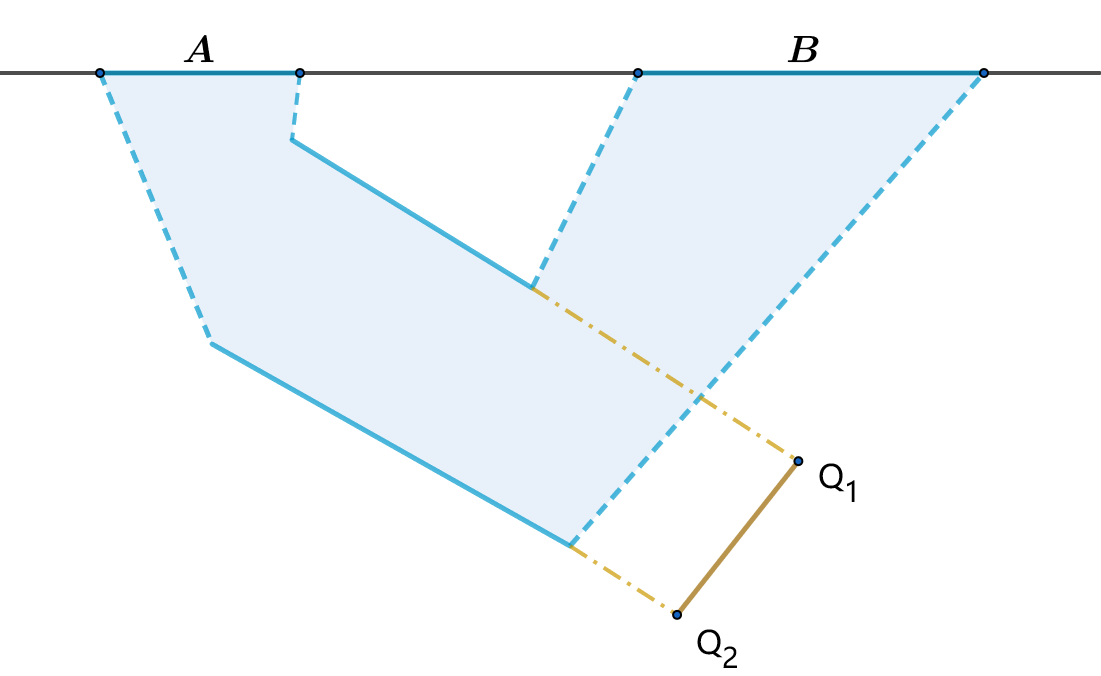}}
    \subfigure[]{
    \includegraphics[width=0.51 \textwidth]{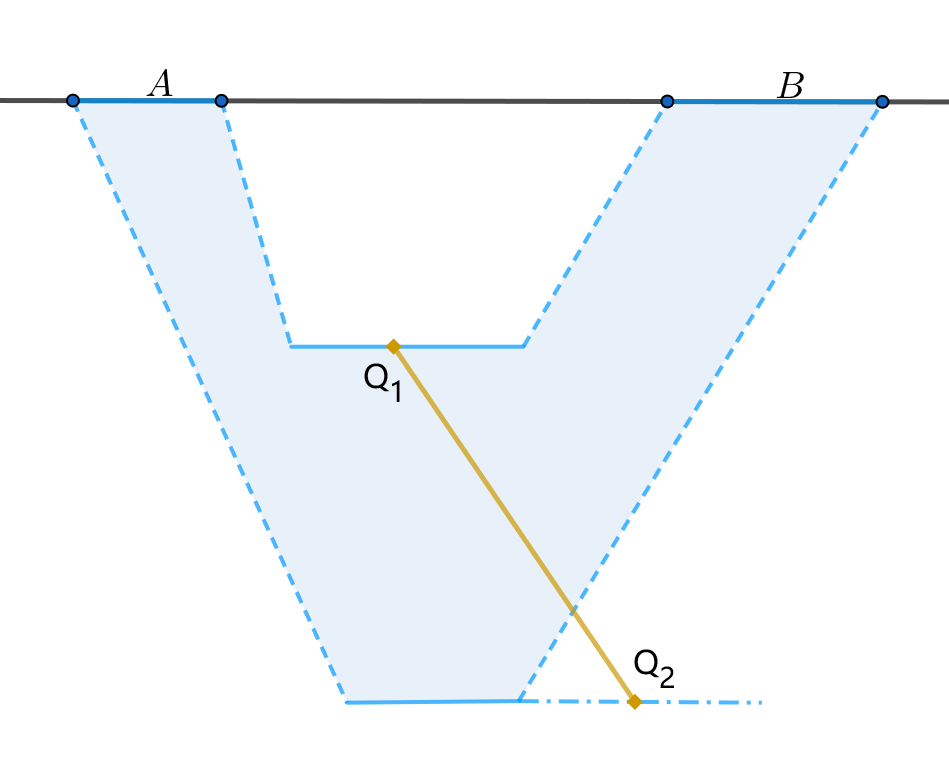}}
    \caption{Configurations of EWCS (brown lines between points $Q_{1}$ and $Q_{2}$) for general boundary two intervals $A$ and $B$ (blue interval), as well as the usual expected connected entanglement wedge (blue region) are shown. Blue dotted lines are null ropes $\g_{(p)}$, blue solid lines are bench $\g$ and the chain lines are part of the whole bifurcating surface $\g_{\xi}$.}
    \label{fig:ewcstwo}
    \end{figure}

Due to the non-local property in $u$ direction of boundary BMS field theory, which can be seen from the two point correlation function \eqref{twopcl}, it's rather unclear that a closed bulk region related to the swing surface $\g_{\cA}$ would exist. One example is the AdS$_{3}$/WCFT holographic model, the "pre-entanglement wedge" is not closed in the $v$ direction due to the non-local feature in $z$ direction of boundary WCFT \cite{Chen:2022fte}.

Related to the second question, there are two special bulk surfaces that we could grow null normal congruence from to construct the boundaries of a bulk region. One is the whole bifurcating surface $\g_{\xi}$ which is unbounded and invariant under the bulk modular flow $\xi$. Another one is the finite bench $\g$ which is a bounded portion of $\g_{\xi}$. Due to the homology condition between swing surface and boundary interval as well as the Noether charge formalism \eqref{surcharge}, the finite bench $\gamma$ may be more basic. 

The above mentioned questions turn out to be closely related to each other in flat$_3$/BMSFT model. We approach these problems from more practical ways rather than the more philosophical homology condition. To be consistent with the presentation style of this section, we just show the observations. 

The existing literature \cite{Basak:2022cjs,Basu:2021awn} only consider the symmetric two intervals on the boundary, where the EWCS would end on the finite bench $\g$. However when considering more general configurations of boundary two intervals, we observed that the endpoints of EWCS can exceed $\g$. We plot several new situations in Figure \ref{fig:ewcstwo}, where the usual expected entanglement wedge \cite{Apolo:2020bld,Basak:2022cjs} and true EWCS are plotted. We can see from the pictures that the not carefully defined connected entanglement wedge would lead to big problems. 

Due to these observations, we see that the whole modular invariant bifurcating horizon $\g_{\xi}$ may be more basic. We would provide more evidence along this perspective through PEE, BPE and bulk modular flow in the next section.

\section{Bulk Causality related to single interval}
\label{section4}

In this section we give a detailed analysis about causality structures related to finite bench $\g$ and infinite bifurcating surface $\g_{\xi}$ of a single boundary interval $\cA$. We use PEE as a useful tool to explore fine correspondence between boundary and bulk modular flow. When familiar with the subtleties in flat$_3$/BMSFT model during the process, we go to the question of finding "entanglement wedge" $\cW_{\cE}[\cA]$ in flat$_3$/BMSFT model. As a by product, we solve the problem of PEE in flat$_3$/BMSFT model stated in section \ref{sec:intro}. For simplicity and without loss of generality we present all the detailed analysis in the Poincar\'e Vacuum. 

Let us slightly generalize the parametrization in \cite{Apolo:2020bld} of swing surface in the Poincare vacuum \eqref{poincarevacuum}. Considering a general boundary field interval $\cA$ with endpoints
\be  \partial \mathcal{A} = \big\{ \big( u_{l},z_{l} \big), \big( u_{r},z_{r} \big) \big\} \label{intervalflat} \ee 
The boundary conditions of the null ropes $\gamma_{l,r}$ emanating from endpoints $\p \cA$ are simply
\be  \gamma_{l,r}: \; u=u_{l,r}, \; z=z_{l,r}   \ee
The length of the spacelike geodesic connected between two null ropes $\g_{l,r}$ is given by
\be L(r_{l},r_{r})=\sqrt{2 r_{r} (u_{l} - u_{r}) + r_{l} (-2 u_{l} + 2 u_{r} + r_{r} (z_{l} - z_{r})^2)}   \label{lengthflat} \ee 
where $r_{l,r}$ are radial coordinates of the points on $\g_{l,r}$. The extreme of \eqref{lengthflat} is found at
\be r_{l}=-r_{r}= - \frac{2 (u_{l} - u_{r})}{(z_{l} - z_{r})^2}.  \label{anylt}  \ee 
From here we can see a necessity to analytically continuate the original Poincare vacuum spacetime with only $r \ge 0$ to the one that also includes negative values of $r$ in order to include just the single interval swing surface $\g_{\cA}$. The bench $\g$ is just a straight line going through the points parametrized by
\be 
t(s)=t_l+(t_r-t_l)s,\quad  x(s)=x_l+(x_r-x_l)s,\quad y(s)=y_l+(y_r-y_l)s \label{pbench}
\ee 
where the left and right endpoints of $\g$ have following expressions,
\begin{small}
\be  (t_{l,r},x_{l,r},y_{l,r} )= \left( 2u_{l,r}-\frac{(u_{l,r}-u_{r,l})(1+4z_{l,r}^{2})}{2(z_{l,r}-z_{r,l})^{2}} ,-\frac{2(u_{l,r}-u_{r,l})z_{l,r}}{(z_{l,r}-z_{r,l})^2} , -2 u_{l,r}+\frac{(u_{l,r}-u_{r,l})(-1+4z_{l,r}^{2})}{2(z_{l,r}-z_{r,l})^{2}} \right)     \ee
\end{small}

\subsection{Bifurcating horizons}

There are several coordinate systems that we would go back and forth when trying to clearly show the causal relations between boundary field theory and bulk gravity theory.
\begin{itemize}
    \item \textbf{ Bondi coordinates} of the original Poincare vacuum: 
    \be \text{bulk}: (u,r,z), \quad \text{boundary}: (u,r \to \infty,z) \ee 
    
    \item \textbf{Cartesian coordinates and Penrose coordinates} of the covering global Minkowski spacetime:
    \be \text{Cartesian}:(t,x,y), \quad \text{Penrose}: (U, V, \Phi), \; (T,X,Y) \label{pencoord} \ee 
   which are related by the standard textbook transformations,
  \be U=\arctan{(t-\sqrt{x^{2}+y^{2}})}, \quad V= \arctan{(t+\sqrt{x^{2}+y^{2}})}, \quad \Phi=\phi=\arctan{\frac{y}{x}} \nn \ee 
  \be T=   V+U , \quad X=  (V-U) \cos{\Phi}, \quad Y=  (V-U) \sin{\Phi}  \label{flatTXY} \ee 
\end{itemize}

\textbf{From boundary to boundary}\\

we first deal with the image of field interval $\cA$ on the future null infinity $\mathscr{I}^{+}$ \footnote{This is a choice of us, which means that we can also choose to map boundary field theory to the past null infinity $\mathscr{I}^{-}$ by putting minus signs in coordinate transformations \eqref{pointranf}.} of Penrose diagram with coordinates $(U,V,\Phi)$. There are several facts about this map:  

\begin{itemize}
    \item  constant $z$ line of field theory would be mapped to constant $\Phi$ line on $\mathscr{I}^{+}$  of the Penrose diagram. So a strip like region which is the causal domain $D[\cA]$ of field interval would map to a corner region on the boundary null cone. In particular, the $z=0$ axis would be mapped to $\Phi=\frac{\pi}{2}$ line. This can be seen from the following transformation, 
    \be \Phi=\arctan{\frac{y}{x}}|_{r\to \pm \infty}=\arctan{\frac{1-4z^2}{4z}} \label{Phi}. \ee
    When $z$ goes from $0$ to $\infty$, $\Phi$ would go from $\frac{\pi}{2}$ to $-\frac{\pi}{2}$. Because \eqref{Phi} is a monotonic decreasing function when $z >0$, then the map is one to one.
    
    \item A symmetric interval about the origin $(u=0,z=0)$ would map to a symmetric interval about the point $( U=0 ,V=\frac{\pi}{2} ,\Phi=\frac{\pi}{2})$ on $\mathscr{I}^{+}$ of the Penrose diagram. This can be seen as follows:
    \begin{align}
        & \sqrt{x^2+y^2}|_{r\to \pm \infty}   =\frac{1}{4}(1+4 z^2) |r|+\frac{2u(1-4z^2)}{1+4z^2}\frac{|r|}{r}, \nn \\
       & \text{when} \; r\to \infty, \quad U   =\arctan{\frac{4 u}{1+4z^2}}, \quad V=\frac{\pi}{2} .\label{rinfty}  
    \end{align}   
    
  \eqref{Phi} and \eqref{rinfty} give us a bijective map from the infinite $(u,z)$ plane where the original BMS field theory live to the whole future null infinity $\mathscr{I}^{+}$ of the compact Penrose diagram. 
    
\end{itemize}
    
\textbf{bench $\g$ and bifurcating surface $\g_{\xi}$}\\

we choose the symmetric boundary interval $\cA$ in \eqref{intervalflat} for convenience,
\be -u_{l}=u_{r}=\frac{l_{u}}{2}, \quad  -z_{l}=z_{r}=\frac{l_{z}}{2} \label{single} \ee 
putting them into \eqref{pbench}, we get the parametrization of the finite bench
\be  (t,x,y)=\left( \lambda, -\frac{l_{u}}{l_{z}}, -\frac{l_{z}^{2}+1}{l_{z}^{2}-1} \lambda \right), \quad |\lambda| < \left| \frac{l_{u}}{2}(1-\frac{1}{l_{z}^{2}}) \right| . \label{bench} \ee 
When the parameter $\lambda$ has parameter range $\lambda \in (-\infty, \infty)$ in \eqref{bench}, the parameter equations denote the whole bifurcating surface $\g_{\xi}$. In the Penrose diagram, $\g_{\xi}$ always end on the spacelike infinity $i^{0}$ with coordinates $(U,V,\Phi)=(-\frac{\pi}{2},\frac{\pi}{2},\pm \frac{\pi}{2})$, which are the results of $|(l_{z}^{2}+1)/(l_{z}^{2}-1) |>1$.  \\

\textbf{bifurcating Killing horizon}\\

The bifurcating Killing horizon $N_{l,r}$ are composed of null congruence emitted from the bifurcating surface $\g_{\xi}$. Locally each null generator of $N_{l,r}$ is perpendicular to $\g_{\xi}$ at the intersection point. They could be parametrized as,
\be  t = \lambda_{1} + \kappa \lambda_{2} \, \text{sgn}(\kappa), \quad  
    x =  -\frac{l_{u}}{l_{z}} \pm \sqrt{\kappa^2-1} \lambda_{2} \, \text{sgn}(\kappa)  , \quad y=\kappa \lambda_{1} +  \lambda_{2} \, \text{sgn}(\kappa) \label{nullflat} \ee 
where $\kappa \equiv -\frac{l_{z}^{2}+1}{l_{z}^{2}-1} $ and $\text{sgn}(\kappa)$ denotes the sign function of parameter $\kappa$. $\lambda_{1}$ parametrize $\g_{\xi}$ similar to \eqref{pbench}, and $\lambda_{2}$ parametrize the null congruence emitted from $\g_{\xi}$. When $\lambda_{2}$ take values in $(0, \infty)$, two future Killing horizons where two null ropes $\g_{l,r}$ sit appear with plus and minus sign in $x$ coordinates of \eqref{nullflat}. When $\lambda_{2}$ take values in $(-\infty, 0)$, two past Killing horizons would appear. 

\begin{figure}
    \centering
    \subfigure[]{
    \includegraphics[width=0.48\textwidth]{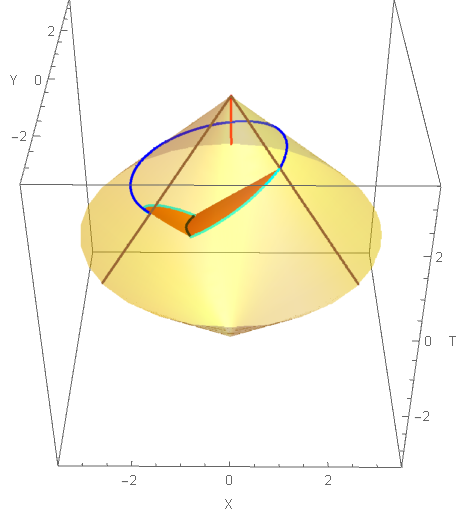} \label{fig:bifhorfint} }
    \subfigure[]{
    \includegraphics[width=0.44\textwidth]{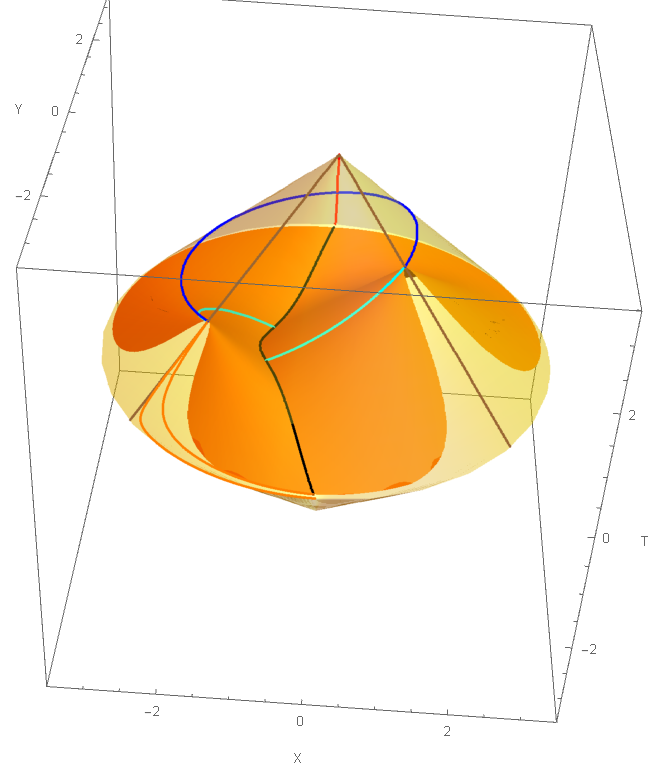}\label{fig:bifhorfwho} }
    \caption{In Penrose diagrams with coordinates $(T,X,Y)$ defined in \eqref{pencoord}, figure \ref{fig:bifhorfint} and \ref{fig:bifhorfwho} show the bifurcating horizons $N_{\g}$ related to finite bench $\g$   \eqref{horiben} and the ones $N_{l,r}$ related to the infinite bifurcating surface $\g_{\xi}$ \eqref{infben} of single boundary interval $\cA$ (blue curve) with $(l_u=1,l_z=2)$ in \eqref{single} separately. In addition to the basic elements appearing in figure \ref{fig:limit1}, we also have null ropes (cyan curves), finite bench $\g$, bifurcating surface $\g_{\xi}$ (black curves) and bifurcating horizon $N_{l,r}$ (orange surfaces). Figure \ref{fig:bifhorfint} has no closed region bounded by the $N_{\g} \cup \mathscr{I}^{+}$, while figure \ref{fig:bifhorfwho} actually form a closed region \eqref{ewflat} which can not be shown perfectly due to limitation on the computational power of Mathematica. We hope the two orange curves can show the limiting behaviors of null congruence to the endpoints of bifurcating surface $\g_\xi$.
    }
    \end{figure}
    
 From equations \eqref{flatTXY} and  \eqref{nullflat} we can draw the Killing horizons in the compact Penrose diagram, see Figure \ref{fig:bifhorfint} and \ref{fig:bifhorfwho}. In accordance with our intuition that the non-local property of BMS field theory would destroy the closeness, the two null surfaces $N_{\g}$ related to the finite bench $\g$ suspend in the Lorentzian Minkowski spacetime with only two points touching the null infinity, so no closed region is formed. We have 
 \be N_{\g} \cap \mathscr{I}^{+}  = \p \cA, \quad \text{where}\; N_{\g} \subset N_{l}\cup N_{r} \label{horiben} \ee 
Unlike the AdS$_3$/WCFT case \cite{Chen:2022fte}, the Killing horizons related to $\g_{\xi}$ touch the boundary of Minkowski spacetime on the whole spacelike infinity $i^{0}$ and two lines on future null infinity $\mathscr{I}^{+}$, so a closed region is formed. We have 
\be \left(N_{l}\cup N_{r}\right) \cap \mathscr{I}^{+}= i^{0} \cup l_{\p \cA} \label{infben} \ee where $l_{\p \cA} \subset \p D[\cA] $ represent part of boundary of $D[\cA]$ which start from endpoints $\p \cA$ and end on the spacelike infinity $i^{0}$. So a special region, we call it $\cW_{\cE}^{f}[\cA]$, bounded by bifurcating Killing horizons and asymptotic boundaries of Minkowski spacetime are formed
\be \cW_{\cE}^{f}[\cA]=\Tilde{J}^{+}(\g_{\xi})= \text{Region Bounded by} \;\; N_{l} \cup N_{r} \cup \mathscr{I}^{+} \cup i^{0} \label{ewflat} \ee 
where $\Tilde{J}^{+}(\g_{\xi})$ denote the bulk causal future of $\g_{\xi}$. There are two important features about the Figure \ref{fig:bifhorfwho}, both are related to the limiting behaviors of the null congruence emitted from $\g_{\xi}$
 \begin{itemize}
  \item All null congruence emitted from finite points of $\g_{\xi}$ would only intersect with asymptotic boundaries on one point \eqref{horiben}, which is one endpoint of boundary interval $\cA$. Mathematically when we have $\lambda_{1}< \infty$ and $\lambda_{2} \to \infty$ in \eqref{nullflat}, then
        \be 
             r   \to |\kappa| \lambda_{2}+  \lambda_{1}\pm \frac{\sqrt{\kappa^2-1}}{|\kappa|}\frac{l_{u}}{l_{z}}= t+  \lambda_{1}\pm \frac{2 l_{u}}{l_{z}^2+1} ,
          U   \to \pm \arctan{ \frac{2 l_{u}}{l_{z}^2+1}}, V  \to \frac{\pi}{2}  \label{endpoint}
        \ee
        Comparing \eqref{single},  \eqref{rinfty} and \eqref{endpoint} we see the validity of the above statement. 
        
        \item The null congruence emitted from the endpoints of $\g_{\xi}$ locating on the spacelike infinity $i^{0}$ would first go around $i^{0}$ until it touches the boundary of causal domain $D[\cA]$, then it would go up following this boundary line until touches the endpoint $\p \cA$ of field interval, see the orange lines in Figure \ref{fig:bifhorfwho}. Clearly there is a critical point for its different behaviors on the Penrose diagram. Actually when $\lambda_{2} \ll \lambda_{1}$, the first part of $t,x,y$ parametrization in \eqref{nullflat} would dominate and the effect of this term is changing the $\Phi$ angle. When $\lambda_{2} \gg \lambda_{1}$, the second part, which shows more explicitly its lightlike property, of $t,x,y$ parametrization in \eqref{nullflat} would dominate. The competition of first part and second part in \eqref{nullflat} tell us where the turning point sit.  
        
    \end{itemize}

\subsection{Decomposition of bulk spacetime} 

In this subsection we analyze the decomposition of the global flat$_3$ in terms of both the future and past bifurcating horizons of $\g_{\xi}$, and make a comparison with AdS/CFT to show the unusual features in flat$_3$/BMSFT model.  

Viewing from the global coordinates of AdS/CFT, the four null Killing horizons of HRT surface related to single interval $\cA$ together would separate the whole Lorentzian AdS spacetime into four non-intersection parts, which nicely match with the boundary causal structure \cite{Headrick:2014cta}. 
  \begin{figure}
      \centering\includegraphics[width=0.35\textwidth]{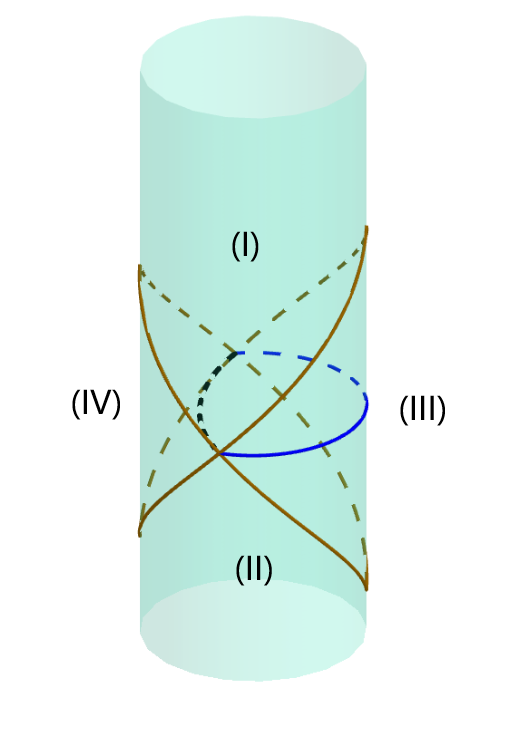}
      \caption{Bulk decomposition \eqref{adsdecop} of global AdS$_3$ with respect to the HRT surface \cite{Headrick:2014cta} is presented. Basic components are boundary interval (blue), HRT surface (dotted black), boundary $\p D[\cA]$ (yellow) of causal domain $D[\cA]$ and $D[\cA^{c}]$ related to the complement interval $\cA^{c}$. Causal regions (I) to (IV) are defined below \eqref{bulk4}. }
      \label{fig:globalads}
  \end{figure}  
 \begin{figure}
    \centering
    \subfigure[]{
    \includegraphics[width=0.46\textwidth]{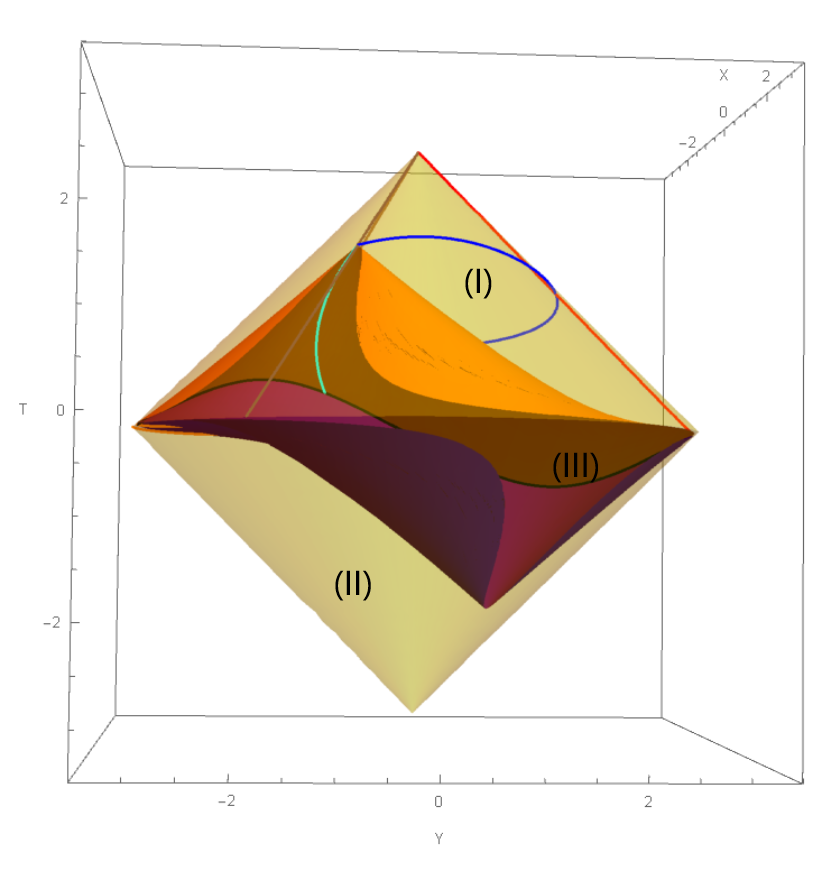} }
    \subfigure[]{
    \includegraphics[width=0.48\textwidth]{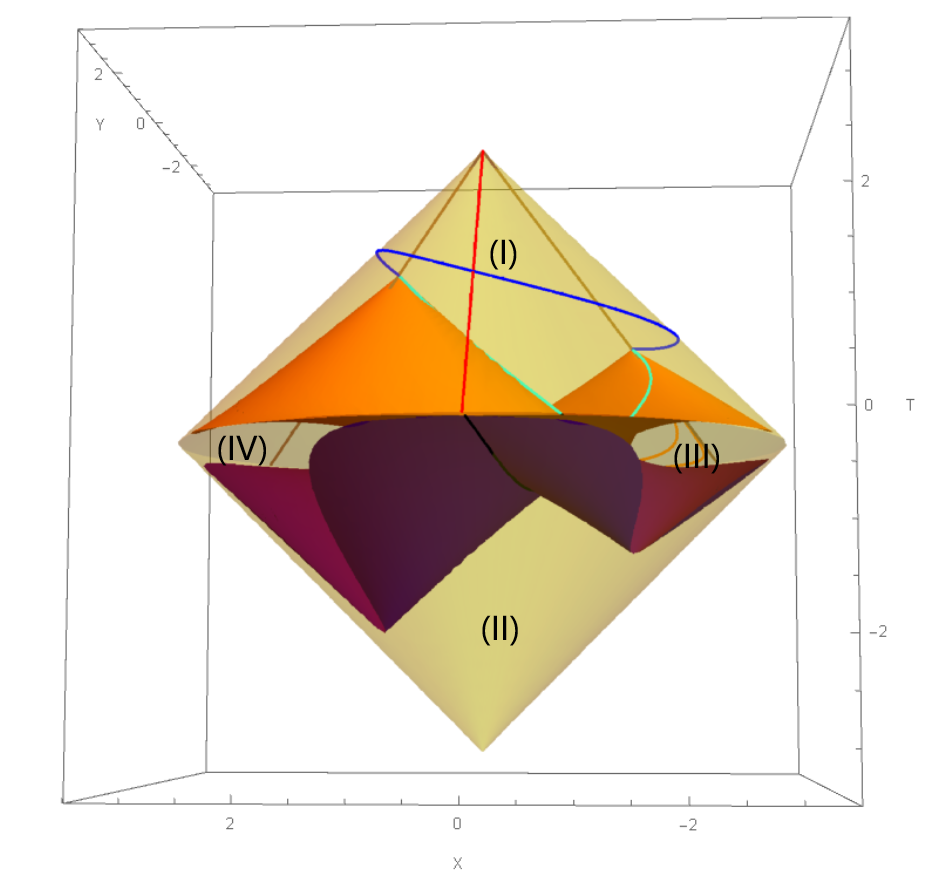} }
    \caption{Different perspectives on the bulk decomposition \eqref{flatbu1} with respect to the bifurcating surface $\g_{\xi}$ of global flat$_3$. In addition to the main elements appearing in figure \ref{fig:bifhorfwho}, we also have two past null bifurcating horizons (purple surfaces) which again actually form closed surfaces like the future ones. Causal regions (I) to (IV) are defined below \eqref{bulk4}. }
    \label{fig:Poincarevacuum}
    \end{figure} Mathematically, we can decompose boundary spacetime $\cB$ as follows:
   \be \cB=D[\cA]\cup D[\cA^{c}]\cup J^{+}[\p \cA] \cup J^{-}[\p \cA] \label{bound4}\ee
where $D[\cA]$ is the boundary causal domain of interval $\cA$ and $J^{\pm}[p]$ denote the causal future and past of point $p$ on $\cB$. This tells us that the full boundary spacetime $\cB$ would decompose into four causally non-overlapping regions: the causal domain of the region $\cA$ and its complement $\cA^{c}$, and the causal future and past of the entangling surface $\p \cA$. For the bulk spacetime $\cM$ we have the decomposition, 
\be \cM=\Tilde{D}[\cR_{\cA}] \cup \Tilde{D}[\cR_{\cA^{c}}] \cup \Tilde{J}^{+}[\g_{\cA}] \cup \Tilde{J}^{-}[\g_{\cA}] \label{adsdecop} \ee where the tilde of corresponding notation denote the bulk one, for example $\Tilde{D}$ is the bulk causal domain and $\cR_{\cA}$ is the spacelike homology surface interpolating between boundary subregion $\cA$ and bulk HRT surface $\g_{\cA}$. The causal split of the bulk into two spacelike and two timelike separated regions from the perspective of $\g_{\cA}$ precisely match the boundary causal decomposition \eqref{bound4} when restrict \eqref{bulk4} to the boundary due to the following relations
\be
 \Tilde{D}[\cR_{\cA}] \cup \cB = D[\cA] \quad
 \Tilde{D}[\cR_{\cA^{c}}] \cup \cB = D[\cA^{c}] \quad 
\Tilde{J}^{\pm}[\g_{\cA}]\cup \cB= J^{\pm}[\p \cA] \label{bulk4}
\ee
 To facilitate the discussion about AdS and flat spacetime in a unified way, we define the following notations, see Figure \ref{fig:globalads} and \ref{fig:Poincarevacuum}

\begin{itemize}
    \item bulk Region (I): Causal future of bifurcating horizon $\g_{\xi}$: $ \Tilde{J}^{+}[\g_{\xi}]$
    \item bulk Region (II): Causal past of bifurcating horizon $\g_{\xi}$: $ \Tilde{J}^{-}[\g_{\xi}]$
    \item bulk Region (III), (IV): Two spacelike region which contain all the points spacelike separated from $\g$ 
\end{itemize}
For flat$_3$/BMSFT model, mathematically we have following relations on the boundary,
\be \cB=\mathscr{I}^{+}=D[\cA] \cup D[\cA^{c}] \label{bdcau1} \ee
where $\cB$ denote the spacetime where BMS field theory lives and 
\be  \cM= \Tilde{J}^{+}[\g_{\xi}] \cup \Tilde{J}^{-}[\g_{\xi}] \cup \text{(III)}  \cup \text{(IV)} \label{flatbu1}\ee 
which satisfy
\be \Tilde{J}^{+}[\g_{\xi}] \cap \p \cM=\mathscr{I}^{+}, \quad \Tilde{J}^{-}[\g_{\xi}] \cap \p \cM=\mathscr{I}^{-}, \quad \text{(III)} \cap \p \cM= i^{0}_{1}, \quad \text{(IV)} \cap \p \cM= i^{0}_{2}  \label{flatbu2} \ee 
where $i^{0}_{1}$ denotes part of the spacelike infinity with parameter range $\Phi \in (-\pi/2,\pi/2) $, $i^{0}_{2}$ denotes part of the spacelike infinity with parameter range $\Phi \in (\pi/2, 3\pi/2) $. Although regions (III) and (IV) can be defined as the bulk causal domains
\be \text{(III)}=\Tilde{D}[\cR_{i^{0}_{1}}], \quad  \text{(IV)}=\Tilde{D}[\cR_{i^{0}_{2}}] \ee 
of the spacelike homology surface $\cR_{i^{0}_{1}}$ and $\cR_{i^{0}_{2}}$ with properties 
\be \p \cR_{i^{0}_{1}}= i^{0}_{1} \cup \g_{\xi}, \quad \p \cR_{i^{0}_{2}}= i^{0}_{2} \cup \g_{\xi}, \quad \cR_{i^{0}_{1}}\cup \cR_{i^{0}_{1}}= \Sigma_{i^{0}}  \ee 
where $\Sigma_{i^{0}}$ is a bulk Cauchy surface of the whole Minkowski spacetime $\cM$, we can not find special meaning of $i^{0}_{1}$, $i^{0}_{2}$ and the corresponding homology surface $\cR_{i^{0}_{1}}$ and $\cR_{i^{0}_{2}}$. Also we have no good idea about how to make physical distinctions between $\cR_{i^{0}_{1}}$ and $\cR_{i^{0}_{2}}$. we summarize main features of causality structures in flat$_3$/BMSFT model by comparisons with AdS/CFT
\begin{itemize}

    \item Both global AdS and global flat$_3$ can be decomposed into four regions according to (I) $\sim$ (IV). In AdS spacetime, (III) is identified as the entanglement wedge $\cW_{\cE}[\cA] $, and the homology surface $\cR_{\cA}$ is a spacelike surface in (III). Boundary interval $\cA$ is a spacelike interval and spacelike separated from the HRT surface. In flat spacetime, (I) is the special region we called $\cW_{\cE}^{f}[\cA]$ \eqref{ewflat}. $\cA$ is a interval viewed from the bulk and locate in the causal future of bifurcating horizon $\g_{\xi}$.

    \item  For AdS/CFT the bulk decomposition of spacetime $\cM$ precisely match the boundary one \eqref{bulk4}. The entanglement wedge of $\cA$ and its complement $\cA^{c}$ have no overlap. While in flat$_3$/BMSFT model, the bulk decomposition have no relation with the boundary one \eqref{flatbu2}. In addition, the special region $\cW_{\cE}^{f}[\cA]$ is exactly the same as the one of the complement interval $\cW_{\cE}^{f}[\cA^{c}]$.
\end{itemize}

\subsection{PEE: intersection of swing surface}

In the following two subsections, we study the PEE correspondence in flat$_3$/BMSFT model using two ways. One is similar to the intersection of HRT surfaces, the other one is the correspondence between boundary and bulk modular flow. This quantity not only provides us more bulk quantities other than swing surface and EWCS to explore, but also gives us a chance to be familiar with the structures of bulk modular flow. As a byproduct, we solve the PEE problem in flat$_3$/BMSFT model, thus giving the foundations about the observed match of BPE \cite{Camargo:2022mme,Basu:2022nyl}.
 
\begin{figure}
    \centering
    \subfigure[]{
    \includegraphics[width=0.45 \textwidth]{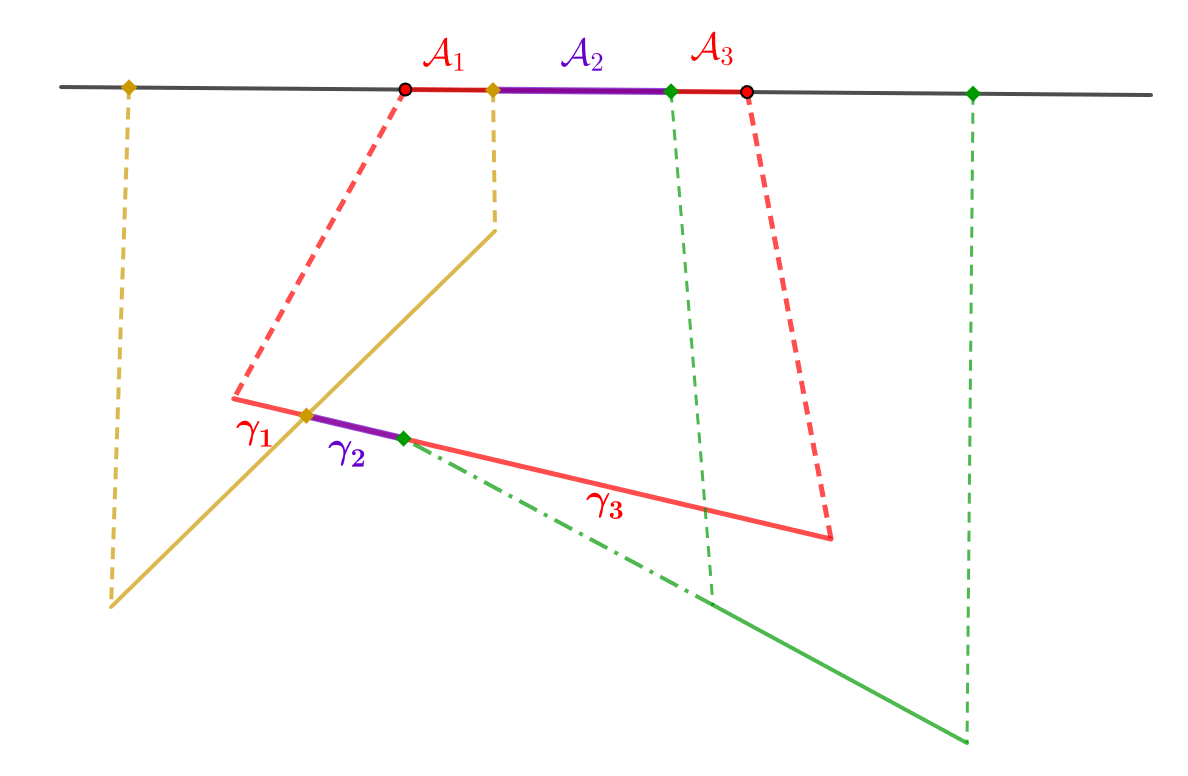}\label{fig:PEEswsf} }
    \subfigure[]{
    \includegraphics[width=0.5 \textwidth]{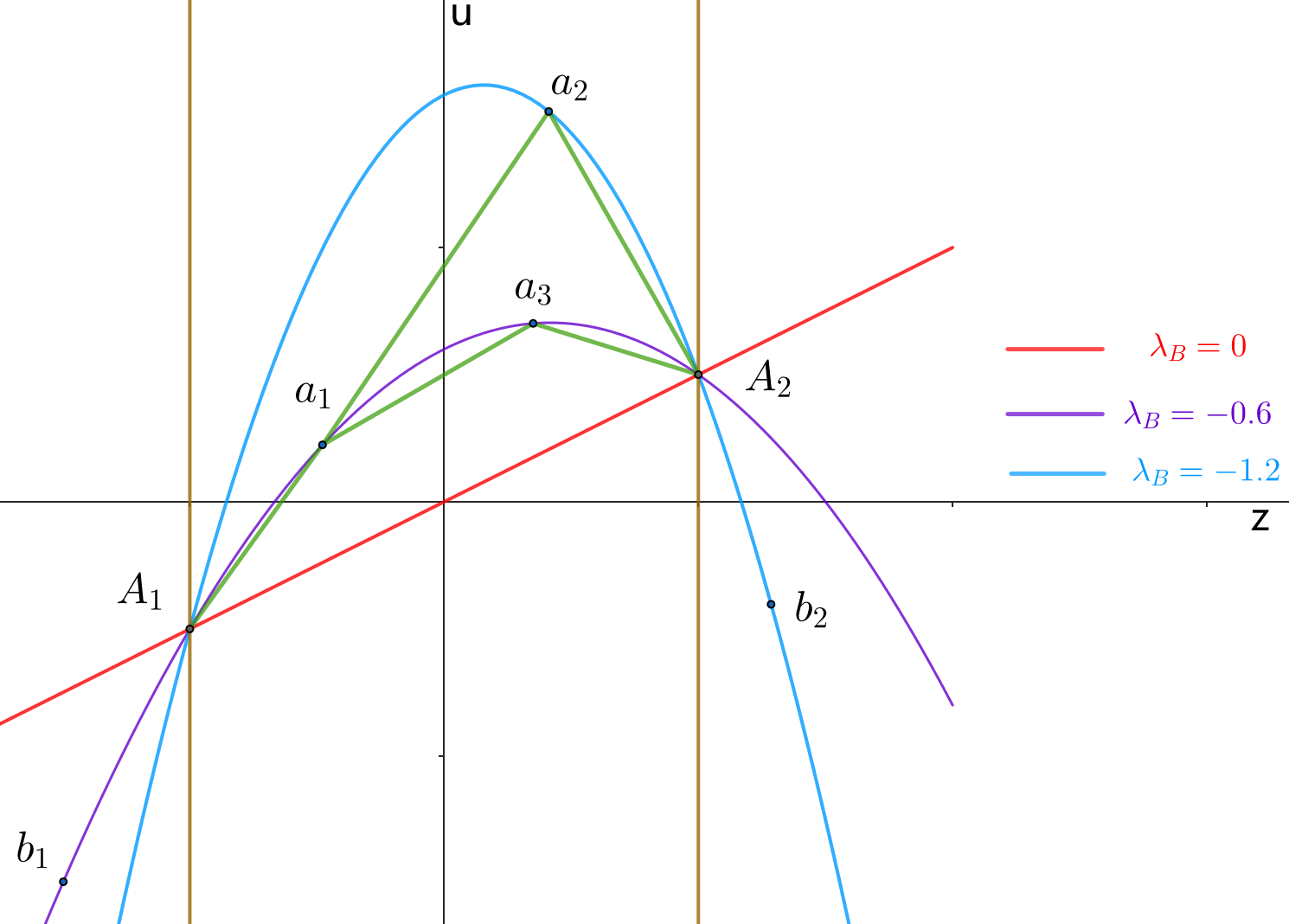}\label{fig:BMSmodflow} }
    \caption{Figure \ref{fig:PEEswsf} shows the swing surface intersection method to determine the PEE correspondence. Bulk geodesic $\g_{2}$ correspondences to boundary subinterval $\cA_{2}$. Figure \ref{fig:BMSmodflow} shows the boundary modular flow line with parametrization consistent with \eqref{bbbdflow}. Points $a_1,b_1,a_3$ and $a_2,b_2$ are on the same modular flow lines separately. }
    \end{figure}

On the field side without loss of generality, we choose the boundary interval $\cA$ to be straight line between two points $A_1, A_2$ with coordinates $(z=-1,u=-\frac{1}{2})$ and $(z=1,u=\frac{1}{2})$, see Figure \ref{fig:BMSmodflow}. Then the PEE for subinterval $a_1 a_2$ with endpoints 
\be \( u=z_{a1},z=\frac{z_{a1}}{2}+\l_{B1}(z_{a1}^2-1) \), \quad  \( u=z_{a2},z=\frac{z_{a2}}{2}+\l_{B1}(z_{a2}^2-1)  \) \label{a1cood}  \ee in interval $A_1A_2$ is given by
\be S_{A_{1}A_{2}}(a_1 a_2)= 2(\lambda_{B1}-\lambda_{B2}) \label{PEEone} \ee 
We can see that if two points of subinterval $a_1a_2$ lie on the same modular flow line, the PEE of this subinterval would be zero. This implies that points on the same modular flow line correspond to exactly one point on the bifurcating surface $\g_{\xi}$, and different modular flow lines correspond to different points on $\g_{\xi}$. In other words, boundary modular flow lines are in one to one correspondence with points on the bifurcating surface $\g_{\xi}$. This is exactly what happens in the bulk. 
    
On the bulk side to find the specific point on the bifurcating surface $\g_{\xi}$ which corresponds to the boundary point $a_1$, we need choose another boundary point $b_1$, see Figure \ref{fig:PEEswsf}, and demand that the bifurcating surface related to $a_1,b_1$ intersect the one related to $A_1,A_2$, see Figure \ref{fig:BMSmodflow}. Using the parametrization \eqref{pbench}, we find that when $a_1$ have coordinates \eqref{a1cood}, the above intersection condition need us to have
\be u_{b1}=\frac{z_{b1}}{2}+\lambda_{B1}(z_{b1}^2-1)), \quad s_1=\frac{1+4 z_{b1} \lambda_{B}}{2+4 z_{a1} \lambda_{B}+4 z_{b1} \lambda_{B} } , \quad s_2=\frac{1}{2}-2 \lambda_{B} \label{peeint} \ee 
where $(u_{b1},z_{b1})$ is the coordinates of $b_1$ point, and $s_{1,2}$ are the parameters in \eqref{pbench} related to intervals $a_1,b_1$ and $A_1,A_2$ separately. Putting \eqref{peeint} into \eqref{pbench} we can see that the intersection point is
\be \left( -\frac{3}{2} \lambda_{B}, -\frac{1}{2},  \frac{5}{2} \lambda_{B}\right) \label{peebulk} \ee 

Thus we proved our above statement about the one to one correspondence between modular flow line denoted by $\lambda_{B}$ and bulk point on bifurcating horizon \eqref{peebulk}, which is consistent with modular flow invariant property of PEE \cite{Wen:2018whg}. Two features need to be emphasized. One is that the two corresponding points $a_{1}$ and $b_{1}$ can both run on the same modular flow line independently, the specific bulk point would not change. This is not the same as AdS/CFT case, where the two points should run synchronously. Another feature is that the intersection point seldom lie on the finite bench of swing surface, which can be seen from the value of $s_{1,2}$ in \eqref{peeint}. The second feature provides us more evidence that only the information about the finite bench is not enough.

\begin{figure}
    \centering
    \subfigure[]{
    \includegraphics[width=0.46 \textwidth]{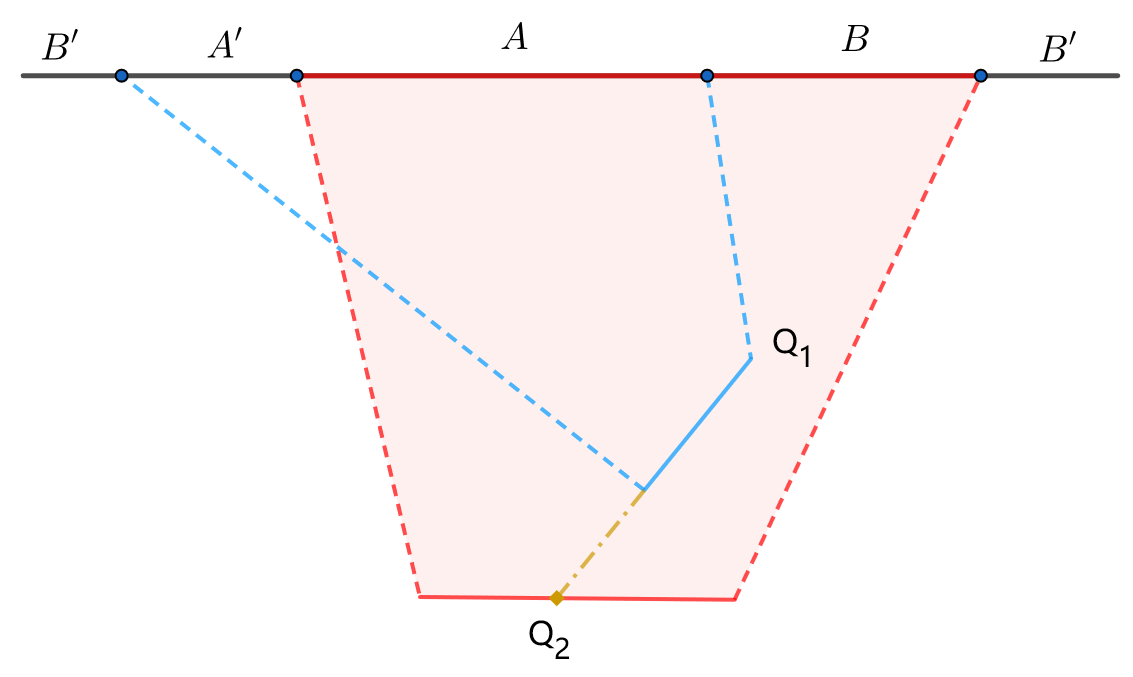}}
    \subfigure[]{
    \includegraphics[width=0.46 \textwidth]{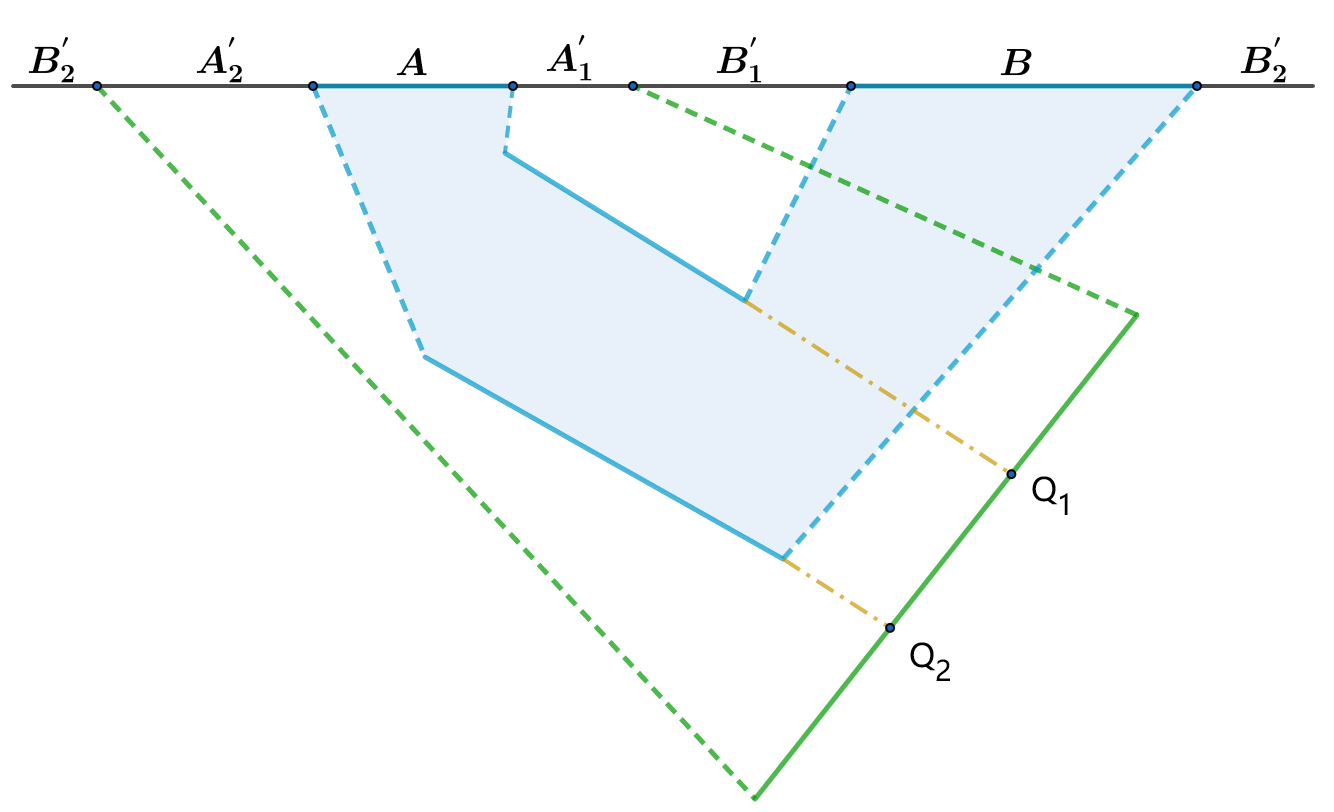}}
    \subfigure[]{
    \includegraphics[width=0.48 \textwidth]{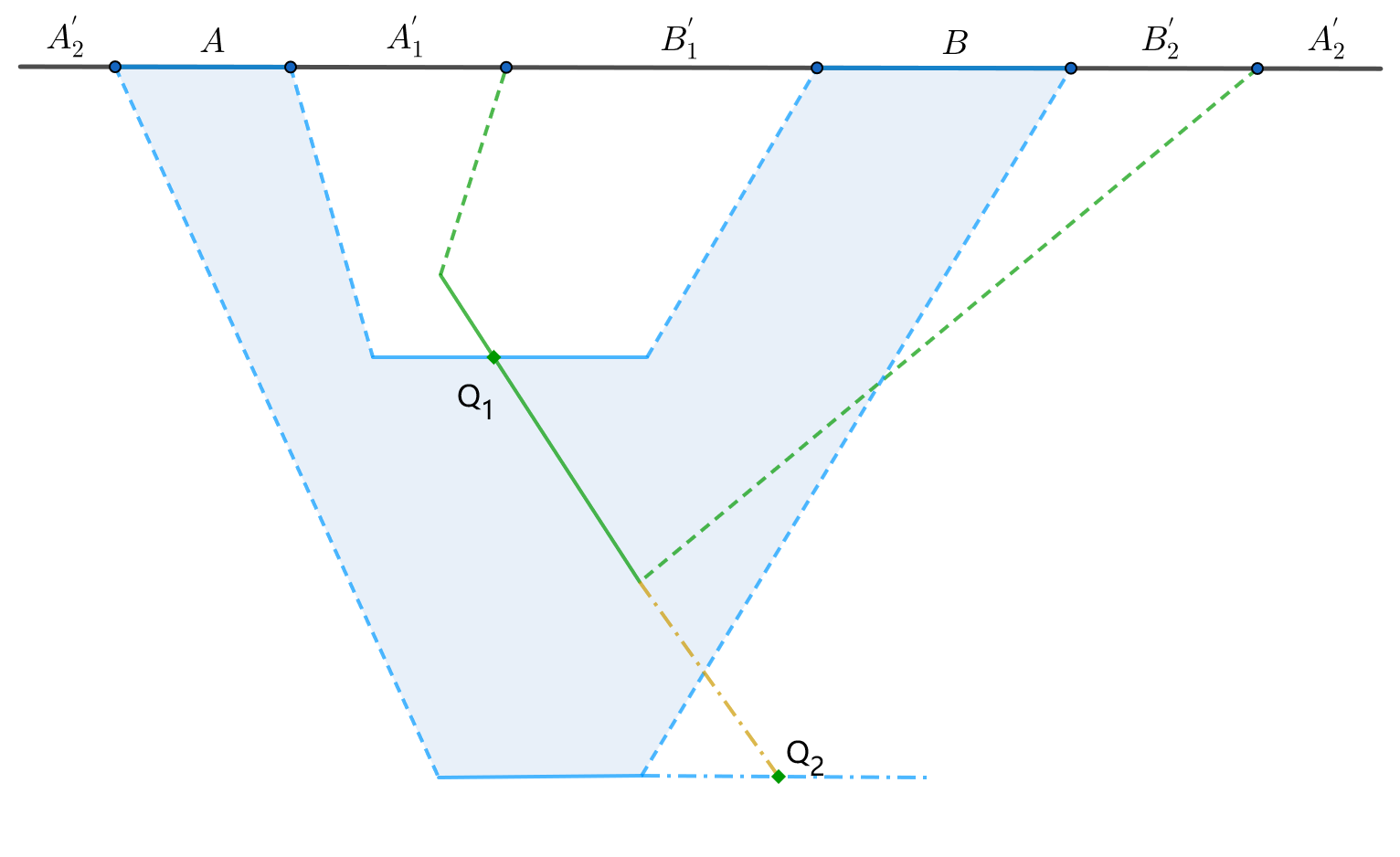}}
    \subfigure[]{
    \includegraphics[width=0.44 \textwidth]{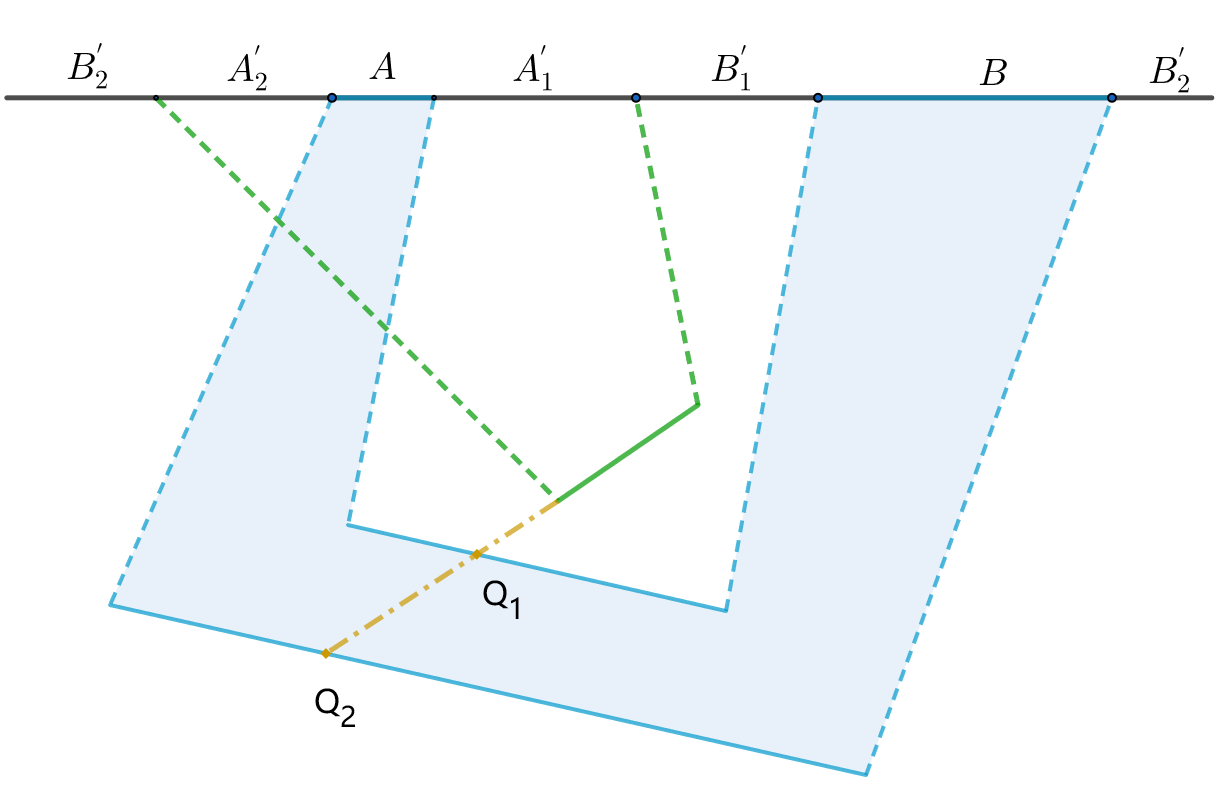}}
    \caption{Configurations about BPE correspondence of general boundary two intervals $A,B$ for both adjacent and non-adjacent cases (but with different colors) in flat$_3$/BMSFT model are presented. Bulk geodesics $Q_1 Q_2$ are the bulk dual of boundary BPE(A:B). Comparisons with those in \cite{Camargo:2022mme,Basu:2022nyl} for preliminary set up and notations are useful. Again the original expected connected entanglement wedge are present to show shortcomings of the original definition.}
    \label{fig:BPEadnad}
    \end{figure}

Because from PEE we can intuitively derive BPE correspondence \cite{Wen:2021qgx}, so we have proved from a more basic step the observations in \cite{Camargo:2022mme,Basu:2022nyl}. We list several unconventional configurations in Figure \ref{fig:BPEadnad} related to adjacent and non-adjacent BPE for completeness. Again we plot the usual expected connected entanglement wedge to show its problem.

\subsection{PEE: boundary and bulk modular flow}
 
In this subsection we use modular flow method to explore the PEE correspondence. Subtleties appear in flat$_3$/BMSFT model compared to AdS/CFT case, which is another manifestation of modular flow property of BMS$_3$ field theory stated in section \ref{sec:intro}. 

We first revisit the modular flow method in standard AdS$_3$/CFT$_2$ case using our notations. This rewriting manifests a freedom of bulk and boundary correspondence as mentioned in section \ref{sectthree}, which is needed for the success of this method in flat$_3$/BMSFT model \footnote{Thank Qiang Wen for discussion about this point.}. We can obtain the modular flow generator for a general interval 
\be \cA= \{ \(-\frac{r_{p}+r_{m}}{2}, -\frac{r_{p}-r_{m}}{2}\),  \(\frac{r_{p}+r_{m}}{2}, \frac{r_{p}-r_{m}}{2}\) \} \label{intA} \ee  in CFT by the coordinate transformations \cite{Casini:2011kv}
\be 
x+t =r_{p} \frac{x_{r}+ t_{r}-1}{x_{r}+ t_{r}+1}, \quad \quad \quad
x-t =r_{m} \frac{x_{r}- t_{r}-1}{x_{r}-t_{r}+1}
\ee 
from Rindler spacetime which have local flow generator 
\be l^{\mu} \propto x_{r} \partial_{t_{r}}+t_{r} \partial_{x_{r}} \ee 
to the causal domain $D[\cA]$ of this interval. Then we get the the modular flow generator $\zeta$ of symmetric interval $\cA$ lying in $t=\frac{r_{p}-r_{m}}{r_{p}+r_{m}} x $ timeslice as follows,
\begin{align}
    & \zeta^{\mu} =r_{m} r_{p} \left[ (r_{p}+r_{m})P_{t}+(r_{p}-r_{m})P_{x}-(r_{p}+r_{m})k^{t}+(r_{p}-r_{m})k^{x}\right] \label{boundflow}  \\
\propto & \( r_{m}^2 r_{p}+r_{p}^2 r_{m}-r_{p}(t-x)^2-r_{m}(t+x)^2    \)\partial_{t}-\( -r_{m}^2 r_{p}+r_{p}^2 r_{m}+r_{p}(t-x)^2-r_{m}(t+x)^2 \)\partial_{x} \nn 
\end{align}  
where $P_{t}, P_{x}, k^{t}, k^{x}$ are the boundary conformal generators: t direction translation, x direction translation, t-component special conformal transformation and x-component special conformal transformation. The explicit expressions of these conformal generators are 
\be \ba 
&P_{t}=\partial_{t}, \quad P_{x}=\partial_{x} \\
k^{t}=(t^2+x^2)\partial_{t}&+2 t x \partial_{x}, \quad  k^{x}=(t^2+x^2)\partial_{x}+2 t x \partial_{t} \label{adsbdfl}
\ea\ee 
The corresponding bulk killing vector fields are 
\be \ba    
&P_{t}=\partial_{t}, \quad \quad \quad \quad  P_{x}=\partial_{x} \\
k^{t}=(t^2+x^2+z^2)\partial_{t}+2 t x \partial_{x}&+2 t z \partial_{z}, \quad  k^{x}=(t^2+x^2-z^2)\partial_{x}+2 t x \partial_{t}+2 z x \partial_{z} \label{adsbbfl}
\ea   \ee 
Using the holographic dictionary between \eqref{adsbdfl} and \eqref{adsbbfl} we can obtain the exact bulk modular flow generator $\xi$ as
\begin{align}
    \xi^{\mu}&= t^{\mu}_{bulk} \partial_{t} + x^{\mu}_{bulk} \partial_{x} +  z^{\mu}_{bulk} \partial_{z}  \label{bulkkilling} \\
&\propto \p_z (-2 (r_m + r_p) t z + 2 (r_p-r_m) x z)+ 
 \p_x (r_m r_p (r_p-r_m)+ (r_p-r_m) (t^2 + x^2 - z^2) \nn \\
& - 2 (r_m + r_p) t x ) + 
 \p_t (r_m r_p (r_m + r_p) + 
    2 (r_p-r_m) t x - (r_m + r_p) (t^2 + x^2 + z^2)). \nn
\end{align}   
We can see that when taking the $z \to 0$ boundary limit, the bulk killing vector field \eqref{bulkkilling} reduces to the boundary modular flow generator \eqref{boundflow}. In the following we choose the interval $\cA$ to lie on the $t=0$ constant time slice withe $r_m=r_p$ in \eqref{intA}. Then from \eqref{bulkkilling} we can easily derive the location of bifurcating surface and bifurcating Killing horizon,
\begin{itemize}
    \item bifurcating Killing horizon: 
    \be -(t^{\mu}_{bulk})^2+(x^{\mu}_{bulk})^2+(z^{\mu}_{bulk})^2=0  \to z=\pm \sqrt{(t\pm R)^2-x^2} \ee 
    
    \item bifurcating surface: 
    \be  t^{\mu}_{bulk}= x^{\mu}_{bulk}= z^{\mu}_{bulk}=0 \ee
    \be \{ t=0, z^2+x^2=R^2\}, \quad \text{Or} \quad \{ t^2=R^2, z=x=0  \} \ee 
\end{itemize}
For explicit manifestation take $r_p=r_m=1$ in \eqref{intA}, we can derive the corresponding parametrization equations of boundary modular flow from \eqref{boundflow},
\be \frac{dx(t)}{dt}=-\frac{2 x(t) t}{1-x(t)^2-t^2} \to x(t)=\frac{1}{2}\( \l_{B}\pm \sqrt{ \l_{B}^2+4 t^2-4} \) \label{bbbdflow} \ee 
where $t$ parametrize one modular flow line and $\lambda_{B}$ distinct different modular flow lines. $t$ and $\lambda_{B}$ together manifest the degree of freedom of 2d plane. Similarly from \eqref{bulkkilling} we get the trajectory of bulk modular flow,
\be
\ba
 x(t)&=\frac{1}{2(1+\lambda_{b1}^2)}\(\lambda_{b2} \pm \sqrt{-4(1-t^2)(1+\lambda_{b1}^2)+ \lambda_{b2}^2 } \), \\
 z(t)&=\frac{ \lambda_{b1}}{2(1+\lambda_{b1}^2)}\(\lambda_{b2}\pm \sqrt{-4(1-t^2)(1+\lambda_{b1}^2)+ \lambda_{b2}^2 } \)
 \label{buflow}
 \ea
 \ee 
where parameters $t,\lambda_{b1},\lambda_{b2}$ together manifest the degree of freedom of 3d bulk. The plus and minus sign in \eqref{bbbdflow} and \eqref{buflow}
denote different branches that we need. When the above mentioned parameters have the relation
\be \lambda_{b1}\to 0, \quad \lambda_{b2}=\lambda_{B}, \ee 
the bulk modular flow line reduce to the boundary one respectively. When they have the relation
\be \l_{b1}=\frac{1}{2} \sqrt{\l_{b2}^2-4}, \label{adsflbd}   \ee
the bulk modular flow line sit on the bifurcating horizons. Choosing a specific co-dimension one plane in 3d bulk by fixing the value of $\lambda_{b2}=\lambda_{B}$ \footnote{Note the plane defined here is not the same as the modular plane in \cite{Wen:2018whg}, which is an explicit manifestation of the freedom mentioned above.}, we can get the boundary and bulk correspondence of PEE, 
\be  \text{boundary:} \( x=\frac{1}{2}(\lambda_{B} \pm \sqrt{\lambda_{B}^2-4}),t=0  \)\leftrightarrow \text{bulk}:\(x=\frac{2}{\l_{B}}, z=\frac{\sqrt{\l_{B}^2-4}}{\l_{B}},t=0\) \label{bmsflowpara}  \ee 
which is consistent with the results in \cite{Wen:2018whg}.  

While for the Poincar\'e vacuum of flat$_3$/BMSFT model, the exact boundary modular flow generator $\zeta$ and the corresponding bulk Killing vector field $\xi$ are \cite{Apolo:2020qjm} as follows,  
\begin{align}
    \zeta & \propto W(z)\partial_{u}+ Y(z)\partial_{z}, \quad
     \xi   \propto W(z)\partial_{u}+ X(z) \partial_{z}-r \partial_{z}X(z) \partial_{r} \label{bbdu1}\\
     X(z)&=   Y(z)-\frac{u}{r}Y''(z)-\frac{1}{r}T'(z), \quad W(z)= T(z)+u Y'(z) \label{bbdu2}  \\
     T(z)&=\frac{2\pi[u_{r}(z-z_{l})^2-u_{l}(z-z_{r})^2]}{(z_{r}-z_{l})^2}, \quad Y(z)=-\frac{2\pi(z-z_{l})(z-z_{r})}{z_{r}-z_{l}}, \quad z\in [z_{l},z_{r}]. \notag
\end{align}
The final expressions for $\zeta$ and $\xi$ are 
\be
\zeta :
\begin{cases} 
&\zeta^{\mu}=\frac{2 \pi}{(z_l-z_r)^2} \left( u_r(z-z_l)^2-u_l(z-z_r)^2+(2z-z_l-z_r)(z_l-z_r)u \right)   \\
&\zeta^{z}=\frac{2 \pi}{(z_l-z_r)^2}  (z-z_l)(z-z_r)(z_l-z_r)  
\end{cases} \label{bbdu3}
\ee
and
\be 
\xi :
\begin{cases}
& \xi^{\mu}= \frac{2 \pi}{(z_l-z_r)^2} \left( u_r(z-z_l)^2-u_l(z-z_r)^2+(2z-z_l-z_r)(z_l-z_r)u \right)\\ 
&\xi^{r}=\frac{2 \pi}{(z_l-z_r)^2} \left[   2(u_r-u_l)+r \left( z_{l}^{2}-z_{r}^{2}+2z(z_r-z_l) \right)   \right]\\ 
&\xi^{z}=\frac{2 \pi}{(z_l-z_r)^2} \left( \frac{2u_l(z-z_r)-2u_r(z-z_l)}{r}+(z-z_l)(z-z_r)(z_l-z_r) \right)
\end{cases} \label{bbdu4}
\ee 
From \eqref{bbdu1}, \eqref{bbdu2} and \eqref{bbdu3}, \eqref{bbdu4} we can see the existence of a consistent boundary limit $\xi|_{r\to \infty}=\zeta$. Again without loss of generality, we set the interval to be $(l_{u}=1,l_{z}=2)$ in \eqref{single}. Then the parametrization equations for boundary and bulk modular flow are
\begin{align}
    & \text{boundary}: \quad u(z)=\frac{z}{2}+(z^2-1) \lambda_{B} \label{bd2} \\
    & \text{bulk}: \;\; u(z)=  \frac{1}{4} \left( z+2 \lambda_{b1} \pm p(\lambda_{b1},\lambda_{b2},z) \right), \;\; r(z)=  \frac{z-2\lambda_{b1} \pm p(\lambda_{b1},\lambda_{b2},z)}{2(1-z^2)} \label{bu2}
\end{align}  
where we have
\be p(\lambda_{b1},\lambda_{b2},z)=\sqrt{z^2+4 \lambda_{b2} z^2-4 \lambda_{b1}z+4 \lambda_{b1}^2-4 \lambda_{b2} }.   \ee 
A careful analysis tell us that when we have relations
\be \{ \lambda_{b2}=-4 \lambda_{B} \lambda_{b1}, \; \lambda_{b1} \to -\infty \}  \Longrightarrow  p(\lambda_{b1},\lambda_{b2},z) \to z-2\lambda_{b1}+4\lambda_{B}(z^2-1)= -\infty \ee 
bulk modular flow trajectory \eqref{bu2} would reduce to the boundary one \eqref{bd2}, where $r(z)$ in \eqref{bu2} goes to $\infty$ that is the boundary limit. When we have relations
\be \l_{b2}=\l_{b1}^2-\frac{1}{4}, \quad z=\frac{1}{2\l_{b1}}  \ee 
bulk modular trajectories lie on the bifurcating horizons, and intersect the bifurcating surface at the point
\be t=\frac{3}{4}\l_{b1}, \quad x=-\frac{1}{2}, \quad y=-\frac{5}{4}\l_{b1} \ee 
Thus when the parameters satisfy
\be \l_{b1}=-2 \l_{B}, \quad \l_{b2}=4 \l_{B}^2-\frac{1}{4}, \quad z=-\frac{1}{4 \l_{B}} \label{flatflbd} \ee
we can get the bulk corresponding point \eqref{peebulk}. Note that in flat$_3$/BMSFT model we need to change the values of both bulk modular flow parameters $\l_{b1}$ and $\l_{b2}$ simultaneously to go from the asymptotic boundary to PEE corresponding point on the bifurcating surface $\g_{\xi}$.

\subsection{Entanglement wedge $\cW_{\cE}^{f}[\cA]$ ?}

\begin{figure}
    \centering
    \subfigure[]{
  \includegraphics[width=0.43\textwidth]{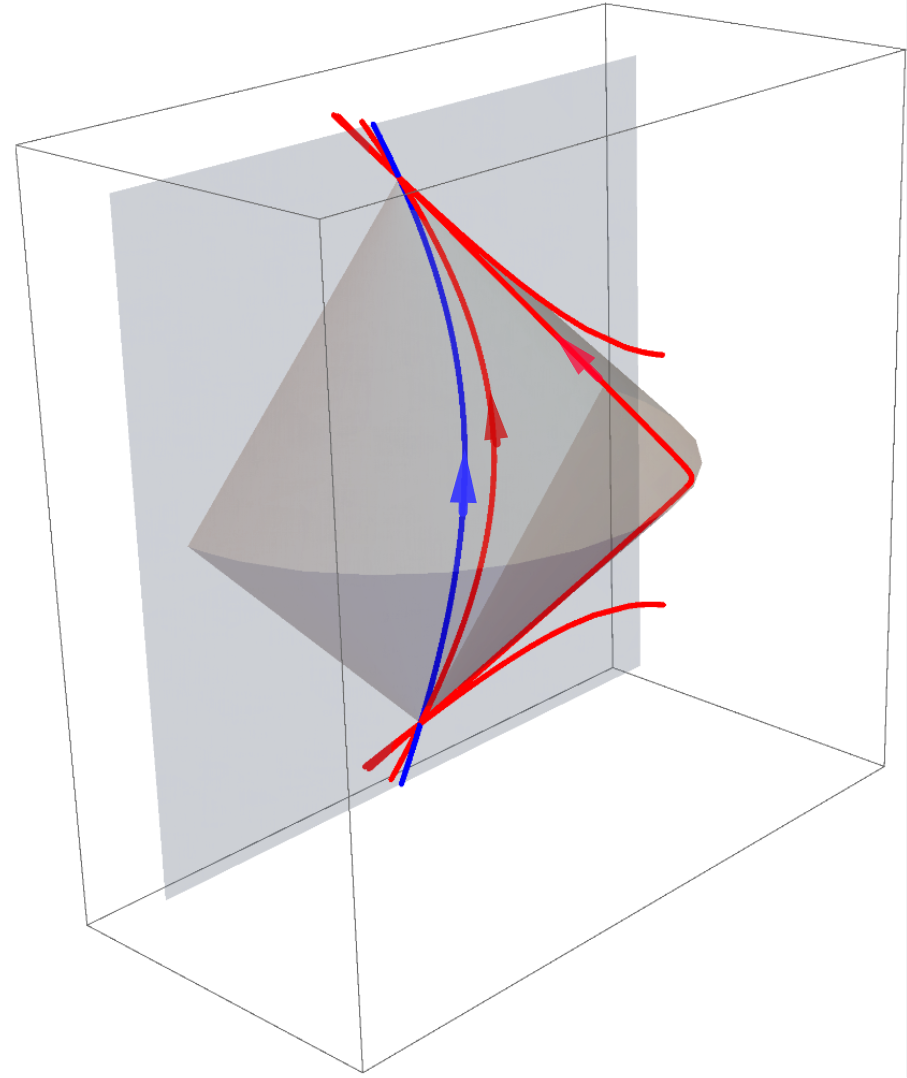} \label{fig:flatflowone} }
    \subfigure[]{   \includegraphics[width=0.47\textwidth]{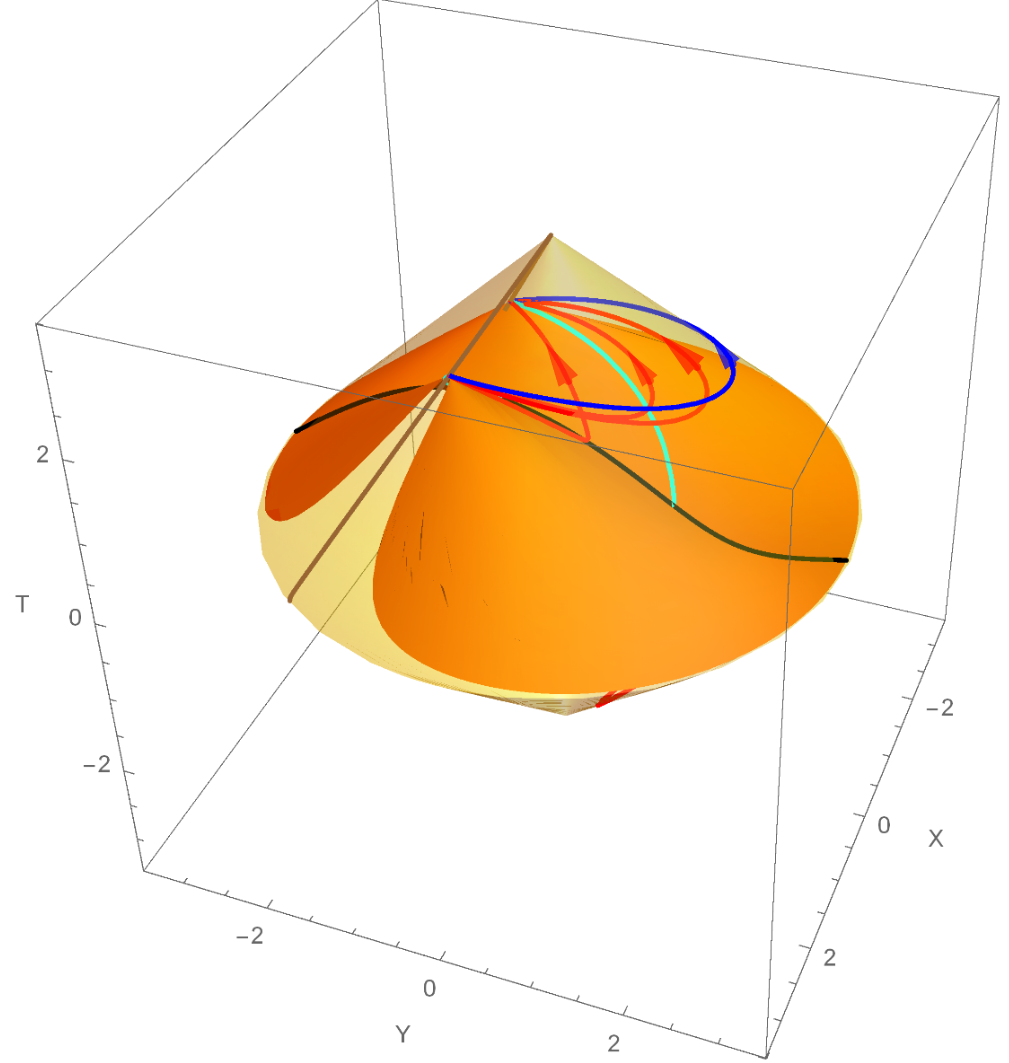} \label{fig:flatflowtwo}  }
    \subfigure[]{
   \includegraphics[width=0.5\textwidth]{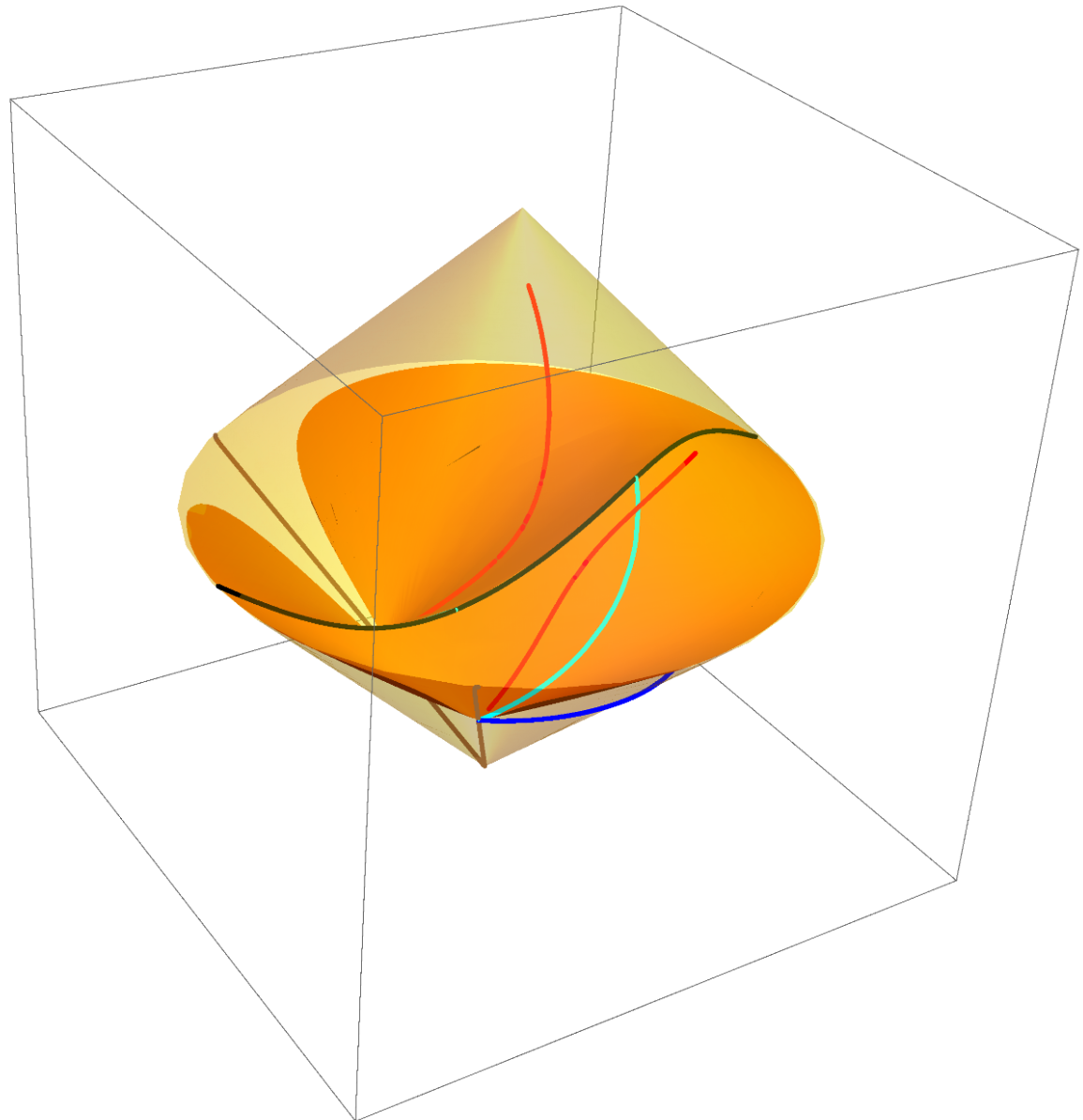} \label{fig:flatflowthree}  }
    \caption{Figure \ref{fig:flatflowone} presents a gradual change of bulk modular flow lines in the Poincar\'e patch of AdS/CFT from asymptotic boundary to bifurcating horizon and finally out of entanglement wedge  $\cW_{\cE}[\cA]$ with parameter data $(0,4),(0.5,4),(\sqrt{3},4),(2,4)$ for $(\l_{b1},\l_{b2})$ in the minus branch of \eqref{buflow}. Figure \ref{fig:flatflowtwo} and \ref{fig:flatflowthree} present a gradual change of bulk modular flow lines in the Poincar\'e vacuum of flat$_3$/BMSFT model from asymptotic boundary to bifurcating horizon and finally out of the special region $\cW_{\cE}^{f}[\cA]$ defined in \eqref{ewflat}. For figure \ref{fig:flatflowtwo} the data are $(-3,0), (-1,-1/3), (0,-0.251)$  for $(\l_{b1},\l_{b2})$ in both branches of \eqref{bu2}; for figure \ref{fig:flatflowthree} the data are $(0,-0.16)$ for $(\l_{b1},\l_{b2})$ in both branches of \eqref{bu2}. In order to see the phenomena  clearly we take a upside down of figure \ref{fig:flatflowthree} compared to figure \ref{fig:flatflowtwo}. The modular time parametrized by $z$ in \eqref{bu2} should go beyond $(-1,1)$ to draw complete pictures shown here.}
    \end{figure}
     
In our parametrization \eqref{bbbdflow} and \eqref{buflow}, there are two features of bulk modular flow in entanglement wedge $\cW_{\cE}[\cA]$ which distinct from those out of entanglement wedge. One is that for both branches the modular time $t$ has parameter range $t\in [-1,1]$ for both boundary and bulk modular flow. Within this finite range of modular time, bulk modular flows reach the bifurcating surface. Out of this range out of the entanglement wedge. Another one is that when $\l_{b1} \geq 0$ exceed \eqref{adsflbd}, the continuous bulk modular flow line will break apart and at the same time go beyond the entanglement wedge of this interval. See figure \ref{fig:flatflowone} for a gradual change with fixed $\l_{b2}=4$ and varying $\l_{b1}$ from $0$ to $2$. 

One complication of the flat$_3$/BMSFT model is that bulk modular flow hit the bifurcating surface out of the range of boundary modular time $z\in [-1,1]$, which can be seen from \eqref{flatflbd}. Apart from this subtlety, similar situations happen as in AdS/CFT case. As we grow parameter $\l_{b1}$ from $-\infty \to -2 \l_{B}$ and at the same time $\l_{b2}$ from $-4 \l_{B} \l_{b1} \to  4 \l_{B}^2-\frac{1}{4}$, the bulk modular trajectories go from the asymptotic boundary to the bifurcating surface. When $\l_{b2}$ exceed the value $ 4 \l_{B}^2-\frac{1}{4}$, the modular lines disconnected into two parts, see Figure \ref{fig:flatflowtwo} and \ref{fig:flatflowthree} for explicitly shown of the process. It turns out that in flat$_3$/BMSFT model the connected bulk modular flow lines can grow from boundary causal domain $D[\cA]$ and go through every point in $\cW_{\cE}^{f}[\cA]$ defined in \eqref{ewflat}, especially the boundary causal domain of the complement interval $D[\cA^{c}]$.

A confusing feature of bulk modular flows in flat$_3$/BMSFT model is that all flow lines are spacelike trajectories, which have spacelike modular time viewing from the global flat$_3$. A related question is how to define the homology surface $\cR_{\cA}$ in this case. They are quantum algebraically and information theoretically related to the semiclassical problems in gravity side. Algebraically the bulk modular Hamiltonian $H_{b}$ can map the algebra of operators $\cA_{B}$ localized in a region $B$ to the algebra of operators $\cA_{B'}$ in the region $B'$,
\be U(s)\cA_{B} U(-s)= \cA_{B'}, \quad U(s)=\r^{i s}=e^{-i H_{b} s} \ee 
along the bulk modular flow lines. When the modular evolution is spacelike from bulk point of view, what is its meaning? Quantum informationally the first quantum correction to holographic entanglement entropy, 
\be S_{\cA}=\frac{Area(\g_{\cA})}{4G}+S_{bulk}+ \cO(1/c) \label{homosurface1} \ee 
need homology surface $\cR_{\cA}$ to compute the reduced density matrix  $\r_{bulk}$. How to define it in this case? We hope more solid study on these key open problems in the future.

\section{Two interval entanglement phase transition and EWN}
\label{section5}

In order to compute the reflected entropy, entanglement negativity, odd entropy or other mixed state entanglement measures from bulk side, we need to have a connected entanglement wedge. When changing the relative configurations of two boundary intervals, there is a phase transition between disconnected entanglement wedge and the connected one. Without specifying clearly what is the connected entanglement wedge from gravity side, all previous calculations regarding the EWCS \cite{Basak:2022cjs,Camargo:2022mme}, including the ones in this paper, are problematic. We give criterion for two intervals phase transition from field theory point of view, which is not trivial already. From the gravity theory point of view, the situations are rather vague. 

\subsection{Entanglement phase transition}

\begin{figure}
    \centering
    \includegraphics[width=0.45\textwidth]{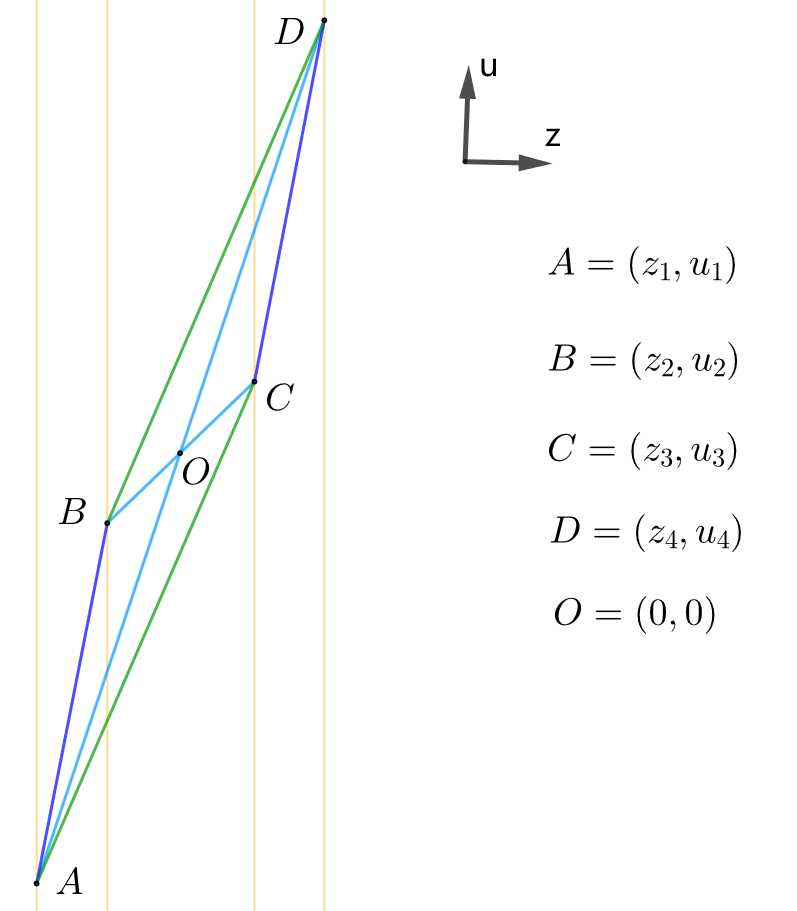}
    \caption{For symmetric configuration of four points $A$, $B$, $C$ and $D$ in BMS$_3$ field theory, there are three competing combination of intervals to become the minimal one. We show the coordinates of these four points and the three pairings with different colors.}
    \label{fig:threeways}
\end{figure}
In this part we consider the entanglement phase transition for two disjoint boundary intervals using formula \eqref{eevac} with $c_L=0$. Compared to the usual CFT case, here we should consider three different pairs rather than two to finally get the minimal one, see Figure \ref{fig:threeways}. If we connect points $B,C$, and $A,D$, the corresponding holographic entanglement entropy is given by
\be
    S_1=c_M \( \frac{u_2-u_3}{z_2-z_3}+ \frac{u_1-u_4}{z_1-z_4}   \)
\ee  
Similarly, we define
\be 
S_2= c_M \( \frac{u_1-u_2}{z_1-z_2}+ \frac{u_3-u_4}{z_3-z_4}   \) 
, \quad S_3=c_M \( \frac{u_1-u_3}{z_1-z_3}+ \frac{u_2-u_4}{z_2-z_4}   \) \ee
The difference between them are
\be
    S_1-S_2= \frac{u}{z(z-1)}, \quad
    S_2-S_3= \frac{u}{z} \quad
    S_1-S_3= \frac{u}{z-1}
\ee
where $u,z$ are cross ratios defined in \eqref{bmscrt} with identifications $z=x,t=u$. Therefore we get
\begin{itemize}
    \item $S_1$ is the minimal one when:
    \be  u<0,\quad z>1 \quad \mathrm{or}\quad u>0,\quad 0<z<1 \label{cl1} \ee
    \item $S_2$ is the minimal one when:
   \be u<0,\quad 0<z<1 \label{cl2} \ee
    \item $S_3$ is the minimal one when:
    \be  u<0,\quad z<0 \quad \mathrm{or}\quad u>0,\quad z>1 \label{cl3} \ee    
\end{itemize}
Taking symmetric configurations for an example,  
\be
    u_a=-u_2=u_3,\;\;z_a=-z_2=z_3,\;\; u_b=-u_1=u_4,\;\;z_b=-z_1=z_4 \label{symintervals}
\ee
with $z_b>z_a$. This configuration ensures that $0<z<1$, which means that $S_3$ can never be the minimal one. Fixing $z_a,z_b$ and $u_a$, there is a critical value
\be  u_c= \frac{u_a z_b }{z_a}\ee 
for $u_b$ such that $S_1=S_2$. When $u_b<u_c$, $S_2$ is the minimal one and vice verse. we can check the difference between the slope of the critical interval $B$ with $u_b=u_c$ and that of interval $A$ is
\be
    \frac{u_c}{z_b}-\frac{u_a}{z_a}=0
\ee
When $S_1$ is the minimal one, we think the configuration have a connected entanglement wedge like the case of AdS/CFT. 

\subsection{Entanglement wedge nesting}

\begin{figure}
    \centering
    \subfigure[]{
\includegraphics[width=0.4\textwidth]{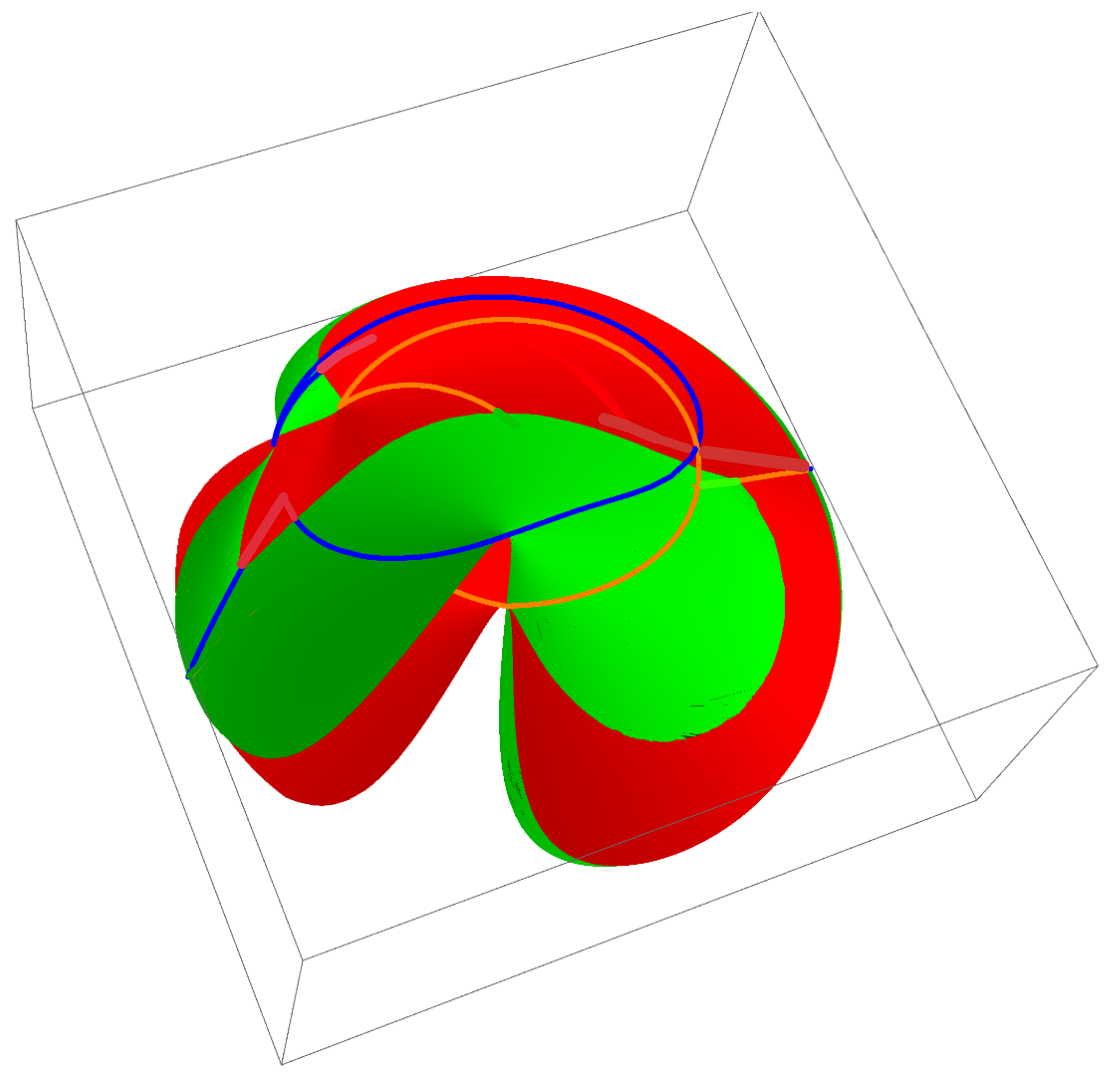} \label{fig:coEWone} }
    \subfigure[]{
\includegraphics[width=0.45\textwidth]{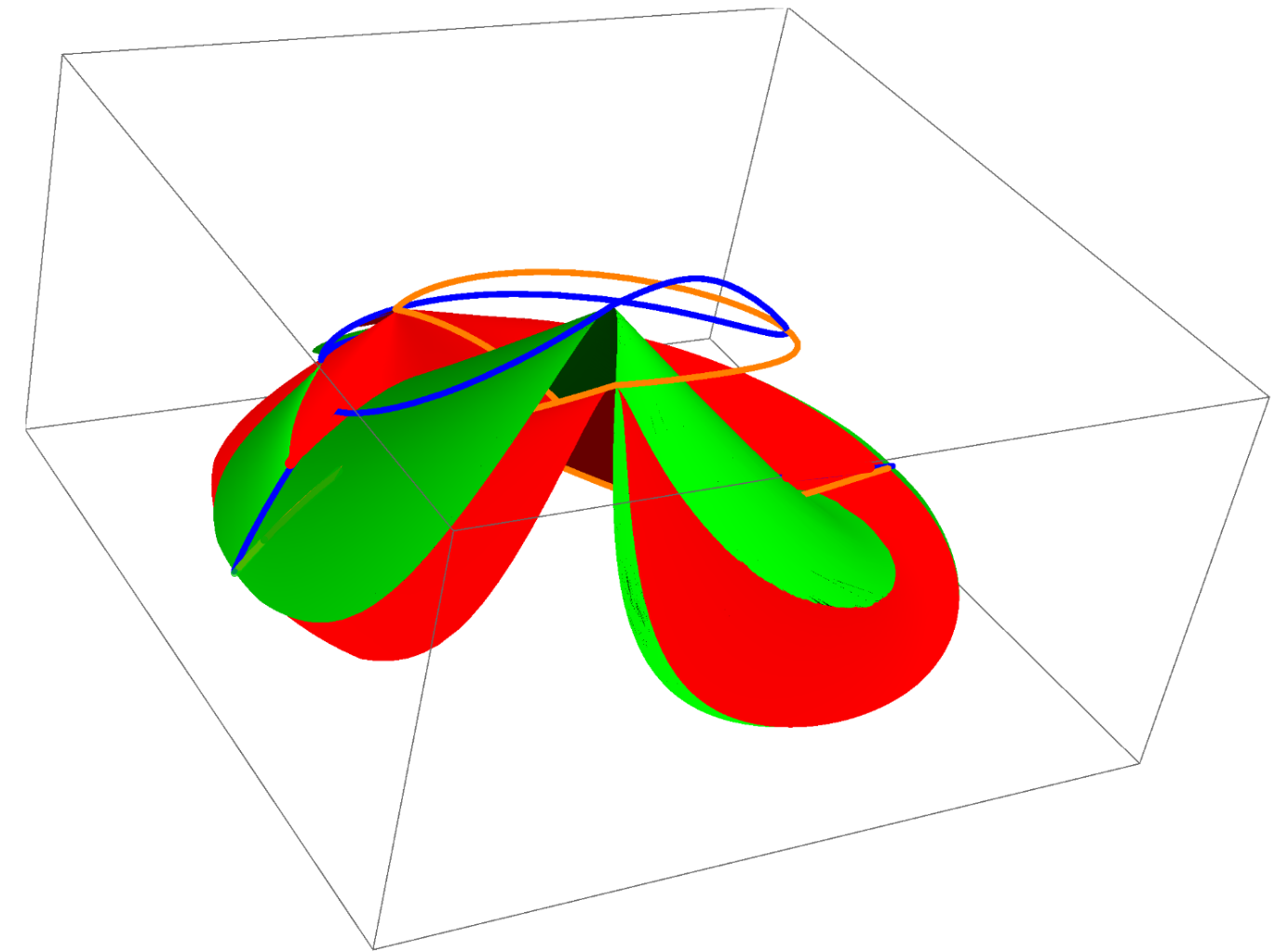} \label{fig:coEWtwo} }
    \subfigure[]{
\includegraphics[width=0.4\textwidth]{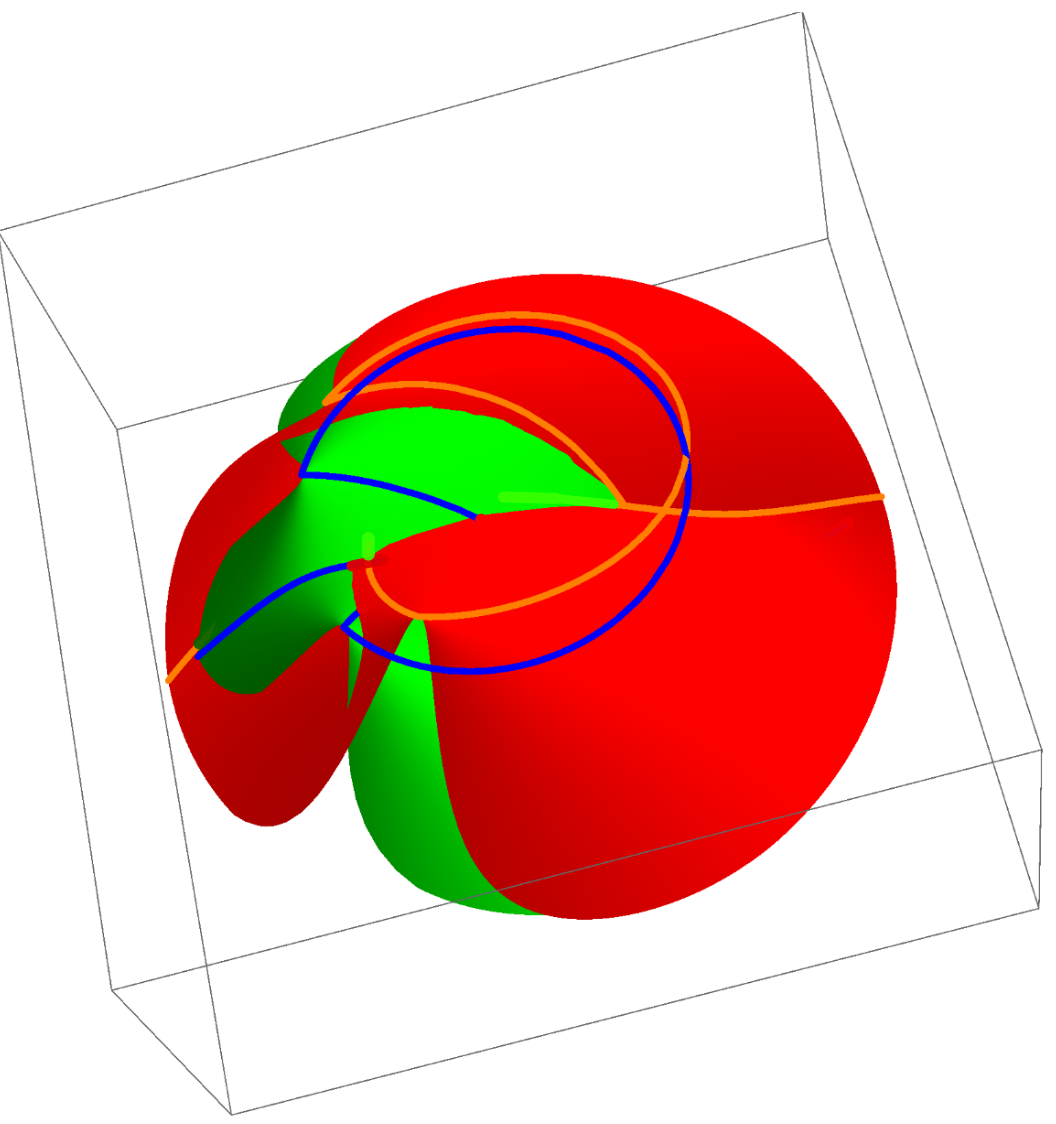} \label{fig:disEWone} }
    \subfigure[]{
\includegraphics[width=0.45\textwidth]{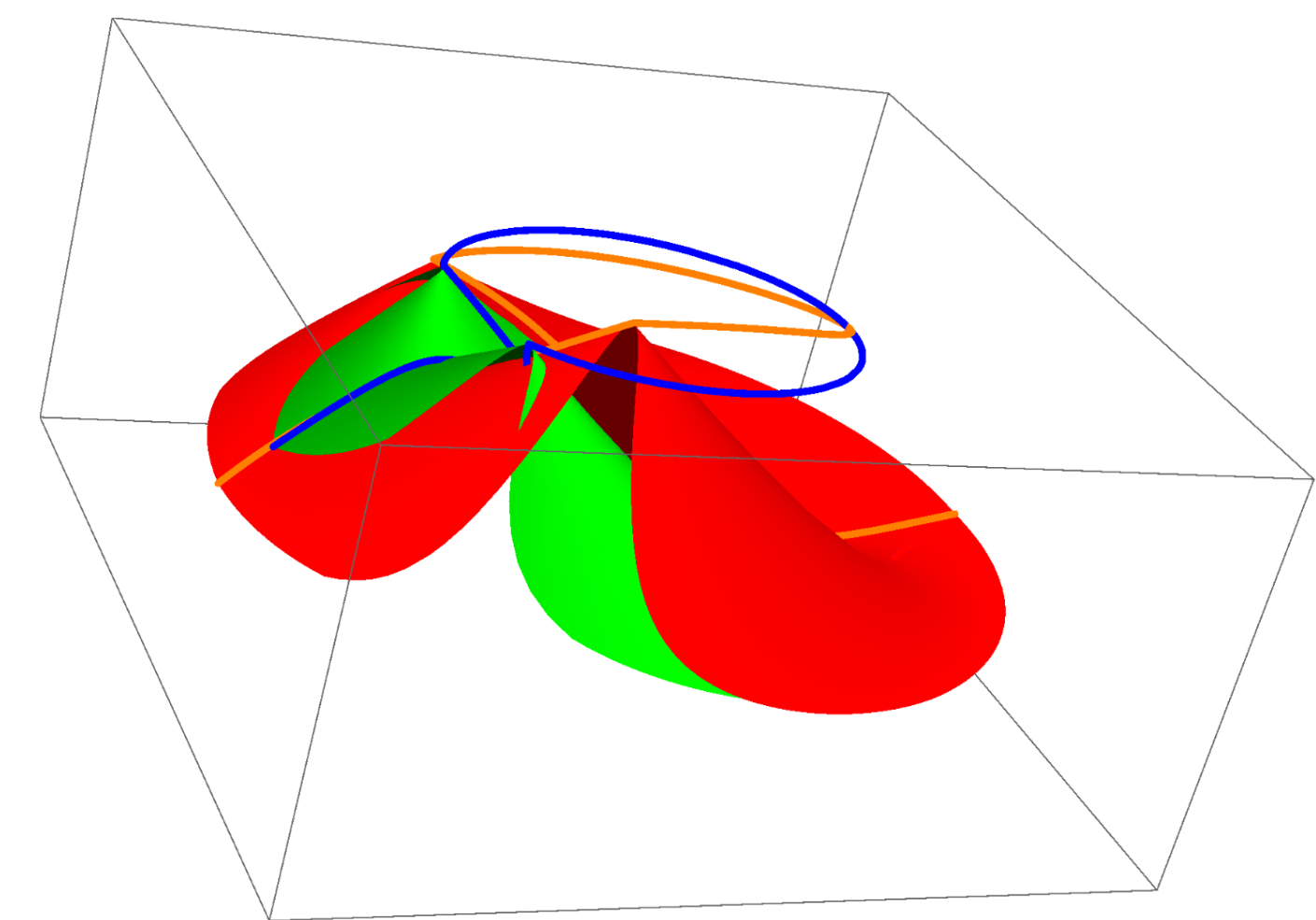} \label{fig:disEWtwo} }
    \caption{Figure \ref{fig:coEWone} and \ref{fig:coEWtwo} show the future bifurcating horizons (Red) of interval (Orange) with $(l_u=1,l_z=2)$ and future bifurcating horizons (Green) of interval (Blue) with $(l_u=4,l_z=2.6)$ from different perspective. This two intervals satisfy \eqref{cl1} and should have a connected entanglement wedge from boundary point of view. Similarly figure \ref{fig:disEWone} and \ref{fig:disEWtwo} show the future bifurcating horizons (Red) of interval (Orange) with $(l_u=1,l_z=2)$ and future bifurcating horizons (Green) of interval (Blue) with $(l_u=1/2,l_z=4)$ from different perspective. This two intervals satisfy \eqref{cl2} and should have disconnected entanglement wedge. Due to our limit abilities, we can not find true difference between the two cases considering the entanglement wedge nesting (EWN) property. }
    \label{fig:coEWNzong}
    \end{figure}

Entanglement wedge nesting (EWN) is a prerequisite of existing a connected entanglement wedge in AdS/CFT.  EWN states that
nested boundary regions should be dual to nested bulk regions, which clearly 
consistent with subregion duality \cite{Casini:2017vbe,Akers:2017ttv} in AdS hologrpahy. According to the last section, region $\cW_{\cE}^{f}[\cA]$ defined in \eqref{ewflat} is a special region owing many similar properties of the entanglement wedge $\cW_{\cE}[\cA]$ in AdS/CFT. However when considering the EWN property of two boundary intervals, this special region failed. Actually, no special region can be found due to our limited capabilities, see figure \ref{fig:coEWNzong}. This is really a big problem in the flat$_3$/BMSFT model, which puts all the previous calculations about EWCS into a dangerous situation.

\section{Conclusions and Open Questions }
\label{section6}

Until now, we have showed the usual and unusual features related to causality structure of bifurcating surface $\g_{\xi}$ in flat$_3$/BMSFT model using the tools from modular flow and various entanglement measures (mainly Reflected entropy and PEE). We studied the two intervals phase transition as well as EWN problem from both field side and gravity side. However due to the observations presented in Section \ref{sectthree}, there are still two big problems special to flat$_3$/BMSFT model intentionally hidden by us. \\

\textbf{The existence of Entanglement wedge?}\\

In flat$_3$/BMSFT model the holographic entanglement entropy does not own just the properties of pure gravity, i.e., the length, but also contains the "direction" character even in pure Einstein gravity. This phenomena appear in all the information measures, reflected entropy, entanglement negativity, odd entropy and PEE. Actually as we argued in subsection \ref{subsecnegt}, this is a general property of flat$_3$/BMSFT model, which physically puts a question mark on the existence or the meaning of entanglement wedge in this model. Historically the success of "It from qubit" program, for example, the subregion-subregion duality \cite{Dong:2016eik}, in Einstein gravity of AdS/CFT holography root in the fact that people can totally geometrize the boundary entanglement entropy $S_{\cA}$. The mathematical object of bulk dual of boundary entanglement entropy is just an area functional with no correction term and no additional character. Maybe the fact that we can't totally geometrize $S_{\cA}$ in flat$_3$/BMSFT model imply us that there is no well defined and useful notion about entanglement wedge in this toy model. Let us make an analogy to make this point clearer. Take "Topological Massive gravity/anomalous CFT"   correspondence as an example, the holographic entanglement entropy contain corrections due to the Chern Simons term \cite{Castro:2014tta}, but the position of RT surface is the same as pure gravity case. There are spacetimes, for example the vacuum AdS$_3$, that are the solutions of both pure Einstein gravity theory and topological massive gravity theory. In these kind of spacetimes, the positions of RT surfaces and related null bifurcating horizons are the same. However the boundary dual CFTs are rather different. How can the same bulk region encoded different boundary information? Actually there are no solid results of entanglement wedge and bulk reconstruction in this duality. Another possibility would be to find more fine structures of the special bulk region $\cW_{\cE}^{f}[\cA]$ in \eqref{ewflat}. We hope more solid works on this problem due to its substantial role in flat$_3$/BMSFT holography. \\

\begin{figure}
    \centering
    \subfigure[]{
    \includegraphics[width=0.44\textwidth]{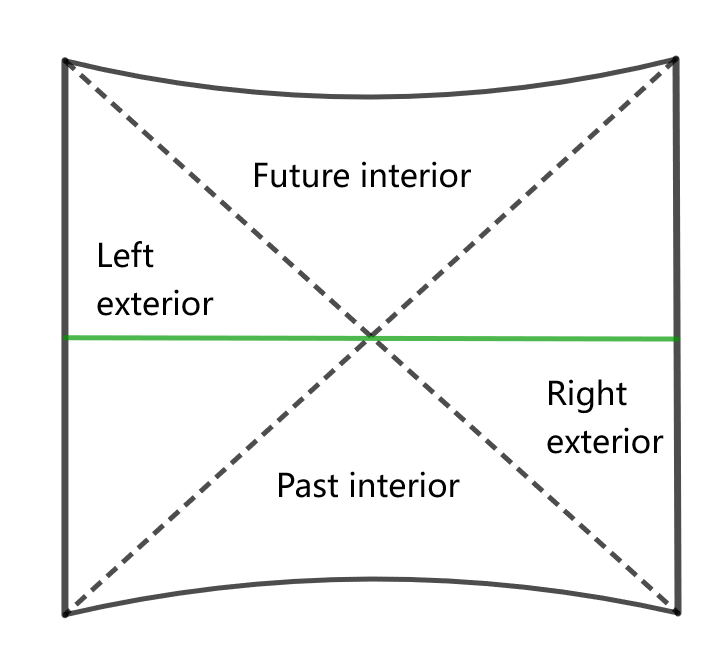} \label{fig:eteBH} }
    \subfigure[]{
    \includegraphics[width=0.46\textwidth]{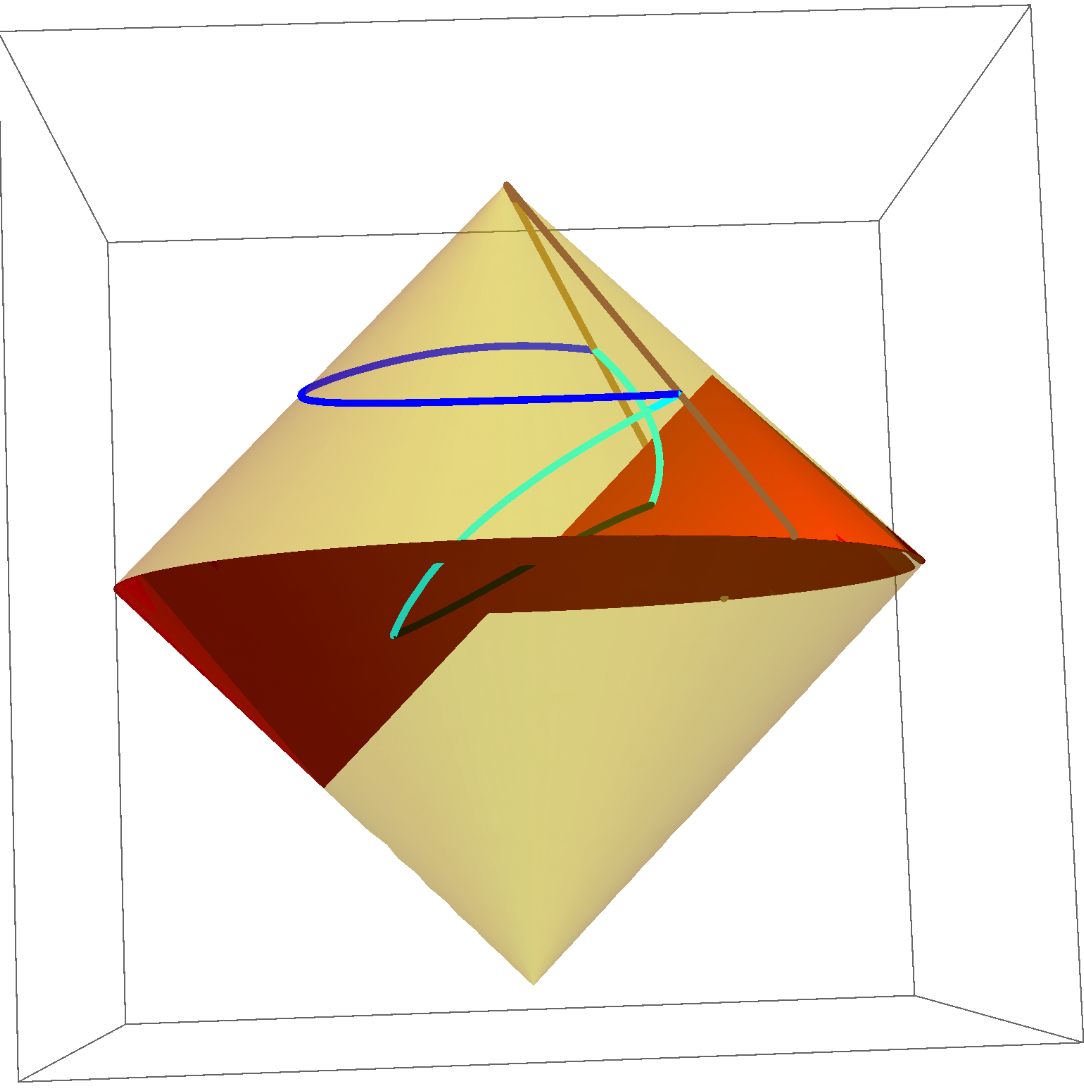} \label{fig:peneflat} }
    \caption{ Figure \ref{fig:eteBH} shows the standard Penrose diagram of eternal black hole in AdS/CFT holography. Figure \ref{fig:peneflat} shows the relative position of swing surface with respect to the boundary of the quotient manifold--the Poincar\'e vacuum. }
    \end{figure}
    
\textbf{Flat holography: which boundary? }\\

Flat spacetime has more complicated asymptotic boundaries than AdS spacetime, thus an important question about flat holography is where the dual boundary field theory lives. Flat$_3$/BMSFT model provides us a vague but unexpected implication on this question.  
    
We made the observation in subsection \ref{bdhorz} and proved it in \eqref{anylt} that the bench $\g$ always penetrate beyond the boundary of original spacetime, see Figure \ref{fig:peneflat}. In other words, we have to analytically continuate the original spacetime to include just the finite bench $\g$. Viewing from gravity, analytic continuation of spacetime is quite normal. However from holographic point of view, no similar things happened before. For example in AdS/CFT holography, RT surface always totally lie within the quotient spacetime, no matter for the Poincar\'e patch or BTZ black holes. This may imply that although field theory live just on the future null infinity $\mathscr{I}^{+}$, we also need information on other asymptotic boundaries especially the past null infinity $\mathscr{I}^{-}$. 

Let's make an analogy with the standard eternal black hole case which is dual to the thermofield double state (TFD state) in AdS/CFT holography \cite{Maldacena:2001kr}, see Figure \ref{fig:eteBH}. The Left and right boundaries are spacelike separated with each other, and their CFT Hamiltonians have no coupling. In TFD state, any interval located within left(right) boundary has RT surface totally sit in left(right) exterior and would not penetrate through horizon\footnote{If the interval is the whole left (right) boundary, then the RT surface is just the bifurcating horizon.}. Only when the boundary interval contains part of both left and right CFTs can the RT surface behave like the green curve in Figure \ref{fig:eteBH}. The flat holography case has similar things to some extent, but is more subtle due to the fact that the future null infinity $\mathscr{I}^{+}$ and the past one $\mathscr{I}^{-}$ are timelike separated, which may communicate or contain same but modulated information.

\appendix

\section{Reflected Entropy}
\label{appendixa}

In this appendix, we give a complete derivation about the reflected entropy $S_{Ref}^{\text{BMS}}$ of two disjoint intervals in BMS field theory. For the steps to be self-contained, we have to show some similar calculations with those in \cite{Basak:2022cjs}. The new thing that is not shown in \cite{Basak:2022cjs} is the step of deriving three point coefficient in the OPE of twist operators \eqref{thrope}, which affects the final results up to a constant \cite{Basu:2022nyl}. See \cite{Chen:2022fte} for the notations, figures and detailed introduction of reflected entropy. 

\subsection{The BMS Semi-classical Block}

The function $\mathcal{F}(x,t)$ in \eqref{ffourpint} can be decomposed into BMS conformal blocks $\mathcal{F}_{\alpha}(x,t) $,
\be \mathcal{F}(x,t)=\sum_{\mathcal{\alpha}} C_{12\mathcal{\alpha}} C^{\mathcal{\alpha}}_{34} \mathcal{F}_{\alpha}(x,t) \label{blexp1} \ee
In the semi-classical limit a closed form for $\mathcal{F}_{\alpha}(x,t)$, which is also consistent with the ultra-relativistic limit of the Virasoro block, is obtained by the null vectors of the BMS algebra as well as the monodromy method \cite{Hijano:2018nhq},
\begin{align}
    \mathcal{F}_{\alpha}(x,t) & \sim \bigg( \frac{x^{\beta-1}}{(1-x^{\beta})^2} \bigg) ^{\Delta_L} e^{t\big( \frac{\beta x^{\frac{\beta}{2}} }{x(x^{\beta}-1)} \xi_{\alpha}-\frac{x^{\beta}(\beta+1)+\beta-1}{x(x^{\beta}-1)} \xi_{L}  \big) } \notag \\
    & \times \bigg( \frac{1-x^{\frac{\beta}{2}}}{1+x^{\frac{\beta}{2}}} \bigg)^{\Delta_{\alpha}} e^{\Delta_{H}\log{x} \big( \frac{2 x^{\frac{\beta}{2}} }{\beta (x^{\beta}-1)} \xi_{\alpha}+\frac{2(x^{\beta}+1)}{\beta (1-x^{\beta})} \xi_{L} \big)}
\end{align}
where $\beta=\sqrt{1-24\frac{\xi_{H}}{c_M}}$, $\Delta_{L,H},\xi_{L,H}$ are the conformal weight and boost charge \eqref{wetchg} related to external light/heavy operators and $\Delta_{\alpha},\xi_{\alpha}$ are the ones related to internal operators in the OPE expansion. This Heavy-Heavy-Light-Light correlator need heavy operators scale freely with the central charge $c_{M}$ and light operators obey $1 \ll \xi_{L}, \Delta_{L} \ll c_{M}$. Analytically continuing the quantum numbers of heavy operators to the light one, we get the BMS block $\cF_{\alpha}$ of four same dimension light operators
\begin{align}
     \log \mathcal{F}_{\alpha}(x,t) & \sim \underbrace{t \frac{1}{(-1+x) \sqrt{x}} \xi_{\alpha} + \log{(\frac{1-\sqrt{x}}{1+\sqrt{x}})} \Delta_{\alpha}}_{will\ contr.\ to\ the\ RE} \notag 
     \\ & +\underbrace{t \frac{2}{1-x} \xi_{L}+ [2 \log{(1-x)}+\big( \frac{2(1+x)}{1-x}\xi_{L}+\frac{2\sqrt{x}}{-1+x} \big) ] \Delta_{L}}_{will\ cancel\ by\ normalization}
     \label{bmsblock1}
\end{align}

\subsection{OPE coefficient and Twist operator dimension} 
 
We assume the primary twist operators in orbifold BMSFT all belong to the singlet version of the highest weight representation of BMS algebra \cite{Hao:2021urq}. Then the twist operators $\sigma_{g_{A}}$, $\sigma_{g_{B}}$ and $\sigma_{g_{A}g_{B}^{-1}}$ have the following dimensions,

\begin{itemize}
    \item $\sigma_{g_{A}}$, $\sigma_{g_{B}}$, $\sigma_{g_{A}^{-1}}$, $\sigma_{g_{B}^{-1}}$:
    \be \xi_{g_{A,B}}=n \xi_{m}=n \frac{c_{M}}{24}(m-\frac{1}{m}), \quad  \Delta_{g_{A,B}}=n \Delta_{m} =n \frac{c_{L}}{24}(m-\frac{1}{m})\ee

    \item $\sigma_{g_{A}g_{B}^{-1}}$:
    \be \xi_{g_{A}g_{B}^{-1}}=2\xi_{n}=2 \frac{c_{M}}{24}(n-\frac{1}{n}), \quad   \Delta_{g_{A}g_{B}^{-1}}=2\Delta_{n}=2 \frac{c_{M}}{24}(n-\frac{1}{n}) \label{thrope} \ee 
\end{itemize}
For the three point OPE coefficients important for the final results of reflected entropy, we claim that 
\be \sigma_{g_{A}^{-1}} \sigma_{g_{B}} = C_{nm}^{\text{BMS}} \sigma_{g_{B} g_{A}^{-1}}+...\quad , \quad  C_{nm}^{\text{BMS}  }=(2m)^{-2\Delta_{n}} \ee
This can be proved by using the same method in CFT \cite{Dutta:2019gen} and WCFT \cite{Chen:2022fte}. We show the main steps here:
\begin{align}
    & \langle \sigma_{g_{A}^{-1}} (x_{1},y_{1}) \sigma_{g_{B}}(x_{2},y_{2}) \sigma_{g_{A} g_{B}^{-1}}(x_{3},y_{3}) \rangle_{\text{BMS}^{\otimes mn}(plane)} \nn \\
   =& e^{S_{L}(\phi)}\underbrace{\Big\lvert
  \partial f'
   \Big\lvert^{-h^{L}_{n}}_{f=s_{1}^{+}} \Big\lvert \partial f' \Big\lvert^{-h^{L}_{n}}_{f=s_{2}^{+}} e^{-h^{M}_{n} \left( \frac{g'+s_{1}^{-}f'' }{f'}\Big\lvert_{ \{f=s^{+}_{1},g=s^{-}_{1}\} } + \frac{g'+s_{2}^{-}f'' }{f'}\Big\lvert_{ \{f=s^{+}_{1},g=s^{-}_{1}\} } \right)  }}_{\left( \cD_{1}\right)} \nn  \\
   \times &  \underbrace{  \langle \sigma_{(\tau_{n}^{0})^{-1}}(s_{1}^{+},s_{1}^{-}) \sigma_{\tau_{n}^{m/2}}(s_{2}^{+},s_{2}^{-}) \rangle_{\text{BMS}^{\otimes n}(plane)} }_{\left( \cD_{2}\right)} \nn \\
     =& \left. \left( \langle \sigma_{g_{A}^{-1}} (x_{1},y_{1}) \sigma_{g_{B}}(x_{2},y_{2}) \rangle_{\text{BMS}^{\otimes m}(plane)}\right)^{n} \right|_{A=B} \times \left( \cD_{1} \cD_{2} \right) \nn \\
     =&  e^{-2\xi_{n} (\frac{y_{32}}{x_{32}}+\frac{y_{31}}{x_{31}}-\frac{y_{21}}{x_{21}} )-2n\xi_{m}\frac{y_{21}}{x_{21}}}
         (2m)^{-2 \Delta_{n}}|x_{32}|^{-2 \Delta_{n} }|x_{31}|^{-2 \Delta_{n}} |x_{12}|^{-2n  \Delta_{m} +2 \Delta_{n}} \label{bmsope1}
   \end{align}
the first equality comes from a BMS symmetry transformation 
\be s^{+}=f=\frac{(x-x_{1})^{1/m}}{(x-x_{2})^{1/m}}, \quad s^{-} = g= \frac{f }{m(x-x_{1})(x-x_{2})} [t x_{12}-x t_{21}-x_{1}t_{2}+x_{2}t_{1} ]\label{mapf1} \ee 
which maps the $mn$ replica sheets to a $n$ replica sheets. The twist operators $\sigma_{(\tau_{n}^{0})^{-1}}(s_{1}^{+},s_{1}^{-})$, $\sigma_{\tau_{n}^{m/2}}(s_{2}^{+},s_{2}^{-})$ have quantum numbers $\Delta_{n}$ and $\xi_{n}$ due to their $n$-cyclic monodromy conditions getting from the above map \eqref{mapf1}, and the explicit values of $s^{\pm}_{1,2}$ are
 \begin{align}
     & s_{1}^{+}=-s_{2}^{+}=\frac{(x_{3}-x_{1})^{1/m}}{(x_{3}-x_{2})^{1/m}} \nn \\
     & s_{1}^{-}=-s_{2}^{-}=\frac{s_{1}^{+}}{m(x_{3}-x_{1})(x_{3}-x_{2})}[t_{3}x_{12}-x_{3}t_{21}-x_{1}t_{2}+x_{2}t_{1} ] 
 \end{align}
From the result (\ref{bmsope1}) we can directly see the OPE coefficient $C_{nm}^{\text{BMS}}=(2m)^{-2\Delta_{n}}$ as claimed. 

\subsection{Reflected entropy of vacuum and thermal state on the plane }

In the holographic BMS field theory we assume that the single block dominance work in the semi-classical limit, and the dominant BMS block in the block expansion of the four point function \eqref{blexp1} is the one with lowest quantum dimensions. For $t$-channel OPE of twist operator $\s_{g_{A}}\s_{g_{B}^{-1}}$, the dominant one is related to the primary twist operator $\sigma_{g_{B}g_{A}^{-1}}$. By taking the Von-Neumann limit $n,m\to1$, the external twist operators $\sigma_{g_{A,B}}$ and the internal one $\sigma_{g_{B}g_{A}^{-1}}$ all become light operators, then (\ref{bmsblock1}) can be used to evaluate the reflected entropy, 
\begin{align}
       & \left \langle \sigma_{{g}_A}(x_1,t_{1}) \sigma_{{g}^{-1}_A}(x_2,t_{2}) \sigma_{{g}_B}(x_2,t_{2}) \sigma_{{g}^{-1}_B} (x_4,t_{4})\right \rangle _{ \text{BMS}^{\otimes mn}} \label{4pf1}  \\ 
      &=\frac{e^{-2 \xi_{g_{A}}\frac{t_{12}}{x_{12}}-2 \xi_{g_{B}}\frac{t_{34}}{x_{34}} } }{ x_{12}^{2\Delta_{g_{A}}} x_{34}^{2\Delta_{g_{B}}}  }  \sum_{\alpha} C_{AB\alpha}^{2} \mathcal{F}_{\alpha}(mnc,\Delta_{i},\xi_{i},\Delta_{\a},\xi_{\a},x,t) \nn \\
      &\approx \underbrace{ \frac{e^{-2 \xi_{g_{A}}\frac{t_{12}}{x_{12}}-2 \xi_{g_{B}}\frac{t_{34}}{x_{34}} } }{ x_{12}^{2\Delta_{g_{A}}} x_{34}^{2\Delta_{g_{B}}}  } }_{\text{cancel out }}  \underbrace{e^{t \frac{2}{1-x} \xi_{L}+ [2 \log{(1-x)}+\big( \frac{2(1+x)}{1-x}\xi_{L}+\frac{2\sqrt{x}}{-1+x} \big) ] \Delta_{L}}}_{\text{cancel out}} \notag \\
      & \times \underbrace{ \left(C_{nm}^{\text{BMS}}\right)^{2} e^{t \frac{1}{(-1+x) \sqrt{x}} \xi_{\alpha} + \log{(\frac{1-\sqrt{x}}{1+\sqrt{x}})} \Delta_{\alpha}} \Big\lvert_{\alpha=\sigma_{g_{B}g_{A}^{-1}}} }_{contribute} \label{bms4ps1}
  \end{align}
after the cancellation of several factors explicitly shown in \eqref{bms4ps1} between numerator and denominator in the evaluation of reflected entropy, the final result of vacuum state on the BMS plane turn out to be
  \begin{align}
  S_{Ref;vac}^{\text{BMS}} &=\lim_{m,n \to 1 } \frac{1}{1-n} \ln{ \frac{\left \langle \sigma_{{g}_A}(x_1) \sigma_{{g}^{-1}_A}(x_2) \sigma_{{g}_B}(x_3) \sigma_{{g}^{-1}_B}(x_4) \right \rangle _{\text{BMS}^{\otimes mn}}}{\left ( \left \langle \sigma_{{g}_m}(x_1) \sigma_{{g}^{-1}_m}(x_2) \sigma_{{g}_m}(x_3) \sigma_{{g}^{-1}_m}(x_4) \right \rangle _{\text{BMS}^{\otimes m} }\right)^n}}\notag \\
   & \approx \lim_{m,n \to 1 } \{ -2\ln{C_{mn}^{\text{BMS}}} +\frac{n+1}{n} \left( \frac{c_M}{12}  \frac{t}{(1-x) \sqrt{x}}  + \frac{c_L}{12}  \log{(\frac{1+\sqrt{x}}{1-\sqrt{x}})} \right) \} \notag \\
       &= \frac{c_M}{6}  \frac{t}{(1-x) \sqrt{x}} +\frac{c_L}{6}  \log{(\frac{1+\sqrt{x}}{1-\sqrt{x}})} \label{bmsrefvac}
  \end{align}   
For the thermal state reflected entropy, we need the correlator of four point twist operators on the thermal cylinder using \eqref{bmst2} and \eqref{bmscylinder}, 
\begin{align}
     & \langle \sigma_{{g}_A}(u_1,\phi_1) \sigma_{{g}^{-1}_A}(u_2,\phi_2) \sigma_{{g}_B}(u_3,\phi_3) \sigma_{{g}^{-1}_B} (u_4,\phi_4) \rangle _{\text{BMS}^{\otimes mn}}^{cylinder} \notag \\
         &=e^{\frac{\xi_{g_{A}} \left( \b_{u} \b_{\phi}+ 2 \pi \left( \b_{u} \sum_{j=1}^{4} \phi_{j}+\b_{\phi} \sum_{j=1}^{4} u_{j} \right) \right)  }{\b_{x}^{2}}} \left( \left( \frac{2 \pi}{\b_{\phi}} \right)^{4} e^{\frac{2 \pi \sum_{j=1}^{4} \phi_{j} }{\b_{\phi}}} \right)^{\Delta_{ g_{A}}} \nn \\
         & \times \langle \sigma_{{g}_A}(x_{1},y_{1}) \sigma_{{g}^{-1}_A}(x_{2},y_{2}) \sigma_{{g}_B}(x_{3},y_{3}) \sigma_{{g}^{-1}_B} (x_{4},y_{4}) \rangle _{\text{BMS}^{\otimes mn}}^{plane} \label{factor2}
\end{align}
The first line of \eqref{factor2} would again cancel out between enumerator and denominator, and the second line of \eqref{factor2} contributes to the final answer. Thus the thermal state reflected entropy can be obtained by the same formula \eqref{bmsrefvac}, but with the cross ratios $x,t$ getting from the map \eqref{bmsfiniteT}. Finally we have
\begin{align}
 S_{ref;thermal}^{BMS} &= \frac{c_L}{6}  \log \left(\frac{1+\sqrt{x}}{1-\sqrt{x}}\right)+
       \frac{c_M  t/6}{(1-x)\sqrt{x}} , \\
  x &=\frac{x_{12} x_{34} }{x_{13} x_{24}} \Big\lvert_{x_{i} \to e^{\frac{2 \pi \phi_{i}}{\beta_{\phi}}}}  \label{factor3} , \\
  \frac{t}{x}&= \left( \frac{t_{12}}{x_{12}}+\frac{t_{34}}{x_{34}}-\frac{t_{13}}{x_{13}}-\frac{t_{24}}{x_{24}} \right) \Big\lvert_{ \{ x_{i} \to e^{\frac{2 \pi \phi_{i}}{ \beta_{\phi}}}, t_{i} \to  -\frac{2 \pi }{\beta_{\phi}} e^{\frac{2 \pi \phi_{i}}{\beta_{\phi}}} \left( \phi_{i} \frac{\beta_{u}}{\beta_{\phi}}+u_{i}  \right) \} } \notag 
\end{align}

\section{$M>0$ Zero Mode Background}
\label{appendixb}

In this appendix we work in the $M>0$ zero mode background. After a similar analysis of the bifurcating horizon behavior and the entanglement phase transition in this solution, we will see the conclusions get in the main context from the analysis of the Poincar\'e vacuum are universal.

\subsection{Bifurcating horizon}

The $M>0$ zero mode backgrounds which can be regarded as the flat limit of BTZ black hole of asymptotically AdS$_3$ spacetime \cite{Barnich:2012xq,Bagchi:2012xr} have metric in Bondi coordinates,
\be ds^2=M du^2-2du dr +J du d\phi+r^2 d\phi^2, \quad u \in (-\infty,\infty),\; r\in(0,\infty),\; \phi \in (0,2\pi) \label{M>0} \ee
For the general interval $\cA$ with $\p \cA= \{ (u_l,\phi_l), \; (u_r,\phi_r)   \}$ 
The length $L(r_l, r_r)$ of spacelike geodesic between the two null ropes $r_l, r_r$ satisfying 
\be \g_{l,r}: \quad u=u_{l,r}, \; \; \phi=\phi_{l,r} \ee 
is not illuminating, so we choose to not present it. The extreme of $L(r_l, r_r)$ can be found at 
\be r_{l}=-r_r=\frac{M u_{21}+\sqrt{M} r_c \phi_{21}+r_c \sinh{\left( \sqrt{M} \phi_{12} \right)}  }{ \cosh{ \left( \sqrt{M} \phi_{12} \right)}-1}  \label{rlr} \ee 
where $r_c$ is the Cauchy horizon \eqref{cauhor} and $u_{ij}=u_i-u_j,\phi_{ij}=\phi_i-\phi_j $.  The parameter equations for the Killing horizon $N_{l,r}$ of the bifurcating surface $\g_{\xi}$ are
\begin{align}
   & t=t_l +\kappa (t_l-t_r)+ \lambda \nn \\
   & x=x_l +\kappa (x_l-x_r)+ \tanh{\sqrt{M} \phi_{l,r} } \lambda \label{mg0norl} \\
    & y=y_l +\kappa (y_l-y_r)+ \cosh{\sqrt{M}^{-1} \phi_{l,r} } \lambda \nn 
\end{align} 
where $(t_{l,r},x_{l,r},y_{l,r})$ are the endpoints of the bench $\g$ that can be obtained by using \eqref{rlr} and \eqref{mgztranf}. When $\lambda=0$, \eqref{mg0norl} would reduce to the parametrization of the bifurcating surface $\g_{\xi}$ for $\kappa \in (-\infty, \infty)$ and the bench $\g$ for $\kappa \in (0,1)$. When $\lambda>0$, \eqref{mg0norl} denote two future bifurcating horizons $N_{l,r}$ with two null ropes $\g_{l,r}$ sitting on; while for $\lambda<0$ these equations parametrize the two corresponding past horizons. Similarly these four bifurcating horizons, which together decompose the global flat$_3$ into four non-intersecting causal regions, would converge to four single points on the future/past null infinity separately in the Penrose diagram. Mathematically, we can map boundary in Bondi coordinates $(u, r \to \infty, \phi )$ to boundary in Penrose coordinates $(U,V,\Phi)$ using \eqref{flatTXY}:
\begin{small}
\begin{align}
    & \sqrt{x^2+y^2}\lvert_{r\to \pm \infty}=\frac{\cosh{(\sqrt{M}\phi) }}{ \sqrt{M}} |r|-\frac{\sqrt{M} }{ \cosh{(\sqrt{M}\phi) } }\left( u+ \frac{J \phi}{2M}+\frac{J \cosh{(\sqrt{M}\phi) }  \sinh{(\sqrt{M}\phi )} }{2 M^{3/2}} \right)\frac{|r|}{r} \nn \\
   & \text{when}\; r\to \infty, 
   \quad U=\arctan{\left(\left( u+\frac{J \phi }{2M}  \right)\frac{\sqrt{M} }{ \cosh{(\sqrt{M}\phi )} }  \right)} , \; V= \frac{\pi}{2}, \; \Phi=\arccos{\left( \frac{\sinh{\phi}}{ \cosh{\phi} } \right)} \label{m00map}
\end{align}
\end{small}

\subsection{Entanglement phase transition}

We consider thermal state entanglement phase transition of BMS field theory using formula \eqref{eethermal} with $c_L=0$ for the same configurations in Figure \ref{fig:threeways}.
The boundary intervals are 
\be
    \partial A=\{(u_1,\phi_1),(u_2,\phi_2)\},\quad \partial B=\{(u_3,\phi_3),(u_4,\phi_4)\}
\ee
Similarly we can define 
\begin{equation}
    \begin{aligned}
&S_1=\sqrt{M}\Big[\Big(u_{23}+\frac{J\phi_{23}}{2M}\Big)\coth\Big(\frac{\sqrt{M}\phi_{23}}{2}\Big)+\Big(u_{14}+\frac{J\phi_{14}}{2M}\Big)\coth\Big(\frac{\sqrt{M}\phi_{14}}{2}\Big)\Big]-\frac{2J}{M}\\    
&S_2=\sqrt{M}\Big[\Big(u_{12}+\frac{J\phi_{12}}{2M}\Big)\coth\Big(\frac{\sqrt{M}\phi_{12}}{2}\Big)+\Big(u_{34}+\frac{J\phi_{34}}{2M}\Big)\coth\Big(\frac{\sqrt{M}\phi_{34}}{2}\Big)\Big]-\frac{2J}{M}\\
&S_3=\sqrt{M}\Big[\Big(u_{13}+\frac{J\phi_{13}}{2M}\Big)\coth\Big(\frac{\sqrt{M}\phi_{13}}{2}\Big)+\Big(u_{24}+\frac{J\phi_{24}}{2M}\Big)\coth\Big(\frac{\sqrt{M}\phi_{24}}{2}\Big)\Big]-\frac{2J}{M}
    \end{aligned}
\end{equation}
The difference between them are
\be
    S_1-S_2= \frac{u}{ \phi (\phi-1)}, \quad
    S_2-S_3= \frac{u}{\phi} \quad
    S_1-S_3= \frac{u}{\phi-1}
\ee
where $u,\phi$ are finite temperature cross ratios. Therefore we get
\begin{itemize}
    \item $S_1$ is the minimal one when:
    \be  u<0,\quad \phi>1 \quad \mathrm{or}\quad u>0,\quad 0<\phi<1  \ee
    \item $S_2$ is the minimal one when:
   \be u<0,\quad 0<\phi<1  \ee
    \item $S_3$ is the minimal one when:
    \be  u<0,\quad \phi<0 \quad \mathrm{or}\quad u>0,\quad \phi>1  \ee    
\end{itemize}
Taking symmetric configurations,  
\be
    u_a=-u_1=u_2,\;\; \phi_a=-\phi_1=\phi_2,\;\; u_b=-u_3=u_4,\;\;\phi_b=-\phi_3=\phi_4
\ee
with $\phi_b>\phi_a$. This configuration ensures that $0<\phi<1$, which means that $S_3$ can never be the minimal one. Fixing $\phi_a,\phi_b$ and $u_a$, there is a critical value
\be  u_c= \frac{(2Mu_a+J\phi_a)\sinh(\sqrt{M}\phi_b)\sinh^{-1}(\sqrt{M}\phi_a)-J\phi_b}{2M} \ee 
for $u_b$ such that $S_1=S_2$. When $u_b<u_c$, $S_2$ is the minimal one and vice verse. we can check the difference between the slope of the critical interval $B$ with $u_b=u_c$ and that of interval $A$ is
\be
    \frac{u_c}{\phi_b}-\frac{u_a}{\phi_a}=\frac{(2Mu_a+J\phi_a)(\phi_a\sinh(\sqrt{M}\phi_b)-\phi_b\sinh(\sqrt{M}\phi_a))}{2M\phi_a\phi_b\sinh(\sqrt{M}\phi_a)} \label{phaseT}
\ee
Since $\phi_b>\phi_a$, we always have $\phi_b\sinh(\sqrt{M}\phi_a)-\phi_a\sinh(\sqrt{M}\phi_b)<0$. Therefore, the sign of \eqref{phaseT} only depends on the sign of $2Mu_a+J\phi_a$.

\acknowledgments

We thank Bin Chen, Wei Song, Qiang Wen, Boyang Yu and Luis Apolo for useful discussions. The work is in part supported by NSFC Grant  No. 12275004, 11735001.

\bibliographystyle{JHEP}
\bibliography{main}

\providecommand{\href}[2]{#2}\begingroup\raggedright\begin{thebibliography}{10}

\bibitem{Maldacena:1997re}
J.M.~Maldacena, \emph{{The Large N limit of superconformal field theories and
  supergravity}}, \href{https://doi.org/10.4310/ATMP.1998.v2.n2.a1}{\emph{Adv.
  Theor. Math. Phys.} {\bfseries 2} (1998) 231}
  [\href{https://arxiv.org/abs/hep-th/9711200}{{\ttfamily hep-th/9711200}}].

\bibitem{Witten:1998qj}
E.~Witten, \emph{{Anti-de Sitter space and holography}},
  \href{https://doi.org/10.4310/ATMP.1998.v2.n2.a2}{\emph{Adv. Theor. Math.
  Phys.} {\bfseries 2} (1998) 253}
  [\href{https://arxiv.org/abs/hep-th/9802150}{{\ttfamily hep-th/9802150}}].

\bibitem{Gubser:1998bc}
S.S.~Gubser, I.R.~Klebanov and A.M.~Polyakov, \emph{{Gauge theory correlators
  from noncritical string theory}},
  \href{https://doi.org/10.1016/S0370-2693(98)00377-3}{\emph{Phys. Lett. B}
  {\bfseries 428} (1998) 105}
  [\href{https://arxiv.org/abs/hep-th/9802109}{{\ttfamily hep-th/9802109}}].

\bibitem{Ryu:2006bv}
S.~Ryu and T.~Takayanagi, \emph{{Holographic derivation of entanglement entropy
  from AdS/CFT}},
  \href{https://doi.org/10.1103/PhysRevLett.96.181602}{\emph{Phys. Rev. Lett.}
  {\bfseries 96} (2006) 181602}
  [\href{https://arxiv.org/abs/hep-th/0603001}{{\ttfamily hep-th/0603001}}].

\bibitem{Hubeny:2007xt}
V.E.~Hubeny, M.~Rangamani and T.~Takayanagi, \emph{{A Covariant holographic
  entanglement entropy proposal}},
  \href{https://doi.org/10.1088/1126-6708/2007/07/062}{\emph{JHEP} {\bfseries
  07} (2007) 062} [\href{https://arxiv.org/abs/0705.0016}{{\ttfamily
  0705.0016}}].

\bibitem{Nishioka:2009un}
T.~Nishioka, S.~Ryu and T.~Takayanagi, \emph{{Holographic Entanglement Entropy:
  An Overview}}, \href{https://doi.org/10.1088/1751-8113/42/50/504008}{\emph{J.
  Phys. A} {\bfseries 42} (2009) 504008}
  [\href{https://arxiv.org/abs/0905.0932}{{\ttfamily 0905.0932}}].

\bibitem{Lewkowycz:2013nqa}
A.~Lewkowycz and J.~Maldacena, \emph{{Generalized gravitational entropy}},
  \href{https://doi.org/10.1007/JHEP08(2013)090}{\emph{JHEP} {\bfseries 08}
  (2013) 090} [\href{https://arxiv.org/abs/1304.4926}{{\ttfamily 1304.4926}}].

\bibitem{Harlow:2018fse}
D.~Harlow, \emph{{TASI Lectures on the Emergence of Bulk Physics in AdS/CFT}},
  \href{https://doi.org/10.22323/1.305.0002}{\emph{PoS} {\bfseries TASI2017}
  (2018) 002} [\href{https://arxiv.org/abs/1802.01040}{{\ttfamily
  1802.01040}}].

\bibitem{Almheiri:2020cfm}
A.~Almheiri, T.~Hartman, J.~Maldacena, E.~Shaghoulian and A.~Tajdini,
  \emph{{The entropy of Hawking radiation}},
  \href{https://doi.org/10.1103/RevModPhys.93.035002}{\emph{Rev. Mod. Phys.}
  {\bfseries 93} (2021) 035002}
  [\href{https://arxiv.org/abs/2006.06872}{{\ttfamily 2006.06872}}].

\bibitem{Bousso:2022ntt}
R.~Bousso, X.~Dong, N.~Engelhardt, T.~Faulkner, T.~Hartman, S.H.~Shenker
  et~al., \emph{{Snowmass White Paper: Quantum Aspects of Black Holes and the
  Emergence of Spacetime}},  \href{https://arxiv.org/abs/2201.03096}{{\ttfamily
  2201.03096}}.

\bibitem{Dutta:2019gen}
S.~Dutta and T.~Faulkner, \emph{{A canonical purification for the entanglement
  wedge cross-section}},
  \href{https://doi.org/10.1007/JHEP03(2021)178}{\emph{JHEP} {\bfseries 03}
  (2021) 178} [\href{https://arxiv.org/abs/1905.00577}{{\ttfamily
  1905.00577}}].

\bibitem{Kusuki:2019zsp}
Y.~Kusuki, J.~Kudler-Flam and S.~Ryu, \emph{{Derivation of holographic
  negativity in AdS$_3$/CFT$_2$}},
  \href{https://doi.org/10.1103/PhysRevLett.123.131603}{\emph{Phys. Rev. Lett.}
  {\bfseries 123} (2019) 131603}
  [\href{https://arxiv.org/abs/1907.07824}{{\ttfamily 1907.07824}}].

\bibitem{Wen:2021qgx}
Q.~Wen, \emph{{Balanced Partial Entanglement and the Entanglement Wedge Cross
  Section}}, \href{https://doi.org/10.1007/JHEP04(2021)301}{\emph{JHEP}
  {\bfseries 04} (2021) 301}
  [\href{https://arxiv.org/abs/2103.00415}{{\ttfamily 2103.00415}}].

\bibitem{Faulkner:2013ana}
T.~Faulkner, A.~Lewkowycz and J.~Maldacena, \emph{{Quantum corrections to
  holographic entanglement entropy}},
  \href{https://doi.org/10.1007/JHEP11(2013)074}{\emph{JHEP} {\bfseries 11}
  (2013) 074} [\href{https://arxiv.org/abs/1307.2892}{{\ttfamily 1307.2892}}].

\bibitem{Jafferis:2015del}
D.L.~Jafferis, A.~Lewkowycz, J.~Maldacena and S.J.~Suh, \emph{{Relative entropy
  equals bulk relative entropy}},
  \href{https://doi.org/10.1007/JHEP06(2016)004}{\emph{JHEP} {\bfseries 06}
  (2016) 004} [\href{https://arxiv.org/abs/1512.06431}{{\ttfamily
  1512.06431}}].

\bibitem{Dong:2016eik}
X.~Dong, D.~Harlow and A.C.~Wall, \emph{{Reconstruction of Bulk Operators
  within the Entanglement Wedge in Gauge-Gravity Duality}},
  \href{https://doi.org/10.1103/PhysRevLett.117.021601}{\emph{Phys. Rev. Lett.}
  {\bfseries 117} (2016) 021601}
  [\href{https://arxiv.org/abs/1601.05416}{{\ttfamily 1601.05416}}].

\bibitem{Faulkner:2017vdd}
T.~Faulkner and A.~Lewkowycz, \emph{{Bulk locality from modular flow}},
  \href{https://doi.org/10.1007/JHEP07(2017)151}{\emph{JHEP} {\bfseries 07}
  (2017) 151} [\href{https://arxiv.org/abs/1704.05464}{{\ttfamily
  1704.05464}}].

\bibitem{Witten:2018zxz}
E.~Witten, \emph{{APS Medal for Exceptional Achievement in Research: Invited
  article on entanglement properties of quantum field theory}},
  \href{https://doi.org/10.1103/RevModPhys.90.045003}{\emph{Rev. Mod. Phys.}
  {\bfseries 90} (2018) 045003}
  [\href{https://arxiv.org/abs/1803.04993}{{\ttfamily 1803.04993}}].

\bibitem{Faulkner:2018faa}
T.~Faulkner, M.~Li and H.~Wang, \emph{{A modular toolkit for bulk
  reconstruction}}, \href{https://doi.org/10.1007/JHEP04(2019)119}{\emph{JHEP}
  {\bfseries 04} (2019) 119}
  [\href{https://arxiv.org/abs/1806.10560}{{\ttfamily 1806.10560}}].

\bibitem{Headrick:2014cta}
M.~Headrick, V.E.~Hubeny, A.~Lawrence and M.~Rangamani, \emph{{Causality \&
  holographic entanglement entropy}},
  \href{https://doi.org/10.1007/JHEP12(2014)162}{\emph{JHEP} {\bfseries 12}
  (2014) 162} [\href{https://arxiv.org/abs/1408.6300}{{\ttfamily 1408.6300}}].

\bibitem{Susskind:1998vk}
L.~Susskind, \emph{{Holography in the flat space limit}},
  \href{https://doi.org/10.1063/1.1301570}{\emph{AIP Conf. Proc.} {\bfseries
  493} (1999) 98} [\href{https://arxiv.org/abs/hep-th/9901079}{{\ttfamily
  hep-th/9901079}}].

\bibitem{Polchinski:1999ry}
J.~Polchinski, \emph{{S matrices from AdS space-time}},
  \href{https://arxiv.org/abs/hep-th/9901076}{{\ttfamily hep-th/9901076}}.

\bibitem{Giddings:1999jq}
S.B.~Giddings, \emph{{Flat space scattering and bulk locality in the AdS / CFT
  correspondence}},
  \href{https://doi.org/10.1103/PhysRevD.61.106008}{\emph{Phys. Rev. D}
  {\bfseries 61} (2000) 106008}
  [\href{https://arxiv.org/abs/hep-th/9907129}{{\ttfamily hep-th/9907129}}].

\bibitem{Arcioni:2003td}
G.~Arcioni and C.~Dappiaggi, \emph{{Holography in asymptotically flat
  space-times and the BMS group}},
  \href{https://doi.org/10.1088/0264-9381/21/23/022}{\emph{Class. Quant. Grav.}
  {\bfseries 21} (2004) 5655}
  [\href{https://arxiv.org/abs/hep-th/0312186}{{\ttfamily hep-th/0312186}}].

\bibitem{deBoer:2003vf}
J.~de~Boer and S.N.~Solodukhin, \emph{{A Holographic reduction of Minkowski
  space-time}},
  \href{https://doi.org/10.1016/S0550-3213(03)00494-2}{\emph{Nucl. Phys. B}
  {\bfseries 665} (2003) 545}
  [\href{https://arxiv.org/abs/hep-th/0303006}{{\ttfamily hep-th/0303006}}].

\bibitem{Strominger:2017zoo}
A.~Strominger, \emph{{Lectures on the Infrared Structure of Gravity and Gauge
  Theory}},  \href{https://arxiv.org/abs/1703.05448}{{\ttfamily 1703.05448}}.

\bibitem{Pasterski:2021raf}
S.~Pasterski, M.~Pate and A.-M.~Raclariu, \emph{{Celestial Holography}},  in
  \emph{{Snowmass 2021}}, 11, 2021
  [\href{https://arxiv.org/abs/2111.11392}{{\ttfamily 2111.11392}}].

\bibitem{He:2014laa}
T.~He, V.~Lysov, P.~Mitra and A.~Strominger, \emph{{BMS supertranslations and
  Weinberg\textquoteright{}s soft graviton theorem}},
  \href{https://doi.org/10.1007/JHEP05(2015)151}{\emph{JHEP} {\bfseries 05}
  (2015) 151} [\href{https://arxiv.org/abs/1401.7026}{{\ttfamily 1401.7026}}].

\bibitem{He:2015zea}
T.~He, P.~Mitra and A.~Strominger, \emph{{2D Kac-Moody Symmetry of 4D
  Yang-Mills Theory}},
  \href{https://doi.org/10.1007/JHEP10(2016)137}{\emph{JHEP} {\bfseries 10}
  (2016) 137} [\href{https://arxiv.org/abs/1503.02663}{{\ttfamily
  1503.02663}}].

\bibitem{Cheung:2016iub}
C.~Cheung, A.~de~la Fuente and R.~Sundrum, \emph{{4D scattering amplitudes and
  asymptotic symmetries from 2D CFT}},
  \href{https://doi.org/10.1007/JHEP01(2017)112}{\emph{JHEP} {\bfseries 01}
  (2017) 112} [\href{https://arxiv.org/abs/1609.00732}{{\ttfamily
  1609.00732}}].

\bibitem{Pasterski:2016qvg}
S.~Pasterski, S.-H.~Shao and A.~Strominger, \emph{{Flat Space Amplitudes and
  Conformal Symmetry of the Celestial Sphere}},
  \href{https://doi.org/10.1103/PhysRevD.96.065026}{\emph{Phys. Rev. D}
  {\bfseries 96} (2017) 065026}
  [\href{https://arxiv.org/abs/1701.00049}{{\ttfamily 1701.00049}}].

\bibitem{Pasterski:2017kqt}
S.~Pasterski and S.-H.~Shao, \emph{{Conformal basis for flat space
  amplitudes}}, \href{https://doi.org/10.1103/PhysRevD.96.065022}{\emph{Phys.
  Rev. D} {\bfseries 96} (2017) 065022}
  [\href{https://arxiv.org/abs/1705.01027}{{\ttfamily 1705.01027}}].

\bibitem{Fotopoulos:2019tpe}
A.~Fotopoulos and T.R.~Taylor, \emph{{Primary Fields in Celestial CFT}},
  \href{https://doi.org/10.1007/JHEP10(2019)167}{\emph{JHEP} {\bfseries 10}
  (2019) 167} [\href{https://arxiv.org/abs/1906.10149}{{\ttfamily
  1906.10149}}].

\bibitem{Pate:2019lpp}
M.~Pate, A.-M.~Raclariu, A.~Strominger and E.Y.~Yuan, \emph{{Celestial operator
  products of gluons and gravitons}},
  \href{https://doi.org/10.1142/S0129055X21400031}{\emph{Rev. Math. Phys.}
  {\bfseries 33} (2021) 2140003}
  [\href{https://arxiv.org/abs/1910.07424}{{\ttfamily 1910.07424}}].

\bibitem{Banerjee:2020kaa}
S.~Banerjee, S.~Ghosh and R.~Gonzo, \emph{{BMS symmetry of celestial OPE}},
  \href{https://doi.org/10.1007/JHEP04(2020)130}{\emph{JHEP} {\bfseries 04}
  (2020) 130} [\href{https://arxiv.org/abs/2002.00975}{{\ttfamily
  2002.00975}}].

\bibitem{Costello:2022upu}
K.~Costello and N.M.~Paquette, \emph{{Associativity of One-Loop Corrections to
  the Celestial Operator Product Expansion}},
  \href{https://doi.org/10.1103/PhysRevLett.129.231604}{\emph{Phys. Rev. Lett.}
  {\bfseries 129} (2022) 231604}
  [\href{https://arxiv.org/abs/2204.05301}{{\ttfamily 2204.05301}}].

\bibitem{Hu:2022bpa}
Y.~Hu and S.~Pasterski, \emph{{Celestial recursion}},
  \href{https://doi.org/10.1007/JHEP01(2023)151}{\emph{JHEP} {\bfseries 01}
  (2023) 151} [\href{https://arxiv.org/abs/2208.11635}{{\ttfamily
  2208.11635}}].

\bibitem{Atanasov:2021cje}
A.~Atanasov, W.~Melton, A.-M.~Raclariu and A.~Strominger, \emph{{Conformal
  block expansion in celestial CFT}},
  \href{https://doi.org/10.1103/PhysRevD.104.126033}{\emph{Phys. Rev. D}
  {\bfseries 104} (2021) 126033}
  [\href{https://arxiv.org/abs/2104.13432}{{\ttfamily 2104.13432}}].

\bibitem{Fan:2021isc}
W.~Fan, A.~Fotopoulos, S.~Stieberger, T.R.~Taylor and B.~Zhu, \emph{{Conformal
  blocks from celestial gluon amplitudes}},
  \href{https://doi.org/10.1007/JHEP05(2021)170}{\emph{JHEP} {\bfseries 05}
  (2021) 170} [\href{https://arxiv.org/abs/2103.04420}{{\ttfamily
  2103.04420}}].

\bibitem{Mizera:2022sln}
S.~Mizera and S.~Pasterski, \emph{{Celestial geometry}},
  \href{https://doi.org/10.1007/JHEP09(2022)045}{\emph{JHEP} {\bfseries 09}
  (2022) 045} [\href{https://arxiv.org/abs/2204.02505}{{\ttfamily
  2204.02505}}].

\bibitem{Strominger:2021lvk}
A.~Strominger, \emph{{w(1+infinity) and the Celestial Sphere}},
  \href{https://arxiv.org/abs/2105.14346}{{\ttfamily 2105.14346}}.

\bibitem{Almheiri:2014lwa}
A.~Almheiri, X.~Dong and D.~Harlow, \emph{{Bulk Locality and Quantum Error
  Correction in AdS/CFT}},
  \href{https://doi.org/10.1007/JHEP04(2015)163}{\emph{JHEP} {\bfseries 04}
  (2015) 163} [\href{https://arxiv.org/abs/1411.7041}{{\ttfamily 1411.7041}}].

\bibitem{Barnich:2011mi}
G.~Barnich and C.~Troessaert, \emph{{BMS charge algebra}},
  \href{https://doi.org/10.1007/JHEP12(2011)105}{\emph{JHEP} {\bfseries 12}
  (2011) 105} [\href{https://arxiv.org/abs/1106.0213}{{\ttfamily 1106.0213}}].

\bibitem{Trautman:1958zdi}
A.~Trautman, \emph{{Radiation and Boundary Conditions in the Theory of
  Gravitation}}, {\emph{Bull. Acad. Pol. Sci. Ser. Sci. Math. Astron. Phys.}
  {\bfseries 6} (1958) 407} [\href{https://arxiv.org/abs/1604.03145}{{\ttfamily
  1604.03145}}].

\bibitem{Donnay:2022wvx}
L.~Donnay, A.~Fiorucci, Y.~Herfray and R.~Ruzziconi, \emph{{Bridging Carrollian
  and celestial holography}},
  \href{https://doi.org/10.1103/PhysRevD.107.126027}{\emph{Phys. Rev. D}
  {\bfseries 107} (2023) 126027}
  [\href{https://arxiv.org/abs/2212.12553}{{\ttfamily 2212.12553}}].

\bibitem{Laddha:2020kvp}
A.~Laddha, S.G.~Prabhu, S.~Raju and P.~Shrivastava, \emph{{The Holographic
  Nature of Null Infinity}},
  \href{https://doi.org/10.21468/SciPostPhys.10.2.041}{\emph{SciPost Phys.}
  {\bfseries 10} (2021) 041}
  [\href{https://arxiv.org/abs/2002.02448}{{\ttfamily 2002.02448}}].

\bibitem{Bagchi:2022emh}
A.~Bagchi, S.~Banerjee, R.~Basu and S.~Dutta, \emph{{Scattering Amplitudes:
  Celestial and Carrollian}},
  \href{https://doi.org/10.1103/PhysRevLett.128.241601}{\emph{Phys. Rev. Lett.}
  {\bfseries 128} (2022) 241601}
  [\href{https://arxiv.org/abs/2202.08438}{{\ttfamily 2202.08438}}].

\bibitem{Donnay:2022aba}
L.~Donnay, A.~Fiorucci, Y.~Herfray and R.~Ruzziconi, \emph{{Carrollian
  Perspective on Celestial Holography}},
  \href{https://doi.org/10.1103/PhysRevLett.129.071602}{\emph{Phys. Rev. Lett.}
  {\bfseries 129} (2022) 071602}
  [\href{https://arxiv.org/abs/2202.04702}{{\ttfamily 2202.04702}}].

\bibitem{Penedones:2010ue}
J.~Penedones, \emph{{Writing CFT correlation functions as AdS scattering
  amplitudes}}, \href{https://doi.org/10.1007/JHEP03(2011)025}{\emph{JHEP}
  {\bfseries 03} (2011) 025} [\href{https://arxiv.org/abs/1011.1485}{{\ttfamily
  1011.1485}}].

\bibitem{Fitzpatrick:2011ia}
A.L.~Fitzpatrick, J.~Kaplan, J.~Penedones, S.~Raju and B.C.~van Rees, \emph{{A
  Natural Language for AdS/CFT Correlators}},
  \href{https://doi.org/10.1007/JHEP11(2011)095}{\emph{JHEP} {\bfseries 11}
  (2011) 095} [\href{https://arxiv.org/abs/1107.1499}{{\ttfamily 1107.1499}}].

\bibitem{deGioia:2022fcn}
L.P.~de~Gioia and A.-M.~Raclariu, \emph{{Eikonal approximation in celestial
  CFT}}, \href{https://doi.org/10.1007/JHEP03(2023)030}{\emph{JHEP} {\bfseries
  03} (2023) 030} [\href{https://arxiv.org/abs/2206.10547}{{\ttfamily
  2206.10547}}].

\bibitem{Bagchi:2010zz}
A.~Bagchi, \emph{{Correspondence between Asymptotically Flat Spacetimes and
  Nonrelativistic Conformal Field Theories}},
  \href{https://doi.org/10.1103/PhysRevLett.105.171601}{\emph{Phys. Rev. Lett.}
  {\bfseries 105} (2010) 171601}
  [\href{https://arxiv.org/abs/1006.3354}{{\ttfamily 1006.3354}}].

\bibitem{Bagchi:2012cy}
A.~Bagchi and R.~Fareghbal, \emph{{BMS/GCA Redux: Towards Flatspace Holography
  from Non-Relativistic Symmetries}},
  \href{https://doi.org/10.1007/JHEP10(2012)092}{\emph{JHEP} {\bfseries 10}
  (2012) 092} [\href{https://arxiv.org/abs/1203.5795}{{\ttfamily 1203.5795}}].

\bibitem{Bondi:1962px}
H.~Bondi, M.G.J.~van~der Burg and A.W.K.~Metzner, \emph{{Gravitational waves in
  general relativity. 7. Waves from axisymmetric isolated systems}},
  \href{https://doi.org/10.1098/rspa.1962.0161}{\emph{Proc. Roy. Soc. Lond. A}
  {\bfseries 269} (1962) 21}.

\bibitem{Sachs:1962wk}
R.K.~Sachs, \emph{{Gravitational waves in general relativity. 8. Waves in
  asymptotically flat space-times}},
  \href{https://doi.org/10.1098/rspa.1962.0206}{\emph{Proc. Roy. Soc. Lond. A}
  {\bfseries 270} (1962) 103}.

\bibitem{Barnich:2012xq}
G.~Barnich, \emph{{Entropy of three-dimensional asymptotically flat
  cosmological solutions}},
  \href{https://doi.org/10.1007/JHEP10(2012)095}{\emph{JHEP} {\bfseries 10}
  (2012) 095} [\href{https://arxiv.org/abs/1208.4371}{{\ttfamily 1208.4371}}].

\bibitem{Bagchi:2012xr}
A.~Bagchi, S.~Detournay, R.~Fareghbal and J.~Sim\'on, \emph{{Holography of 3D
  Flat Cosmological Horizons}},
  \href{https://doi.org/10.1103/PhysRevLett.110.141302}{\emph{Phys. Rev. Lett.}
  {\bfseries 110} (2013) 141302}
  [\href{https://arxiv.org/abs/1208.4372}{{\ttfamily 1208.4372}}].

\bibitem{Barnich:2015mui}
G.~Barnich, H.A.~Gonzalez, A.~Maloney and B.~Oblak, \emph{{One-loop partition
  function of three-dimensional flat gravity}},
  \href{https://doi.org/10.1007/JHEP04(2015)178}{\emph{JHEP} {\bfseries 04}
  (2015) 178} [\href{https://arxiv.org/abs/1502.06185}{{\ttfamily
  1502.06185}}].

\bibitem{Hijano:2018nhq}
E.~Hijano, \emph{{Semi-classical BMS$_{3}$ blocks and flat holography}},
  \href{https://doi.org/10.1007/JHEP10(2018)044}{\emph{JHEP} {\bfseries 10}
  (2018) 044} [\href{https://arxiv.org/abs/1805.00949}{{\ttfamily
  1805.00949}}].

\bibitem{Jiang:2017ecm}
H.~Jiang, W.~Song and Q.~Wen, \emph{{Entanglement Entropy in Flat Holography}},
  \href{https://doi.org/10.1007/JHEP07(2017)142}{\emph{JHEP} {\bfseries 07}
  (2017) 142} [\href{https://arxiv.org/abs/1706.07552}{{\ttfamily
  1706.07552}}].

\bibitem{Apolo:2020bld}
L.~Apolo, H.~Jiang, W.~Song and Y.~Zhong, \emph{{Swing surfaces and holographic
  entanglement beyond AdS/CFT}},
  \href{https://doi.org/10.1007/JHEP12(2020)064}{\emph{JHEP} {\bfseries 12}
  (2020) 064} [\href{https://arxiv.org/abs/2006.10740}{{\ttfamily
  2006.10740}}].

\bibitem{Apolo:2020qjm}
L.~Apolo, H.~Jiang, W.~Song and Y.~Zhong, \emph{{Modular Hamiltonians in flat
  holography and (W)AdS/WCFT}},
  \href{https://doi.org/10.1007/JHEP09(2020)033}{\emph{JHEP} {\bfseries 09}
  (2020) 033} [\href{https://arxiv.org/abs/2006.10741}{{\ttfamily
  2006.10741}}].

\bibitem{Bagchi:2016bcd}
A.~Bagchi, R.~Basu, A.~Kakkar and A.~Mehra, \emph{{Flat Holography: Aspects of
  the dual field theory}},
  \href{https://doi.org/10.1007/JHEP12(2016)147}{\emph{JHEP} {\bfseries 12}
  (2016) 147} [\href{https://arxiv.org/abs/1609.06203}{{\ttfamily
  1609.06203}}].

\bibitem{Basu:2021awn}
D.~Basu, A.~Chandra, V.~Raj and G.~Sengupta, \emph{{Entanglement wedge in flat
  holography and entanglement negativity}},
  \href{https://doi.org/10.21468/SciPostPhysCore.5.1.013}{\emph{SciPost Phys.
  Core} {\bfseries 5} (2022) 013}
  [\href{https://arxiv.org/abs/2106.14896}{{\ttfamily 2106.14896}}].

\bibitem{Basak:2022cjs}
J.K.~Basak, H.~Chourasiya, V.~Raj and G.~Sengupta, \emph{{Reflected entropy in
  Galilean conformal field theories and flat holography}},
  \href{https://doi.org/10.1140/epjc/s10052-022-11129-8}{\emph{Eur. Phys. J. C}
  {\bfseries 82} (2022) 1169}
  [\href{https://arxiv.org/abs/2202.01201}{{\ttfamily 2202.01201}}].

\bibitem{Camargo:2022mme}
H.A.~Camargo, P.~Nandy, Q.~Wen and H.~Zhong, \emph{{Balanced partial
  entanglement and mixed state correlations}},
  \href{https://doi.org/10.21468/SciPostPhys.12.4.137}{\emph{SciPost Phys.}
  {\bfseries 12} (2022) 137}
  [\href{https://arxiv.org/abs/2201.13362}{{\ttfamily 2201.13362}}].

\bibitem{Wen:2018whg}
Q.~Wen, \emph{{Fine structure in holographic entanglement and entanglement
  contour}}, \href{https://doi.org/10.1103/PhysRevD.98.106004}{\emph{Phys. Rev.
  D} {\bfseries 98} (2018) 106004}
  [\href{https://arxiv.org/abs/1803.05552}{{\ttfamily 1803.05552}}].

\bibitem{Basu:2022nyl}
D.~Basu, \emph{{Balanced Partial Entanglement in Flat Holography}},
  \href{https://arxiv.org/abs/2203.05491}{{\ttfamily 2203.05491}}.

\bibitem{Hao:2021urq}
P.-x.~Hao, W.~Song, X.~Xie and Y.~Zhong, \emph{{A BMS-invariant free scalar
  model}},  \href{https://arxiv.org/abs/2111.04701}{{\ttfamily 2111.04701}}.

\bibitem{Bagchi:2014iea}
A.~Bagchi, R.~Basu, D.~Grumiller and M.~Riegler, \emph{{Entanglement entropy in
  Galilean conformal field theories and flat holography}},
  \href{https://doi.org/10.1103/PhysRevLett.114.111602}{\emph{Phys. Rev. Lett.}
  {\bfseries 114} (2015) 111602}
  [\href{https://arxiv.org/abs/1410.4089}{{\ttfamily 1410.4089}}].

\bibitem{Calabrese:2009qy}
P.~Calabrese and J.~Cardy, \emph{{Entanglement entropy and conformal field
  theory}}, \href{https://doi.org/10.1088/1751-8113/42/50/504005}{\emph{J.
  Phys. A} {\bfseries 42} (2009) 504005}
  [\href{https://arxiv.org/abs/0905.4013}{{\ttfamily 0905.4013}}].

\bibitem{Chen:2022fte}
B.~Chen, Y.~Liu and B.~Yu, \emph{{Reflected entropy in AdS$_{3}$/WCFT}},
  \href{https://doi.org/10.1007/JHEP12(2022)008}{\emph{JHEP} {\bfseries 12}
  (2022) 008} [\href{https://arxiv.org/abs/2205.05582}{{\ttfamily
  2205.05582}}].

\bibitem{Barnich:2010eb}
G.~Barnich and C.~Troessaert, \emph{{Aspects of the BMS/CFT correspondence}},
  \href{https://doi.org/10.1007/JHEP05(2010)062}{\emph{JHEP} {\bfseries 05}
  (2010) 062} [\href{https://arxiv.org/abs/1001.1541}{{\ttfamily 1001.1541}}].

\bibitem{Barnich:2012aw}
G.~Barnich, A.~Gomberoff and H.A.~Gonzalez, \emph{{The Flat limit of three
  dimensional asymptotically anti-de Sitter spacetimes}},
  \href{https://doi.org/10.1103/PhysRevD.86.024020}{\emph{Phys. Rev. D}
  {\bfseries 86} (2012) 024020}
  [\href{https://arxiv.org/abs/1204.3288}{{\ttfamily 1204.3288}}].

\bibitem{Faulkner:2013ica}
T.~Faulkner, M.~Guica, T.~Hartman, R.C.~Myers and M.~Van~Raamsdonk,
  \emph{{Gravitation from Entanglement in Holographic CFTs}},
  \href{https://doi.org/10.1007/JHEP03(2014)051}{\emph{JHEP} {\bfseries 03}
  (2014) 051} [\href{https://arxiv.org/abs/1312.7856}{{\ttfamily 1312.7856}}].

\bibitem{Casini:2011kv}
H.~Casini, M.~Huerta and R.C.~Myers, \emph{{Towards a derivation of holographic
  entanglement entropy}},
  \href{https://doi.org/10.1007/JHEP05(2011)036}{\emph{JHEP} {\bfseries 05}
  (2011) 036} [\href{https://arxiv.org/abs/1102.0440}{{\ttfamily 1102.0440}}].

\bibitem{Casini:2017vbe}
H.~Casini, E.~Test\'e and G.~Torroba, \emph{{Markov Property of the Conformal
  Field Theory Vacuum and the a Theorem}},
  \href{https://doi.org/10.1103/PhysRevLett.118.261602}{\emph{Phys. Rev. Lett.}
  {\bfseries 118} (2017) 261602}
  [\href{https://arxiv.org/abs/1704.01870}{{\ttfamily 1704.01870}}].

\bibitem{Akers:2017ttv}
C.~Akers, V.~Chandrasekaran, S.~Leichenauer, A.~Levine and
  A.~Shahbazi~Moghaddam, \emph{{Quantum null energy condition, entanglement
  wedge nesting, and quantum focusing}},
  \href{https://doi.org/10.1103/PhysRevD.101.025011}{\emph{Phys. Rev. D}
  {\bfseries 101} (2020) 025011}
  [\href{https://arxiv.org/abs/1706.04183}{{\ttfamily 1706.04183}}].

\bibitem{Castro:2014tta}
A.~Castro, S.~Detournay, N.~Iqbal and E.~Perlmutter, \emph{{Holographic
  entanglement entropy and gravitational anomalies}},
  \href{https://doi.org/10.1007/JHEP07(2014)114}{\emph{JHEP} {\bfseries 07}
  (2014) 114} [\href{https://arxiv.org/abs/1405.2792}{{\ttfamily 1405.2792}}].

\bibitem{Maldacena:2001kr}
J.M.~Maldacena, \emph{{Eternal black holes in anti-de Sitter}},
  \href{https://doi.org/10.1088/1126-6708/2003/04/021}{\emph{JHEP} {\bfseries
  04} (2003) 021} [\href{https://arxiv.org/abs/hep-th/0106112}{{\ttfamily
  hep-th/0106112}}].

\end{thebibliography}\endgroup
\end{document}